\documentclass{article}

\usepackage{ML}

\usepackage[utf8]{inputenc} 
\usepackage[T1]{fontenc}    
\usepackage{hyperref}       
\usepackage{url}            
\usepackage{booktabs}       
\usepackage{amsfonts}       
\usepackage{nicefrac}       
\usepackage{microtype}      
\usepackage{graphicx}
\usepackage[square,sort,comma,numbers]{natbib}
\usepackage{doi}
\usepackage{subcaption}
\usepackage{amssymb, amsmath, amsthm}
\usepackage{algorithm}
\usepackage{algpseudocode}
\usepackage{appendix}
\usepackage{enumitem}
\usepackage{graphbox}
\usepackage[english]{babel}
\usepackage{amsthm}

\title{Spherical and Hyperbolic Toric Topology-Based  Codes On Graph Embedding for Ising MRF Models: \\ Classical and Quantum Topology Machine Learning }

\date{} 					

\author{ Vasiliy Usatyuk \thanks{Corresponding author, L@Lcrypto.com}\\
South-West State University, T8 LLC
\and Denis Sapozhnikov\\
T8 LLC
\and Sergey Egorov  \\
South-West State University }

\renewcommand{\undertitle}{An early draft preprint}
\renewcommand{\shorttitle}{Codes on the graph based Ising models under Spherical, Torical Topologies: Topology ML
 }

\hypersetup{
	pdftitle={SPHERICAL AND HYPERBOLIC TORIC TOPOLOGY-BASED CODES ON GRAPH EMBEDDING FOR ISING MRF MODELS: CLASSICAL AND QUANTUM TOPOLOGY MACHINE LEARNING },
	pdfsubject={},
	pdfauthor={Vasiliy Usatyuk and Sergey Egorov },
	pdfkeywords={LDPC code parity-check matrix, matrix factorization, Automorphism, Toric Manifold, Quasitoric manifold},
}

\begin{document}
	\maketitle

\newcommand{\matA}{\mathbf{A}}
\newcommand{\matB}{\mathbf{B}}
\newcommand{\matC}{\mathbf{C}}
\newcommand{\matD}{\mathbf{D}}
\newcommand{\matE}{\mathbf{E}}
\newcommand{\matF}{\mathbf{F}}
\newcommand{\matG}{\mathbf{G}}
\newcommand{\matH}{\mathbf{H}}
\newcommand{\matI}{\mathbf{I}}
\newcommand{\matK}{\mathbf{K}}
\newcommand{\matL}{\mathbf{L}}
\newcommand{\matM}{\mathbf{M}}
\newcommand{\matN}{\mathbf{N}}
\newcommand{\matO}{\mathbf{O}}
\newcommand{\matP}{\mathbf{P}}
\newcommand{\matQ}{\mathbf{Q}}
\newcommand{\matR}{\mathbf{R}}
\newcommand{\matS}{\mathbf{S}}
\newcommand{\matT}{\mathbf{T}}
\newcommand{\matU}{\mathbf{U}}
\newcommand{\matV}{\mathbf{V}}
\newcommand{\matW}{\mathbf{W}}
\newcommand{\matX}{\mathbf{X}}
\newcommand{\matY}{\mathbf{Y}}
\newcommand{\matZ}{\mathbf{Z}}
\newcommand{\matg}{\mathbf{g}}

\newcommand{\calA}{\mathcal{A}}
\newcommand{\calB}{\mathcal{B}}
\newcommand{\calC}{\mathcal{C}}
\newcommand{\calD}{\mathcal{D}}
\newcommand{\calE}{\mathcal{E}}
\newcommand{\calF}{\mathcal{F}}
\newcommand{\calG}{\mathcal{G}}
\newcommand{\calH}{\mathcal{H}}
\newcommand{\calI}{\mathcal{I}}
\newcommand{\calJ}{\mathcal{J}}
\newcommand{\calK}{\mathcal{K}}
\newcommand{\calL}{\mathcal{L}}
\newcommand{\calM}{\mathcal{M}}
\newcommand{\calN}{\mathcal{N}}
\newcommand{\calO}{\mathcal{O}}
\newcommand{\calP}{\mathcal{P}}
\newcommand{\calQ}{\mathcal{Q}}
\newcommand{\calR}{\mathcal{R}}
\newcommand{\calS}{\mathcal{S}}
\newcommand{\calT}{\mathcal{T}}
\newcommand{\calU}{\mathcal{U}}
\newcommand{\calV}{\mathcal{V}}
\newcommand{\calW}{\mathcal{W}}
\newcommand{\calX}{\mathcal{X}}
\newcommand{\calY}{\mathcal{Y}}
\newcommand{\calZ}{\mathcal{Z}}

\newcommand{\bbA}{\mathbb{A}}
\newcommand{\bbB}{\mathbb{B}}
\newcommand{\bbR}{\mathbb{R}}
\newcommand{\bbZ}{\mathbb{Z}}
\newcommand{\bbE}{\mathbb{E}}
\newcommand{\bbH}{\mathbb{H}}

\newcommand{\veca}{\mathbf{a}}
\newcommand{\vecb}{\mathbf{b}}
\newcommand{\vecc}{\mathbf{c}}
\newcommand{\vecd}{\mathbf{d}}
\newcommand{\vece}{\mathbf{e}}
\newcommand{\vecf}{\mathbf{f}}
\newcommand{\vecg}{\mathbf{g}}
\newcommand{\vech}{\mathbf{h}}
\newcommand{\veci}{\mathbf{i}}
\newcommand{\vecj}{\mathbf{j}}
\newcommand{\veck}{\mathbf{k}}
\newcommand{\vecl}{\mathbf{l}}
\newcommand{\vecm}{\mathbf{m}}
\newcommand{\vecn}{\mathbf{n}}
\newcommand{\veco}{\mathbf{o}}
\newcommand{\vecp}{\mathbf{p}}
\newcommand{\vecq}{\mathbf{q}}
\newcommand{\vecr}{\mathbf{r}}
\newcommand{\vecs}{\mathbf{s}}
\newcommand{\vect}{\mathbf{t}}
\newcommand{\vecu}{\mathbf{u}}
\newcommand{\vecv}{\mathbf{v}}
\newcommand{\vecw}{\mathbf{w}}
\newcommand{\vecx}{\mathbf{x}}
\newcommand{\vecy}{\mathbf{y}}
\newcommand{\vecz}{\mathbf{z}}

\newcommand{\vecalpha}{\boldsymbol{\alpha}}
\newcommand{\vecbeta}{\boldsymbol{\beta}}
\newcommand{\veceta}{\boldsymbol{\eta}}
\newcommand{\vectheta}{\boldsymbol{\theta}}
\newcommand{\vecphi}{\boldsymbol{\phi}}
\newcommand{\vecpsi}{\boldsymbol{\psi}}
\newcommand{\vecrho}{\boldsymbol{\rho}}
\newcommand{\vectau}{\boldsymbol{\tau}}
\newcommand{\vecmu}{\boldsymbol{\mu}}
\newcommand{\veceps}{\boldsymbol{\epsilon}}
\newcommand{\vecxi}{\boldsymbol{\xi}}
\newcommand{\vecPhi}{\boldsymbol{\Phi}}
\newcommand{\vecDelta}{\boldsymbol{\Delta}}

\newcommand{\matDelta}{\boldsymbol{\Delta}}
\newcommand{\matEta}{\boldsymbol{\eta}}
\newcommand{\matOmega}{\boldsymbol{\Omega}}
\newcommand{\matPhi}{\boldsymbol{\Phi}}
\newcommand{\matPsi}{\boldsymbol{\Psi}}
\newcommand{\matTheta}{\boldsymbol{\Theta}}
\newcommand{\matLambda}{\boldsymbol{\Lambda}}
\newcommand{\matSigma}{\boldsymbol{\Sigma}}
\newcommand{\matzero}{\mathbf{0}}
\newcommand{\IndexSetI}{\mathcal{I}}
\newcommand{\grad}{\mathcal{\nabla}}

\newcommand{\vecone}{\mathbf{1}}
\newcommand{\veczero}{\mathbf{0}}

\def\maximize{\mathop{{\mathgroup\symoperators maximize}}}
\def\Maximize{\mathop{{\mathgroup\symoperators Maximize}}}
\def\minimize{\mathop{{\mathgroup\symoperators minimize}}}

\def\approach{\mathop{{\mathgroup\symoperators \longrightarrow}}}
\def\defineoperator{\mathop{{\mathgroup\symoperators =}}}
\newcommand{\define}{\defineoperator^{\text{def}}}

\newcommand{\Tr}{\text{Tr}}
\newcommand{\trace}{\text{trace}}
\newcommand{\diag}{\text{diag}}
\newcommand{\gradWJ}{\nabla_{\scriptscriptstyle{\matW}}\calJ}
\newcommand{\const}{\text{constant}}
\newcommand{\fracpartial}[2]{\frac{\partial #1}{\partial  #2}}

\newcommand{\defeq}{\stackrel{\text{def}}{=}}

\newcommand{\Xh}{\widehat{X}}

\begin{abstract}

The paper introduces an application of Information Geometry for describing the ground states of Ising models (Markov Random Fields) using parity-check matrices of Cyclic, Quasi-Cyclic codes on Toric and Spherical topologies. This approach establishes a connection between machine learning and error-correcting coding, specifically with regards to automorphism and the size of the circulant of the quasi-cyclic code, which determines the number of energy minima (ground states) of the quantum system. The proposed approach has implications for developing new embedding methods based on Trapping sets (TS(a,0)-codewords and TS(a,b)-pseudocodewords) optimized error-correcting codes using Statistical Physics and Number Geometry. The paper also demonstrates a relationship between the k-state Ising model's (Potts model) ground condition and lattices from number geometry, offering insights into fundamental principles governing complex systems like quantum computation and quantum error-correction, computer vision and NLP. The proposed code based embedding can be applied as factorization to diverse datasets from various domains, suggesting its effectiveness and versatility for different types of data and applications, such as quantum and classical computer vision, social networks, and  optimal control problem data. Moreover, the paper establishes a direct connection between DNN architecture and error-correcting coding by demonstrating how state-of-the-art DNN architectures (ChordMixer, Mega, Mega-chunk, CDIL, ...) from the long-range arena can be equivalent to specific types (Cage-graph, Repeat Accumulate) of block and convolutional LDPC codes.
The results of this research hold significant implications for the classification of codes and their application to quantum atomic models of chemicals on the spherical topology from the periodic system of Mendeleev. The study demonstrates that quasi cyclic codes and those with a complex structure of automorphisms correspond to certain types of chemical elements, particularly mixed automorphism Shu-Lin-Fossorier QC-LDPC code representing Carbon.  The Quantum Approximate Optimization Algorithm (QAOA) utilized in the Sherrington-Kirkpatrick Ising model can be considered analogous to back-propagation loss function landscape in training DNN. This similarity creates a comparable problem with Trapping sets pseudo-codeword, resembling the Belief propagation (BP) soft decoding process. Furthermore, the layer depth (p) in QAOA correlates to the number of decoding BP iterations in the Wiberg (covering) decoding tree.
 The research findings have far-reaching implications, including but not limited to the classification of codes, development of novel embedding methods, construction of DNNs, and modeling physical and chemical properties of materials based on code representations. Overall, this work has the potential to advance multiple fields, from Information Theory, DNN architecture design (sparse and structured prior graph topology), efficient hardware design for Quantum and Classical DPU/TPU (graph, quantize and shift register architect.) to Materials Science, KAM-theory method (Trapping sets gauge for Sym./AntiSym. wavefunc) for embedding of finite dimension coded based torus into the infinite dimensional Hilbert space and beyond.

\end{abstract}


\section{Introduction}
\label{sec:intro}

Low-density parity-check (LDPC) codes are a class of error-correcting codes that have found widespread applications in modern communication systems. These codes were first introduced by Robert G. Gallager in his doctoral thesis in 1963, where he described a coding scheme based on sparse parity-check matrices ~\cite{{G65}}. However, the potential of LDPC codes remained largely unrealized for several decades until they were rediscovered and generalized by several researchers in the last decade 1970s and early 1980. In particular, the breakthrough works of Zyablov V. V. and Pinsker M.S. ~\cite{ZyabPins75}  , Tanner R.M. ~\cite{Tanner81}  and Margulis G.A. ~\cite{Marg82}    contributed significantly to the early and later stages development of LDPC codes. Their research provided important insights into the pure mathematical nature, complexity and performance properties of these codes, setting the stage for further research. Exceeding the discovery of Generalize and Non-binary LDPC codes and QC (mixed/hidden) automorphism LDPC codes possessing structural which efficiently depict dynamic systems filled with symmetry, encompassing supersymmetric and symmetrical Hamiltonian dynamical systems.

The renewed interest in LDPC codes in the 1990s was driven by the development of energy efficient high-throughput soft decoding message passing algorithms and code construction methods that could efficiently: correct errors, encrypt, compress, interpolate, extrapolate, classify and upscale on varied type of communication/data/signal channels, including sparse filter bank, ML/DNN sparse graph models, sparse source codes, normal graph models for loss landspace space decurvature and large automorphism (symmetry) local convex/concave gauge.

The researchers:  Aaronson S.,  Abrikosov A.A., Ay N.,  Aganagic M., Agrell E., Arnold V.I., Arikan E., Ajtai M., Anzin M.M., Alferov Z., Aharonov D., Amari S., Alcubierre M., Bar-Natan D., Barg A., Baranovsky E.P., Barbaresco F., Balatsoukas-Stimming A., Baldi M., Banihashemi A.H., Berrou C., Belfiore J.C., Ben Arous G., Bengio Y.,  Boutillon E., Bocharova I.E., Bohr N., Born M., Brink S.T., Brigo D., Brown K.R.,  Butler B., Bin L., Burnaev E., Buchmann  J. A., Calderbank A.R., Chertkov M., Choromanska A., Chugg K., Cole C., Costello J. D., Coskun I., Chollet F., Cichocki A., Cvitanovic P., Dasbach O.T., Declerq D., Delaunay B. N., Djordjevic I.B., Divsalar D., Dolecek L., Donoho D.L., Donaldson S.K., Dwork C., Eleftheriou E.S., Einstein A.,  Flanagan M.F., Ferris A.J., Felstrom A.J.,  Forney G.D., Fossorier M. P. C., Fontein F., Fuja T.E.,   Fukuda K., Frolov A.,  Gander W.,   Galbraith S.,   Gama  N., Graff S.M., Grothendieck A., Gromov M.L., Gunnar E. C., Gibbs J.W.,  Ginzburg V., Grant A.J., Ghrist R., Gross W.G., Gross W.M., Goda H.,  Goldwasser S., Goldreich O., Goryachkin O.V., Gukov S., Hawking S.T., Halevi S., Heisenberg W., Harrell E., Hacon C., Hoholdt T.,  Hofer H., Hinton G., Hwang Jun-Muk, Iriyeh H., Isakov S.B., Johannesson R.,   Jordan P., Johnson S., Justesen J., Kannan R., Kamenev M., Kawamata Y., Kapitsa P., Kalachev G.V.,  Kolmogorov A.N.,    Kontsevich M.,  Kantorovich L., Korobov N.M., Kozintsev I. V.,   Kilby J.S., Kelley C.A., Kurkoski B., Kudryashov B.D., Kschischang F.R., Khalitov R., Kroemer H.,   Lagarias J.C., Leontief W., LeCun Y., Leland M., Lentmaier M.,  Levenshtein V.I., Liva G., Liu G., Likhobabin E., Luby V., Litsyn S., Ling C., Lyubashevsky V.,   Lovász L.,  Lobachevsky N., Kannan  R., Klein F.,   Kitaev A.Y., Korkin A. N., Khot S.,   Kiyomoto S., Mark C. B., MacKay D.J., Maclagan D., Manton J. H.,  Martín V.,  Martinet J., McEliece R.J., Mezard M., Meshulam R., Melnikov V.K., Meier A., Milenkovic O.,   Micciancio D., Marriott P., Montanari A., Moser J. K.,  Mori S., Murin D.M., Mustață M.,  Mow W.H., Nazarov L. E., Neal R.M., Nebe G., Nguyen P.Q.,  Nielsen F., Nikolenko S.,   Noether E., Oseledets I., Olmos P.M., Ovinnikov A., Panteleev P., Perelman G., Pennec X.,  Pearl J., Peikert C., Perret L., Pfister H.D., Polyansky Y., Polyansky N., Poincaré H.,  Poulin D., Pontryagin L., Poulliat C., Pöschel J., Pinzari G., Rapoport L.B., Read N., Riemann B., Richardson T., Rieck Y., Ramchandran K., Ratcliffe J. G., Regev O., Ryshkov S.S., Rybin P., Ryan W.E., Rohlin V.A., Robinson M., Rosen A.,  Ruan Y.,   Savin V., Serra A., Shlegel C.,  Schnorr C.-P.,  Schrödinger E., Schmidhuber J., Sloane N.J.A., Schutski R., Shu Lin, Sigel P.H., Sipser M., Sinai Y.G., Skachek V., Sturmfels B.,   Stehlé M., Smarandache R., Sourlas N., Spielman D.A., Stepanov M., Shevchenko V.N.,  Schneider M., Teichmüller P.J.O., Tanaka T., Tal I., Tian G., Thaddeus M., Tse D., Trifonov P., Ikeda S.,  Urbanke R.L., Ungerboeck G., Vardy A., Vasic B., van Handel R., Vetrov D., Venkov B.B.,Venkov B.A.,  Verdú S., Vontobel P., Vorobyev I., Voronoi G.F., Viterbo E., Vikalo H.,   Vaikuntanathan V.,   Wesel R.D., Wibberg N., Weil A.,  Wornell G.W., Yedidia J.S.,  Yau Shing-Tung, Yang Z.,   Zamarashkin N., Zigangirov K.Sh., Zolotarev E.I., Zémor G., Zhou Z., Zhang S., Zhang H. and many others, have contributed to expanding the scope of LDPC codes and demonstrating their effectiveness in a wide range of applications, sometimes without even realizing it. These researchers have explored the application of LDPC codes (equivalent to them under serial processing Turbo/Polar codes, Tensor Networks, Lattice) in various domains  complex dynamic systems.  By applying LDPC codes in these contexts, they have shown the versatility and efficacy of LDPC codes beyond traditional communication applications.

Quasi-cyclic low-density codes have found extensive use in encryption systems, error correction, source coding, high efficient filter bank and compressed sensing methods. Their effectiveness in achieving channel throughput with linear complexity can be attributed to the local nature of the belief propagation method and the high throughput achieved through the use of a shift register (due to Quasi-cyclic automorphisms). The successful application of methods for synthesizing these codes on finite lengths and optimizing decoders has enabled their widespread use. However, their topological properties and potential applications as a priori models of neural networks and trajectories (attractors) of complex dynamic systems remain largely unexplored. It is paradoxical that although the theory of Information Geometry particularly and Information Theory in general  has recently experienced significant development and has found numerous applications in communication, it has surpassed the development of several other areas, including material science, machine learning, mathematical parts of economics, biology, sociology, psychology, and military wargaming.

\textbf{The deliberate selection of LDPC codes and it cycle topology (Trapping Sets) as a fundamental concept in this article underlines its significance. Within the realm of mathematics, the cornerstone resides in the pivotal notion of LDPC codes, particularly focusing on cycles. The nesting of loops, which defines the complexity class in complexity theory, further highlights the importance of cycles. Cycles in Topology plays a crucial role by allowing us to investigate qualitative properties of geometric shapes that remain invariant irrespective of distances, angles, areas, or volumes. In this context, the key concept of mathematics brings us back to cycles, specifically orthogonalities within arbitrary algebras in arbitrary metric spaces. The extrema of functions in optimization theory are derivative characteristics arising from the structure of the optimization landscape shaped by (topology) cycles. When a LDPC codes cycle possesses a specific topology, it gives rise to a codeword (Trapping Sets, TS(a,0), permanent) along with a code distance that facilitates error correction. When these cycles cluster together, they impede precise error correction and give rise to pseudocodewords(TS(a,b), Bethe permanent), where any algorithm require unravel them (by decorrelation or full processing). As their length increases, the dynamics of linear size trapping sets (union of TS(a,0) and TS(a,b)) can be elucidated through Covariance evolution method, exhibiting behavior reminiscent of physical systems at higher temperatures. It is worth noting that the low-temperature dynamics are entirely determined by the sublinear size (relative to the whole graph size) of cycles which can be estimated by Importance Sampling (Bethe free energy under normal graph model, Bethe entropy, Bethe permanent).}
\textbf{However, a mere discrete representation in the form of cycles and their topology alone is inadequate. It is essential to map this discrete representation to the domain of real (complex) numbers, forming a complex modulation constellation in comminucation theory. In mathematics, the term used is indeed "lattice," while in communication theory, it is referred to as a "constellation." A quadratic form, which was previously studied in the context of differential equations, has now been more well know and recognized as a lattice or optimal grids (for exaple Korobov grid ).}

\textbf{Our research focuses on two important components: the topology of cycles under Spherical and Toric Topologies, specifically the trapping set comprising codewords TS(a,0) and pseudocodewords TS(a,b), as well as the mapping of codewords to a lattice based on established principles. Symmetries (graph and it representor parity-check matrix) and all properties (Abelization, ideal lattice/positive defined and etc) arising when the system tends to the ground states.}

The Kolmogorov-Arnold-Moser (KAM) theorem is a result in the field of dynamical systems that provides conditions for the persistence of quasi-periodic motions under small perturbations of integrable Hamiltonian systems. In particular, the KAM theorem states that for a large class of Hamiltonian systems with two degrees of freedom, if the unperturbed system has a dense set of invariant tori and the perturbation is sufficiently small, then there exists a positive measure set of quasi-periodic motions that persist under the perturbation. More precisely, the theorem asserts that, for a non-degenerate Hamiltonian system with two degrees of freedom, if the frequency map associated with the system is sufficiently close to an integrable case, then there exists a positive measure set of quasi-periodic motions that can survive under small perturbations of the Hamiltonian. The theorem has important applications in celestial mechanics, plasma physics, and condensed matter physics, among other areas of science.

The KAM theorem is typically applied to Hamiltonian systems, while the Ising model is a statistical mechanical model. Therefore, it is not directly applicable to the Ising model. However, there are extensions of the KAM theorem to certain classes of non-Hamiltonian systems that exhibit some degree of integrability, such as the nonlinear Schrödinger equation and the Korteweg-de Vries equation. These extensions have been used to study the persistence of quasi-periodic solutions in various physical systems, including spin chains with long-range interactions that can be modeled using the Ising model. In particular, there have been studies on the KAM theory for the XXZ Heisenberg spin chain, which is a generalization of the Ising model. The XXZ model has an additional parameter that controls the strength of the interaction between spins. For certain values of this parameter, the XXZ model exhibits integrability and possesses a large number of conserved quantities. In recent years, there have been several works applying KAM theory to the XXZ model to study the stability of its quasi-periodic solutions under perturbations.

The Fermion KAM (FKAM, anti-Symmetrical Wave-function KAM) theorem is an extension of the classical KAM theorem to systems of interacting fermions. In particular, it provides conditions for the persistence of a Fermi surface under perturbations of the Hamiltonian that describes the system. The Fermi surface is a key concept in the theory of interacting fermions, and it plays a central role in many important physical phenomena, such as superconductivity, magnetism, and quantum Hall effects, ~\cite{Isa81}. The FKAM theorem addresses the question of whether the Fermi surface survives under small perturbations of the Hamiltonian that describe the interactions between the particles. The idea behind the FKAM theorem is to use a similar approach to study the stability of the Fermi surface under perturbations of the Hamiltonian. The main technical challenge is that the Fermi surface is a highly nontrivial object, and its stability depends crucially on the topology of the energy bands near the Fermi level. The FKAM theorem has important implications for our understanding of interacting fermions in condensed matter physics and other areas of physics. It provides a rigorous foundation for many theoretical predictions about the behavior of these systems, and it can be used to guide the design of new materials with novel properties. Hyperbolic toroidal surfaces have some interesting topological properties that are relevant for understanding the behavior of electrons near the Fermi surface. For example, they can have nontrivial winding numbers that determine the number of times a particle must wind around the surface before returning to its original position. These winding numbers are related to the number of particles with a given momentum and spin orientation that populate the Fermi surface. One major difference between our proposed modification and FKAM is the approach used to model the behavior of particles. While FKAM relies on techniques from Fourier analysis to extract information from the oscillations of a cantilever, we propose to use symmetrical sparse coding based on non-orthogonal basis functions and gauge algebraic structures with short Trapping sets(TS) elimination criteria (TS(a,b) pseudocodewords and TS(a,0) codewords).

The trapping set, denoted as $TS(a,b)$, in a bipartite graph refers to a subgraph that consists of cycles and connects to a specific number of variable nodes ($a$) and odd degree check nodes ($b$). For instance, the $TS(5,3)$ trapping set is generated when three 8-cycles overlap, while the $TS(4,4)$ trapping set is formed by the cycle 8 within the bipartite Tanner graph.  These cycles play a crucial role in describing the local curvature of the graph. Interestingly, these cycles in the bipartite graph exhibit a close relationship with simplicial complexes. Simplicial complexes are major Topology Data Analysis structures composed of simplices, which are higher-dimensional generalizations of triangles. The presence of cycles in the bipartite graph can be linked to the formation of simplicial loops or higher-dimensional simplices in simplicial complexes. This connection between local cycles topology (Trapping Sets)  in bipartite graphs and simplicial complexes allows us to understand the local curvature properties of the graph from a higher-dimensional perspective. By examining the cycles and their relationships within the bipartite Tanner graph, we can gain insights into the geometric structure and topological properties of the associated simplicial complex. Using symmetrical sparse coding based on non-orthogonal basis functions and gauge algebraic structures offers several advantages when compared to other methods. These techniques allow for more flexible representations of data and enable the fulfillment of the minimal energy principle for Spherical und hyperbolic toric Hamiltonian dynamic systems. Non-orthogonal basis functions are capable of capturing intricate dependencies between variables without introducing additional noise or trivially recovering non-linear dependencies (as seen in cycles related to codewords, TS(a,0)) that might be missed(overcomplicated) by orthogonal basis functions like Fourier or wavelet bases. Moreover, gauge algebraic structures provide a framework for describing physical phenomena using gauge fields and the associated group symmetries. This approach allows for a comprehensive understanding and description of various physical processes. The utilization of symmetrical sparse coding, which is based on non-orthogonal basis functions and gauge algebraic structures, offers notable advantages over alternative methods. These techniques provide greater flexibility in representing data and facilitate the fulfillment of the minimal energy principle for Spherical und hyperbolic toric Hamiltonian dynamic systems. Non-orthogonal basis functions play a pivotal role in capturing intricate dependencies between variables. Unlike orthogonal basis functions such as Fourier or wavelet bases, non-orthogonal functions can capture complex relationships without introducing additional noise or trivially recovering non-linear dependencies. This is particularly evident in cycles related to codewords, denoted as TS(a,0), where non-linear dependencies are efficiently (localy) captured and globally recovered using message passing. Trapping Sets improving under Toroidal Hyperbolic and Spherical Gauge algebraic structures offer a powerful framework for describing physical phenomena by employing gauge fields and associated group symmetries. By incorporating these structures, we gain a efficient, comprehensive understanding and description of various physical processes. The use of gauge algebraic structures allows us to analyze, predict and interpret physical phenomena in a manner that encompasses their fundamental symmetries and underlying principles.

The Quantum Approximate Optimization Algorithm (QAOA) applied to the Sherrington-Kirkpatrick Ising model shares certain similarities with back-propagation optimization used in training Deep Neural Networks (DNNs), albeit without the Hessian. One of the key similarities arises from the presence of Trapping Sets pseudo-codewords, which resemble the Belief Propagation (BP) soft decoding process. In both QAOA and DNN training, there is a risk of encountering trapping sets, which are configurations that can lead to suboptimal solutions or convergence issues. This similarity highlights the challenge of finding optimal solutions in both optimization processes. Additionally, the depth parameter (p) in QAOA corresponds to the number of iterations performed during decoding using the Wiberg (covering) decoding tree. Similarly, increasing the layer depth in DNNs can be seen as performing more iterations of the back-propagation algorithm, allowing for potentially improved optimization and convergence. By drawing these analogies between QAOA and DNN training, we can leverage concepts and techniques developed in coding theory and optimization algorithms to enhance the performance and convergence properties of quantum optimization algorithms such as QAOA. This cross-pollination of ideas can lead to advancements in both quantum optimization and deep learning fields. QAOA utilized in the Sherrington-Kirkpatrick Ising model exhibits similarities to the back-propagation optimization employed in training DNNs. The presence of trapping sets and the relationship between the layer depth in QAOA and the number of BP iterations in DNNs highlight potential areas of convergence and shared challenges. Leveraging knowledge from coding theory and optimization algorithms can contribute to the advancement of both quantum optimization and deep learning methodologies.

Square matrices are extensively used in machine learning as they serve as a fundamental component for representing linear and non-linear correlations within data. These matrices play a vital role in various models, serving as building blocks for different applications. One prominent example is the kernel matrix employed in Support Vector Machines (SVM) or Gaussian Processes. The kernel matrix captures the pairwise similarity or distance between data points, allowing SVMs and Gaussian Processes to model complex relationships in the data. Similarly, in graph or network representation, the affinity matrix represents the pairwise similarities or connections between nodes, aiding in tasks such as clustering, community detection, or link prediction. Another application of square matrices in machine learning is seen in Transformers, where attention mechanisms are utilized. The attention matrix enables the model to focus on relevant parts of the input sequence during processing, facilitating effective information exchange across different dimensions or positions. Moreover, square matrices are crucial for optimizing solution spaces in machine learning tasks such as graph matching or network architecture optimization. By manipulating the elements of these matrices, it becomes possible to optimize objective functions and find optimal solutions for complex problems. Sparse factorization of square matrices is a powerful technique that decomposes a given matrix into a product of two or more sparse matrices. This factorization aids in reducing computational complexity and improving numerical stability during matrix operations. Sparse factorization techniques have been widely employed in various domains, including recommendation systems, image processing, and network analysis. Non-linear sparse factorization can indeed be seen as a technique for creating low-dimensional embeddings that effectively capture the underlying structure of high-dimensional data. By incorporating methods from topology data analysis and information geometry, we can gain insights into the curvature of the data and develop models that accurately represent this curvature. As a result, classification, interpolation, and regression tasks become more efficient and accurate due to the reduced complexity. Moreover, the incorporation of stochastic dynamic systems allows us to account for the inherent uncertainty present in real-world data. This enables us to make predictions that are robust to noise and other sources of variability, enhancing the reliability of our models. In the specific context of non-linear sparse factorization, our study focuses on the application of toroidal and spherical topologies to Markov Random Field binary and q-ary Ising models. These models give rise to low-density quasi-cyclic codes and codes with hidden automorphism. The demonstration of these concepts and their implications will be provided in greater detail in subsequent sections of the article.

ML and DNN part of this research introduces a novel approximation technique that leverages QC-LDPC codes and random LDPC codes with unknown automorphism from Hyperbolic Torical Topology under the Ising model ground state. The approach improves upon the Extrinsic Message Degree Distribution (EMD) and employs an Approximate Extrinsic Message Degree Distribution (ACE) to broke short Trapping Sets. By factorizing the approximated square matrix into full-rank sparse matrices, this technique achieves effective decomposition.To determine the non-zero entry positions, the Simulated Annealing method with EMD optimization (\cite{USA18}) and Progressive-Edge Grown (\cite{Hu05}) with ACE optimization (\cite{TiJoViWe})  are utilized. To evaluate the effectiveness of the proposed technique, tests were conducted on both synthetic and real-world square matrices. The results were compared against those obtained using TSVD, LDPC code parity-check matrix, and Sparse Factorization (SF). The findings indicate that for approximating sparse square matrices with unknown non-zero positions, the proposed Hyperbolic Toric Topology-based LDPC codes outperform SF and TSVD when provided with the same number of non-zero entries. Proposed code based approach a cost-effective method for approximating square matrices with unknown non-zero positions and has demonstrated better performance compared to other techniques.

One crucial aspect of the Code-based Topology approach is its application to the analysis of Deep Neural Network (DNN) architectures. This approach establishes a direct connection between DNN architecture and error-correcting coding by demonstrating the equivalence between state-of-the-art DNN architectures, such as ChordMixer, Mega, Mega-chunk, CDIL, from the long-range arena and specific types of block and convolutional LDPC codes, namely Cage-graph and Repeat Accumulate codes. By establishing this correspondence, we can leverage the rich theory and techniques developed in error-correcting coding to analyze and understand the behavior of DNN architectures. This includes investigating properties like performance, robustness, and resilience to noise or adversarial attacks. The equivalence between DNN architectures and specific types of LDPC codes allows us to draw analogies and insights from coding theory to better comprehend the functioning of neural networks. This connection opens up opportunities for utilizing coding theory concepts, such as decoding algorithms, error-correcting capabilities, and optimization techniques, to enhance the design and training of DNN architectures.

We organize the rest of the paper as follows. Section 2 introduces Multi-Edge Type QC-LDPC codes, main properties, construction and decoding methods and explores their connection Quantum system problem and  hypergraph and multi-graph (Tensor Network) representations. Additionally, the section discusses the application of LDPC codes on various graph structures, specifically focusing on their utilization in deep neural networks (DNNs) such as CNNs, RNNs, and state-of-the-art Transformers architectures. 

Section 3 introduces the development of QC codes and QC-LDPC codes as a ground state solution for the Ising model, based on Hyperbolic Toric Topology. The section also explores special cases involving Abelianization, as well as the connection between Ising models and Number Geometry Lattice.  We delve into the Quantum Approximate Optimization Algorithm (QAOA) and its application in the Sherrington-Kirkpatrick Ising model. We draw an analogy between QAOA and back-propagation optimization, commonly used in training deep neural networks (DNNs), albeit without the inclusion of the Hessian matrix. This similarity gives rise to a comparable challenge involving Trapping sets pseudo-codeword, reminiscent of the soft decoding process of Belief Propagation (BP). Furthermore, the layer depth (p) employed in QAOA exhibits a correlation with the number of iterations during decoding using the Wiberg (covering) decoding tree.

In Section 4, includes the application of spherical topology solutions to elements derived from the periodic system of Mendeleev. The study reveals that cyclic codes and those exhibiting a complex structure of automorphisms correspond to specific types of chemical elements. Notably, the mixed automorphism Shu-Lin-Fossorier QC-LDPC code represents Carbon. Furthermore, in this section, we demonstrate that the MET QC-LDPC Graph Model represents pumped up Fermions while exhibiting behaviors that are characteristic of Bosons. Moreover, we present an analysis of Topology Data Analysis (TDA) Fundamentals and the outcomes of applying it to the Klein bottle topology in computer vision CNN. Additionally, we explore the relationship between spatially coupled LDPC (convolutional) and QC-LDPC codes topology, highlighting their resemblance to twisted (tailbitted) toroidal codes resembling the Klein bottle structure. We demonstrate the connection between Abelization Cases and the ideal lattice in Number Geometry, specifically in relation to Ising Models. We investigate how the topology of cycles and their unions, Trapping set, contribute to the formation of the loss surface landscape. Additionally, we analyze the Trapping sets enumerator under different temperatures, considering both lower and higher temperatures in the context of Bayesian inference and Quantum Ising systems.

In Section 5, we explore the application of sparse LDPC codes and MET QC-LDPC codes in low-rank matrix factorization using a Trapping Sets breaking methods known as Progressive Edge Grown with Approximate Extrinsic Mesage Degree optimization (Approximate EMD, ACE) and Simulated Annealing with EMD optimization. We present the experimental settings and results obtained by applying this approach to diverse datasets from various domains. These findings highlight the effectiveness and versatility of the method for different types of data and applications, including biology, politics, social networks, and optimal control problem data. 

Section 6 of the article presents the conclusions drawn from the research, while Section 7 (Appendix A) provides a detailed description of the sources and data used for the sparse factorisation in Section 5, specifically focusing on special type of embedding - factorization of  the square matrices.

\newpage

\section{Multi-Edge (Protograph, multigraph and hypergraph) Type QC-LDPC codes and it connection to DNN: CNN, RNN, Transformer}

\begin{figure}
\centering
\includegraphics[width=4.2in]{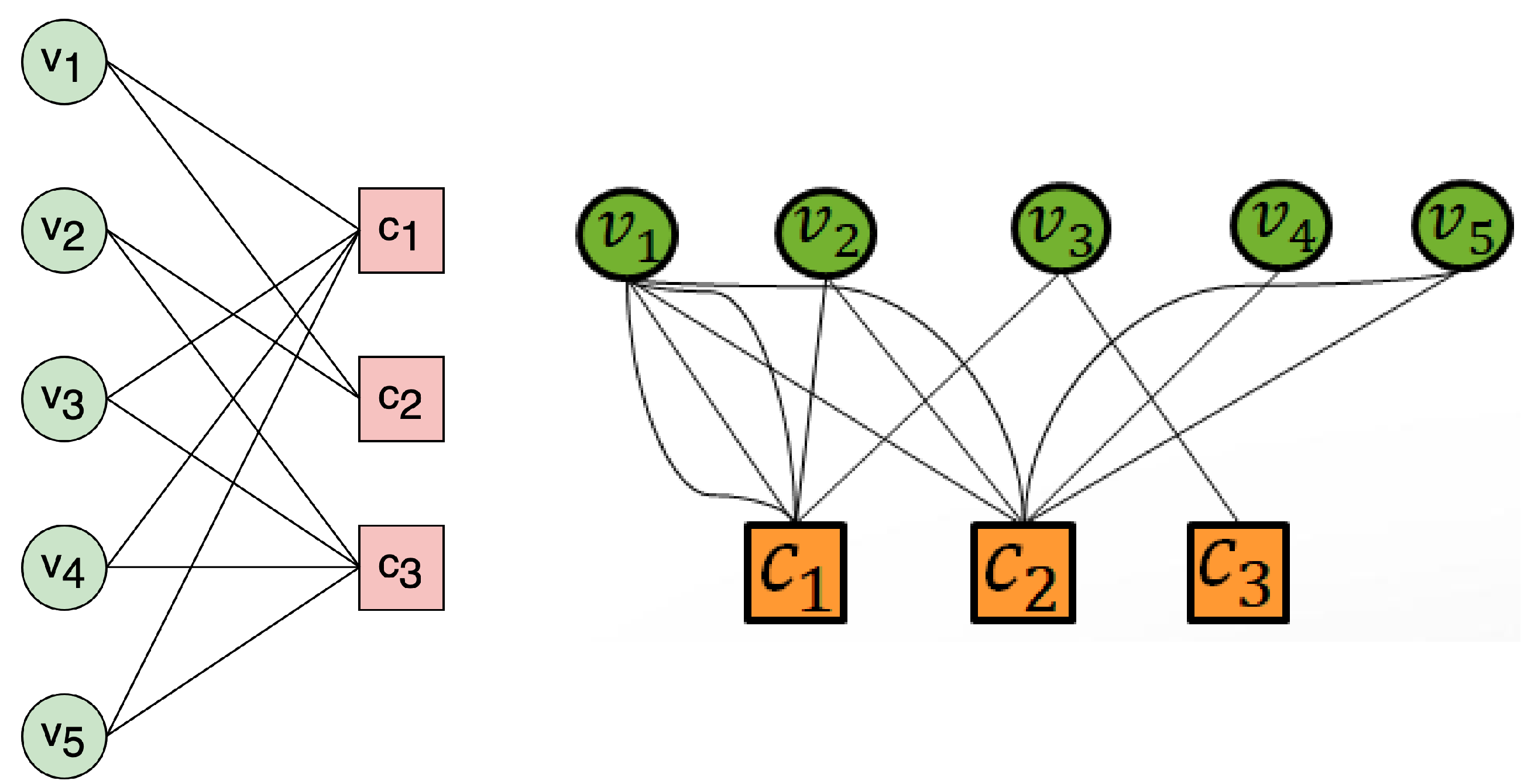} 
  \caption{(left) Tanner Graph of $H$ parity-check matrix (right) Multi-graph of $H_2$ parity-check matrix}
  \label{proto}
\end{figure}

LDPC (Low-Density Parity-Check) codes are a type of linear error-correcting codes that can be represented by a sparse parity-check matrix. Let's denote the length of the codeword as N and the number of information bits as K. LDPC codes use an (N, K) block code scheme. A parity-check matrix that specifies parity check equations of a code can be represented as a Tanner graph. The Tanner graph corresponding to the parity-check matrix $H$ is shown below, \cite{RR08,RyanShu09}:
	
\begin{equation} 
	H= \begin{bmatrix}  
	1 & 0 & 1 & 1 & 1  \\
	1 & 1 & 0 & 0 & 0  \\
	0 & 1 & 1 & 1 & 1  \\ \end{bmatrix}  
\end{equation}

 on ~Fig.~\ref{proto}, left.

Quasi-Cyclic Low-Density Parity-Check (QC-LDPC) codes are a class of linear block error-correcting codes. They can be represented by a parity-check matrix H, where the matrix has a special structure known as quasi-cyclic. Let's consider an (N, K) QC-LDPC code. Here, N represents the code length (number of codeword bits), and K represents the number of message bits (length of the original message). The remaining bits in the codeword (N - K) are parity bits. The Tanner graph of a QC-LDPC code is described by a parity-check matrix $H$ consisting of square blocks that could be either zero matrices or circulant permutation matrices. The $L\times L$ circulant permutation matrix $P=(P_{ij})$ is defined:

\begin{equation} 
P_{ij}=
\begin{cases}
1,\quad\text{if } i+1\equiv j \mod L\\
0,\quad \text{otherwise}.
\end{cases}
\end{equation} 

Here, $P^k$ is the circulant permutation matrix (CPM) that shifts the identity matrix $I$ to the right by $k$ times for any $k$, $0\le k\le L-1$. To simplify notation, we denote the zero matrix by $I^{-1}$. The set $\{-1, 0, 1,\ldots, L-1\}$ is denoted by $A_{L}$.
Suppose the matrix $H$ of size $L \times L$ is defined as follows:
\begin{equation} 
H=\left[\begin{array}{cccc} {P^{a_{11} } } & {P^{a_{12} } } & {\cdots } & {P^{a_{1n} } } \\ {P^{a_{21} } } & {P^{a_{22} } } & {\cdots } & {P^{a_{2n} } } \\ {\vdots } & {\vdots } & {\ddots } & {\vdots } \\ {P^{a_{m1} } } & {P^{a_{m2} } } & {\cdots } & {P^{a_{mn} } } \end{array}\right] = \left[\begin{array}{cccc} {I^{a_{11} } } & {I^{a_{12} } } & {\cdots } & {I^{a_{1n} } } \\ {I^{a_{21} } } & {I^{a_{22} } } & {\cdots } & {I^{a_{2n} } } \\ {\vdots } & {\vdots } & {\ddots } & {\vdots } \\ {I^{a_{m1} } } & {I^{a_{m2} } } & {\cdots } & {I^{a_{mn} } } \end{array}\right] ,
\end{equation} 

where $a_{i,j} \in A_{L}$ and $L$ is the circulant size of $H$. We call a code $C$ with parity-check matrix $H$ a QC-LDPC code.

Let $E(H) = (E_{ij}(H))$ be the exponent matrix of $H$ given by:

\begin{equation} 
E(H)=\left[\begin{array}{cccc} {a_{11} } & {a_{12} } & {\cdots } & {a_{1n} } \\ {a_{21} } & {a_{22} } & {\cdots } & {a_{2n} } \\ {\vdots } & {\vdots } & {\ddots } & {\vdots } \\ {a_{m1} } & {a_{m2} } & {\cdots } & {a_{mn} } \end{array}\right],
\end{equation} 

where $E_{ij}(H) = a_{ij}$. The protograph mother matrix or base graph $M(H)$ is a $m\times n$ binary matrix obtained from replacing $-1$'s and other integers by $0$ and 1, respectively, in $E(H)$.

The Tanner graph of matrix $H$ forms a cycle if the value of the equation $\sum_{i=1}^{2l}(-1)^i a_i\equiv 0 \mod L$ is satisfied. 

\begin{figure}
\centering
\includegraphics[width=63mm, viewport=30.00mm 194.40mm 136.75mm 277.00mm]{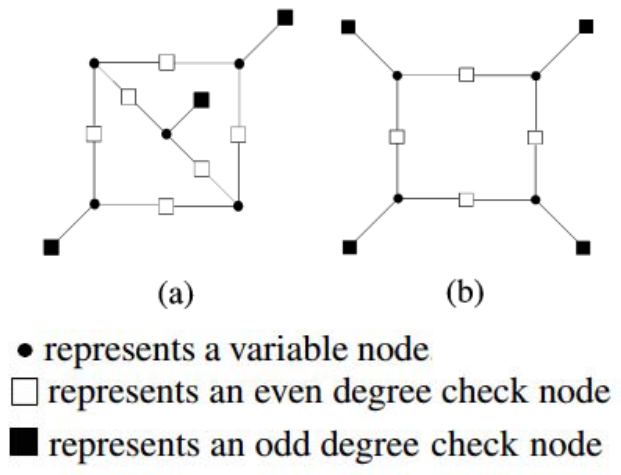} 
  \caption{Graphical representation of  Trapping sets: a) TS(5,3), b) TS(4,4)}
  \label{Trapping sets}
\end{figure}

\textbf{Let's consider a sub-graph of the matrix $H$ Tanner graph formed by it's cycles or cycle's overlap. Such a sub-graph includes $a$ variable nodes and $b$ odd degree checks named as trapping set $TS(a,b)$,  ~\cite {Vasic09} }. In Fig. \ref{Trapping sets}, we can see the TS(5,3) generated by the overlap of three 8-cycles and the TS(4,4) formed by cycle 8 in the Tanner graph. The risk of a trapping set depends on the quantity of variable nodes that can result in decoding failure in the event of errors. When there are 3 errors in odd degree check nodes, the TS(5,3) is more dangerous than the TS(4,4), which produces only 4 errors in variable nodes if there are 4 errors in odd degree check nodes and can be interpreted as a weight 4 pseudocodeword. Breaking the worst cycles can improve the weight spectrum of a pseudocodeword and reduce the likelihood of an error-floor. Its risk is closely tied to the decoder and may be altered by changing the decoder's parameters. \textbf{The minimum codeword that determines the code distance (Hamming distance) $d_{min}$ of the LDPC code corresponds to the $TS(a, 0)$, where $a=d_{min}$. }

Trapping set (TS) spectrum is a generalization of the weight spectrum of the code in the case of soft iterative decoding (local system approximation of MAP/ML estimation) by the message-passing decoders (Belief propagation), ~\cite{Vasic09} . By improving the Extrinsic Message Degree (EMD), we can improve the weight spectrum of the LDPC code (codewords spectrum), ~\cite{Ti04}. However, the TS spectrum enumeration problem is much more complex than the weight spectrum enumeration problem,  \cite{DMS03}. The paper has shown the NP complexity of enumerating TS in LDPC codes, \cite{VeSuWo22}. The number of check nodes singly connected to the variable nodes included in the cycle is the EMD of the cycle in the Tanner graph. The EMD value of a code is a crucial aspect because each cycle represents a trapping set. 

A better EMD spectrum for a LDPC code with the same code distance (weight enumerator spectrum) under message-passing decoder provides better error-correcting properties BER/FER. A code with a higher EMD spectrum value will lead to a decrease in the probability of TS pseudocodewords decoding failure. We can eliminate equal-probability trapping sets by building codes with larger minimal EMD values. The TS$(a,b)$ harm depends simply on the value $harm = b/a$, subgraph error probability, and TS multiplicity. However, the decoder's parameters determine the likelihood of TS errors. We can adjust the weight of pseudo-codewords by varying these variables, ~\cite{Ji05} .

The articles suggest a progressive edge-growth technique and its modified version for constructing LDPC codes. These methods aim to avoid the formation of cycles by utilizing an Approximate EMD (ACE) approach,  ~\cite{Ti04,Hu05,Di16}.

The article proposes the utilization of a simulated annealing method for constructing LDPC codes. This method incorporates fast EMD estimation techniques, which are superior to the progressive edge-growth (PEG) method when evaluated using the ACE metric ~\cite{USA18}. The application of simulated annealing (SA) enables an improvement in coding gain characteristic by effectively breaking trapping sets, as demonstrated in  ~\cite{USA23}.

A QC-LDPC code and bipartite Tanner-graph can be generalized to a multi-graph. The multi-graph corresponding to the parity-check matrix:

\begin{equation} 
H_2=\left(\begin{array}{ccccc} {I^{1} +I^{2} +I^{7} } & {I^{9} } & {I^{23} } & {0} & {0} \\ {I^{12} +I^{37} } & {I^{19} } & {0} & {I^{32} } & {I^{11} +I^{12} } \\ {0} & {0} & {I^{33} } & {0} & {0} \end{array}\right)
\end{equation} 

is shown in Figure \ref{proto} (right), where $I$ is the circulant permutation matrix used at multi-edge (protograph) QC-LDPC codes. In the multi-graph representation, the parity-check matrix is represented as a collection of edges connecting vertices (circulant of weight more than 1). The multi-graph representation is more intuitive from a graphical perspective, compare to hypergraph representation as it directly shows the connections between the variables and constraints.

Multi-edge type low-density parity-check (LDPC) codes are a class of error-correcting codes that use multiple types of edges to connect variable nodes and check nodes. These codes were initially proposed by Richardson and Urbanke and later extended by various researchers. The degree distribution of these codes is defined using a joint distribution of the degrees of variable and check nodes, where the variable node degrees are partitioned into multiple classes. This allows for more flexibility in designing the code and improving its performance. To visualize the structure of multi-edge type LDPC codes, researchers have proposed the use of protographs. A protograph is a graphical representation of the code that shows the connectivity pattern between variable and check nodes. David MacKay and Radford M. Neal first introduced protographs in the context of regular LDPC codes. Dariush Divsalar later proposed a protograph-based approach for designing multi-edge type LDPC codes. The Discalar's protograph allows for a simplified representation of the code structure, making it easier to analyze and design the code. The use of protographs has become a popular tool in the design and analysis of LDPC codes. These two approaches (Richardson-Urbanke's MET LDPC ~\cite{Ri02} and MacKay-Neal-Divsalar Protograph LDPC ~\cite{Di05})  are different ways of describing the same type LDPC code.

\subsection{Bayesian statistic at ML and Statistical Physics: local energy method, reparameterization trick}

Let's consider example of LDPC code parity-check matrix $H$ application to encoding message $m$ in error correction. Mention that nonlinear mapping and encoding represented embedding in ML.
\begin{equation} 
x=H^{-1}\times m,
\end{equation} 
where $m$ - transmitted message, $H$- parity-check matrix, $x$-encoded message, $ H^{-1} \times  H^{T}=0$ .
In error correction encoding satisfied zero syndrome condition:
\begin{equation} 
H \times x=0.
\end{equation} 

After encoding we send encoded message $x$ through noisy channel, for example AWGN-channel:

\begin{equation} 
y = x + \eta,
\end{equation} 
where $y$ -  received noisy transmitted encoded message, $ \eta $   
 channel noise distributed according some distribution, for example $ p(\eta)= N(0, \sigma^2).$

Let's using  knowledge of parity-check matrix $H$ and type of distribution $p(\eta)$ to recovery x. 

Maximum a posteriori (MAP) estimation minimizes bit error at error correction, it minimises the probability of each $i$ symbol error at $x$, $x=\{x_1, x_2, ... , x_i\}$:
\begin{equation} 
p(x_i|y)= \max_{x_i} P_{x_i|y} (x_y|y)
\end{equation} 

Maximum Likelihood (ML) estimation minimizes error at whole codeword, it finds the closest codeword to $x$:

\begin{equation} 
p(x|y)= \max_{x} P_{x|y} (x|y).
\end{equation}

MAP estimation, unlike ML, is not invariant under reparameterization. Parameterization equvivalent  as we show later, introducing a new Jacobian (use another lattice bound region, Korobov grid) that impacts on the location of the extremes.

\begin{figure}
\centering
\includegraphics[ width=5in]{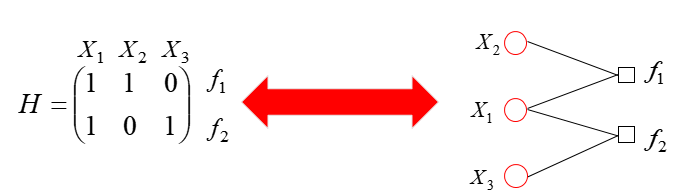} 
  \caption{Parity-check matrix and Tanner-graph for marginalization (MAP/ML estimation)}
  \label{tanner}
\end{figure}

Maximum a posteriori estimation require solution of exponential complex problem:

\begin{equation} 
p(x_i|y)= \sum_{x \slash x_i} p(x|y).  
\end{equation} 
For example for parity-check matrix and bipartite graph from figure  \ref{tanner}, joint probability density of a Tanner-graph without cycles can be divided (factorized) into multipliers (product rule) in accordance with a given structure of the factor graph (marginalization problem):
\begin{equation} 
p(x_1|y)= \sum_{x \slash x_1} p_1(f(x_1,x_2,x_3))=
\sum_{x_2} \sum_{x_3} p(f_1(x_1,x_2) f_{2}(x_1,x_3)) =
\sum_{x_2} p(f_1 (x_1,x_2))  \sum_{x_3} p(f_2(x_1,x_3)).
 \end{equation}

With an increase in the number of elements, the problem becomes exponentially complex, but the ability to divide into multipliers (factorize) inside the summation of probabilities (generalized distributed law), allows you to significantly simplify the calculation of marginals \cite{AjiMcE00}. 
Moreover, neglecting the cycles in the graph, we can obtain an approximation of this MAP(ML) estimation. It call Local rule (local system parameterization) and such approximation called Belief Propagation (Message Passing Method). Such approximation introduced pseudocodewords which make wrong closest codeword due neglecting the cycles in the graph for decrease complexity.

On the other hand, consider a quantum system in an entangled state with spin $\sigma$ of $2^N$ possible configuration, $N$-qbit:
\begin{equation} 
\langle \Psi \rangle = \sum_{\sigma_{i}}  c_{\sigma_{i}}| \sigma_{i }\rangle 	\rightarrow p(\sigma) = \lvert c_\sigma \rvert^{2} .
\end{equation} 

For each choice basis we have wave function state vector $ 
\lvert \Psi \rangle =  \sum_{x}  \Psi (x \lvert x \rangle $  from probability distribution $p(x)= \lvert \Psi (x)  \rvert .$

Estimation problem for quantum system is calculation of minimal eigenvector and values under exponential large Hilber space (similar as marginalization problem at ML/MAP estimation),  $\Psi \in H$:

\begin{equation} 
\frac{\langle \Psi  \lvert  H \lvert \Psi \rangle }{\langle \Psi \lvert \Psi \rangle   }.
\end{equation}

 \begin{figure}
\centering
\includegraphics[ width=6in]{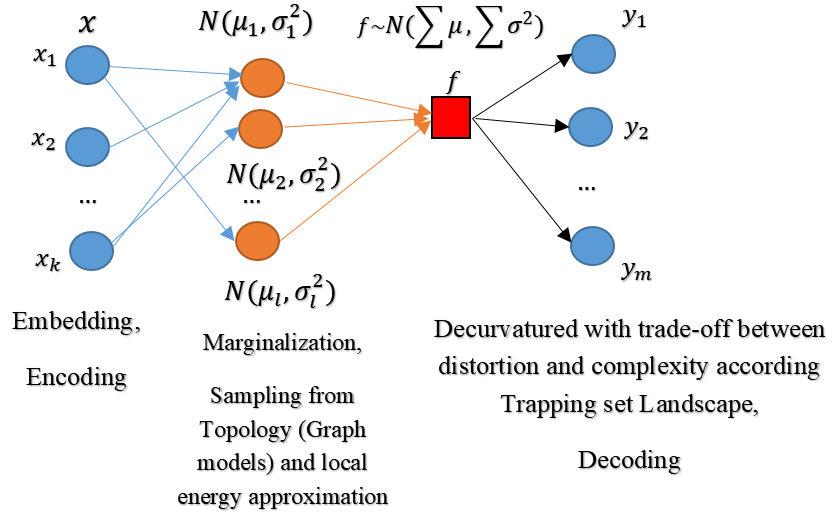} 
  \caption{Illustration of Graph model Embedding and marginalization as ML/MAP estimation (Bayesian inference)}
  \label{GraphEmb}
\end{figure}

For approximate solution of quantum eigenvalue problem use reparametrization:
\begin{equation} 
\langle H  \rangle  = \sum_{x y } H_{x y }   \Psi(y) =  \sum_{x}  \lvert  \Psi(x) \rvert^2  \sum_{y}    H_{x  y \frac{\Psi(y)}{\Psi(x)} },
\end{equation} 
where first sum is states distribution, second sum is local energy of spin configuration.

The quantum physic general distribution law in a local system can be more clearly demonstrated using the Heisenberg  matrix quantum mechanics (gauged). It was developed by Werner Heisenberg, Pascual Jordan und Max Born in the late 1920s and provided a powerful mathematical framework for understanding quantum system phenomena. In the matrix formulation, physical quantities such as position, momentum, and energy are represented matrices. These matrices act on vectors that represent the quantum states of a system. The elements of these matrices, known as matrix elements, describe the transition probabilities between different quantum states.
To calculate the transition probability from state $n$ to state $n-a$, and then from state $n-a$ to state $n-b$, we multiply the frequency for the change of energy from state $n$ to state $n-a$ by the frequency for the change of energy from state $n-a$ to state $n-b$. Mathematically, this can be expressed as:
$P(n \leftarrow n-b) = |<n-b|n>|^2 = |<n-b|n-a><n-a|n>|^2$. Here, $|n>$ represents the quantum state $n$, and $|n-a>$ and $|n-b>$ represent the quantum states $n-a$ and $n-b$, respectively. The symbol $<n-b|n>$ denotes the inner product between the states $|n-b>$ and $|n>$, and $|<n-b|n>|^2$ represents the modulus squared of this inner product, which gives the probability of the transition. Heisenberg originally devised this equation to enable himself to multiply two measurements of the same kind (amplitudes), so it happened not to matter in which order they were multiplied. Heisenberg noticed, however that if he tried to use the same schema to multiply two variables, such as momentum, $p$, and displacement, $q$, then "a significant difficulty arises", \cite{Waerd67}. It turns out that multiplying a matrix of $p$ by a matrix of $q$ gives a different result from multiplying a matrix of $q$ by a matrix of $p$ (due to non-commutativity of multiplication, under conjugated $p$, $q$ variables, see Noether's theorem, Pontryagin duality). It only made a tiny bit of difference, but that difference could never be reduced below a certain limit, and that limit involved Planck's constant, $\hbar$. The order of multiplication actually corresponds to a different way of traversing the graph (by the scheduler), which makes it possible to eliminate the approximation error (local energy) as a result of cycles (Trapping sets).

For graph without cycle (Tree, trivial Topology) it exact equation, unfortunately quantum system without symmetric cycles, codewords of large Hamming distance in channel coding, cannot be long time entangle and are easily collapsed, \cite{Kita97,Kita297,Kita03,Freed98,Bravyi98,Gleb21}. Therefore, quantum systems always contain cycles to counteract entropy. 
Main problem how to construct aprior reparametrization code on the graph with trade-off between complexity and error of estimation, Fig. \ref{GraphEmb}. The connection between machine learning and quantum learning occurs through the collapse of the quantum state, the module of the phase function, and local energy approximation surface (Trapping set loss function surface), \cite{Kout2022,LiHuTay18,Choro14}.

Considering quantum variation monte-carlo and classical variation inference problems we used simple (bipartite) Tanner-graphs without cycles, now let's complicate the graph model and show the connection with its dual model - tensor networks. Variable nodes $v_i$ can exist not only in prime fields but also in Lie groups, ${v}_i \in { GLG}$ or other algebraic structures, ~\cite{KoB16}. Moreover, multi-graphs need not be bipartite and can have multiple stages, such as in the Forney normal graph, ~\cite{For01}. These generalizations enable us to consider non-linear codes or their equivalents, such as tensor networks ~\cite{Fe14},~\cite{Me18}, ~\cite{For18}. This approach enables the consideration of a wide range of graph code types, such as hidden LDPC codes, using a special type of scheduler. These scheduler types include Polar code (Successive Cancellation Scheduler, ~\cite{Fer13,Fos15}, \cite{For01} ) and Convolution Turbo code (serially concatenated codes, ~\cite{Ji06}). This approach is highly advantageous as it allows for the utilization of the same hardware processing device for processing Turbo, Polar, and LDPC codes effectively. We aim to generalize this approach by utilizing LDPC code approximation graphs for various Deep Neural Network (DNN) architectures. This enables us to leverage the benefits of LDPC codes in increasing sparsity and decreasing the number of bits required for learning parameters.

Let's consider the hypergraph $G = (V, E)$ where $V = \{1, 2, 3, 4, 5\}$ and $E = \{\{1, 2, 3, 4\}, \{1, 5\}, \{3, 5\}\}$. In this hypergraph, $V$ represents the set of vertices and $E$ represents the set of edges. The hypergraph $G$ can be mapped to a planar graph $G'$ as shown in Figure 1.

The adjacency tensor of the hypergraph is denoted as $A = mat(G)$, which is defined as follows:

\begin{equation} 
A_{i_1i_2\dots i_m} = 
\begin{cases}
\frac{1}{(m-1)!}, & \text{if } \{v_{i_1}, v_{i_2}, \dots, v_{i_m}\} \in E \\
0, & \text{otherwise}
\end{cases}
\end{equation} 

The size of the matrix representation of $G$ is $5\times 5\times 5$. For the graph $G'$, the adjacency matrix is given by:

\begin{equation} 
A = mat(G') = 
\begin{pmatrix}
0 & 1 & 1 & 1 & 1 \\
1 & 0 & 1 & 1 & 0 \\
1 & 1 & 0 & 1 & 1 \\
1 & 1 & 1 & 0 & 0 \\
1 & 0 & 1 & 0 & 0 \\
\end{pmatrix}
\end{equation} 

The matrix $A = mat(G')$ represents the relations (correlations) between pairs of vertices:

\begin{equation} 
A_{i_1i_2} = 
\begin{cases}
1, & \text{if } (i_1, i_2) \in E \\
0, & \text{otherwise}
\end{cases}
\end{equation} 

\begin{figure}
\centering
\includegraphics[width=3.2in]{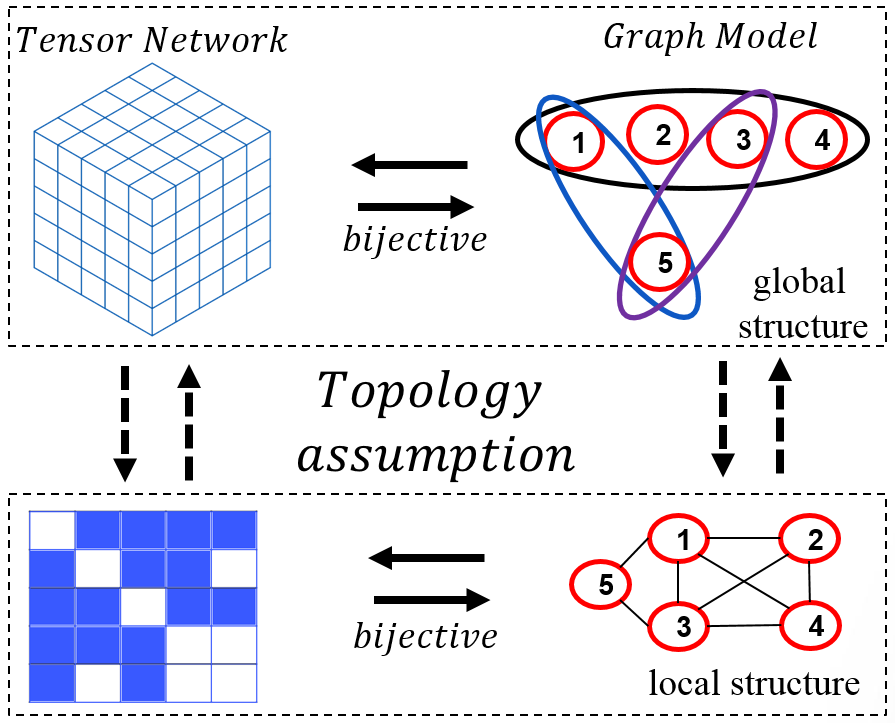} 
  \caption{Connection of Tensor Networks and Graph models}
  \label{Fig2''}
\end{figure}

The adjacency tensor and the graph model are different representations of the relations in the data. A high correlation in the data implies a large number of graph automorphisms, a low rank of the tensor network, and symmetry in the geometric representation. The physical knowledge of dynamic processes provides groups of such symmetry (automorphisms), for example, $SO(2)$ or $U(1)$ for electromagnetic waves. This simplifies graph marginalization and tensor network contraction because it allows for an effective mapping of global structures through local weakly connected representations, capturing the local structure \cite{USA_UK20,USA_UK23}. 
Understanding the properties of the global and local structures formed by a certain class of algebras (quantized by a group) is the main focus of topology (topology data analysis as well), as shown in \ref{Fig2''}.  Topology assumptions, specifically in the context of proposed coded based gauge theory, provide us with a powerful tool to select the appropriate low-dimensional orthogonal subspace for embedding data. By leveraging this capability, we can effectively reconstruct high-dimensional dynamics while maintaining controlled distortion. Once we have identified the optimal subspace, we can leverage it to reconstruct high-dimensional dynamics. This process involves mapping the complex interactions and relationships within the high-dimensional data onto the selected subspace. By doing so, we effectively capture the essential information required for reconstruction while minimizing distortion.The ability to control the level of distortion is crucial in preserving the integrity and accuracy of the reconstructed data. By carefully managing the trade-off between dimensionality reduction and fidelity preservation, we can strike a balance that ensures both efficiency and reliability in the reconstruction process. Lower bound of distortion defined by codewords - Trapping sets(a,0).  Practical distortion defined by overal Trapping set enumerators which contain codewords-Trapping sets(a,0) and pseudocodewords-Trapping sets(a,b).

\textbf{Proposal: It is possible to embed code on a graph (reparameterize) within hyperbolic toric (spherical) geometry, allowing for controllable distortion. The automorphism of a  code determines the size of its circulant, which in turn determines the number of energy minima (ground states) in the corresponding quantum system. The degree of distortion is influenced by the enumerator of trapping sets (union of codewords TS(a,0) and pseudocodewords TS(a,b)) and has a lower bound by codewords (TS(a,0)).}

The duality between code on the graph and tensor networks allows us to harness the advantages of both theories, ~\cite{Ro19} . Code on the graph construction theory enables us to optimize weight distribution by utilizing structured high symmetry (automorphism) block matrices (hardware efficient), thereby reducing complexity. Additionally, it helps improve the Trapping Sets (TS) spectrum through techniques such as Covariance evolution (~\cite{Am09} , topology equivalent channels assumption based on Luby analytical solution ~\cite{Lu97}  ) for linear size TS and Importance Sampling for sublinear size TS, ~\cite{Ri06}.

In conjunction with this, we can leverage mutual information to decrease the number of bits required for representing soft information. This includes quantizing learnable parameters, such as converting them from floating-point representation to a few bits, at high temperatures, hard decision, one bit quantization. By employing these strategies, we can achieve more efficient data representation, ultimately enhancing the overall performance and effectiveness of the DNN.

The study of tensor networks provides a powerful framework for analyzing and understanding the properties of hypergraphs and their corresponding graph models. Tensor networks offer a formalism to represent and manipulate high-dimensional data structures efficiently. In the context of hypergraphs, tensor networks allow us to capture the complex relationships and correlations between vertices in a compact and structured manner.

In the case of the adjacency tensor $A = mat(G)$, the tensor network representation enables us to explore the underlying structure of the hypergraph $G$. The adjacency tensor $A$ encodes the connectivity information between the vertices of the hypergraph. By organizing the tensor elements according to the indices representing the vertices, we can exploit the algebraic operations on tensors to analyze the relationships among the vertices.

Tensor network techniques, such as tensor contraction and graph marginalization, provide valuable tools for extracting meaningful insights from hypergraphs. Contraction operations allow us to combine and aggregate information from different vertices, revealing patterns and correlations within the data. This process is particularly useful for identifying clusters or communities within the hypergraph, as well as detecting important central nodes. Graph marginalization in the tensor network framework allows for the extraction of relevant information by collapsing or summing over certain vertices or edges. This operation helps to simplify the complexity of the hypergraph and focuses on the essential features of the data. By marginalizing out less significant vertices or edges, we can reduce the dimensionality of the problem and uncover the underlying global and local structures more effectively. Tensor network methods also enable the exploration of symmetry and invariance properties in hypergraphs. The presence of symmetries often corresponds to specific transformations or operations that leave the underlying structure unchanged. Tensor network techniques can leverage these symmetries to enhance computation and analysis efficiency. For example, certain tensor network architectures explicitly incorporate symmetries to achieve a more compact representation and exploit the redundancy in the data.

\begin{figure}
\centering
\includegraphics[width=4.in]{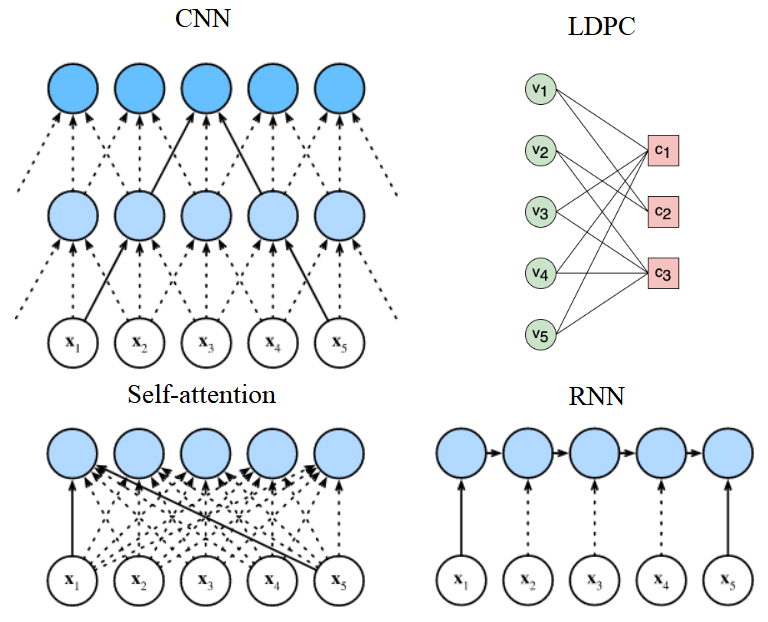} 
  \caption{Comparison of connection structures of LDPC, CNN, RNN, Transformer(Self-attention)}
  \label{Architecture}
\end{figure}

A bipartite graph is a graph whose nodes can be divided into two disjoint sets such that every edge connects a node in one set to a node in the other set. This structure is found in many machine learning models, including convolutional neural networks (CNNs), recurrent neural networks (RNNs), transformers (self-attention), and low-density parity-check (LDPC) codes, \ref{Architecture}.

In CNNs, the input image is represented as a matrix where each pixel corresponds to a node in the bipartite graph. The convolution operation can be seen as an edge between adjacent pixels in the image, forming a bipartite graph. Similarly, in RNNs, each time step corresponds to a node, and the connections between them form a bipartite graph.

Transformers utilize self-attention to compute sequence representation. Each element in the sequence corresponds to a node in the bipartite graph, and the self-attention mechanism computes attention scores between each pair of nodes, forming edges in the bipartite graph.

LDPC codes also have a bipartite graph structure. The variable nodes correspond to one set of nodes, while the check nodes correspond to another set of nodes. The edges between them are determined by the parity-check matrix. By iteratively passing messages between these two sets of nodes, an LDPC decoder can infer the original message sent through a noisy channel.

 \begin{figure}
\centering
\includegraphics[width=6.in]{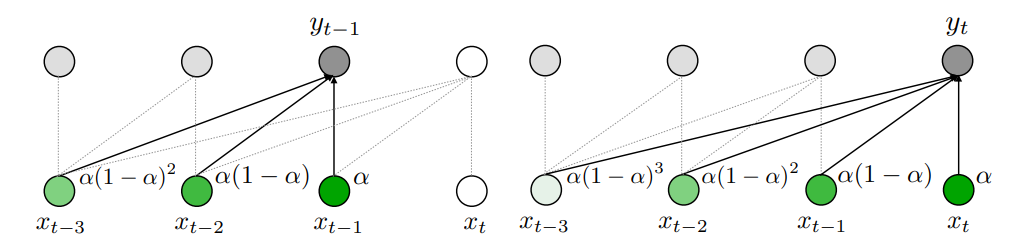} 
  \caption{Illustration of the exponential moving average (EMA) approach, which averages the input values X with weights decaying exponentially over timesteps. Moving average equipped gated
attention (Mega)}
  \label{Mega}
\end{figure}

 \begin{figure}
\centering
\includegraphics[width=6.in]{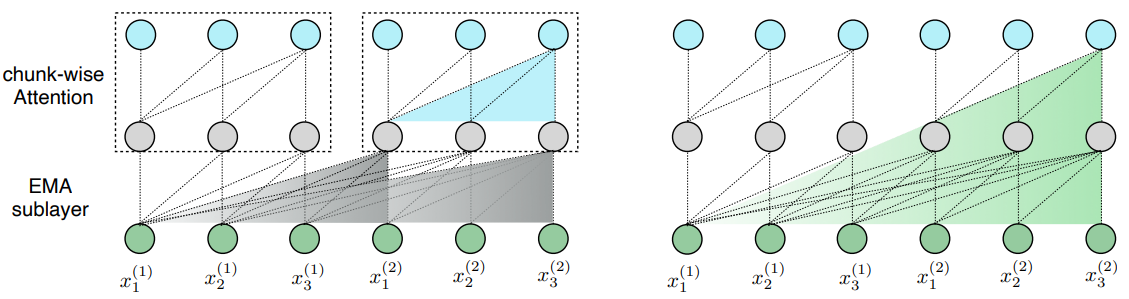} 
  \caption{Illustration of the Mega-chunk (Mega with Linear Complexity) model with two chunks of length 3}
  \label{Mega_chunk}
\end{figure}

To demonstrate a direct connection of DNN architecture to  error-correcting coding, we will consider several state-of-the-art neural network architectures from the long-range arena benchmark for Transformer, ~\cite{LRA21}. The first architecture is MEGA and MEGA-chunk from the article ~\cite{Ma22}." The EMA approach is illustrated in this article, which averages input values X with weight decaying exponentially over timesteps, Fig. \ref{Mega}. The Mega-chunk model with two chunks of length 3 is represented in the Fig. \ref{Mega_chunk}. A bipartite graph representation shows that such a neural network is equivalent to a parity-check matrix of protograph:
\begin{equation} 
H_{MEGA} = 
\begin{pmatrix}
1 & 1 & 1 & 1 \\
0 & 1 & 1 & 1 \\
0 & 0 & 1 & 1 \\
0 & 0 & 0 & 1 \\
\end{pmatrix}
.
\end{equation} 

The Mega and Mega-chunk Attention models use an Generalized Irregular Repeat Accumulate (GeIRA) protograph from ~\cite{DivMcel98, Li05}, as their basis. The Exponential averaging alpha filter used in the MEga and Mega-chunk Attention models is intended to break trapping sets and improve the accuracy of the model by preventing it from getting stuck at local minimum solutions.
 \begin{figure}
\centering
\includegraphics[width=6.in]{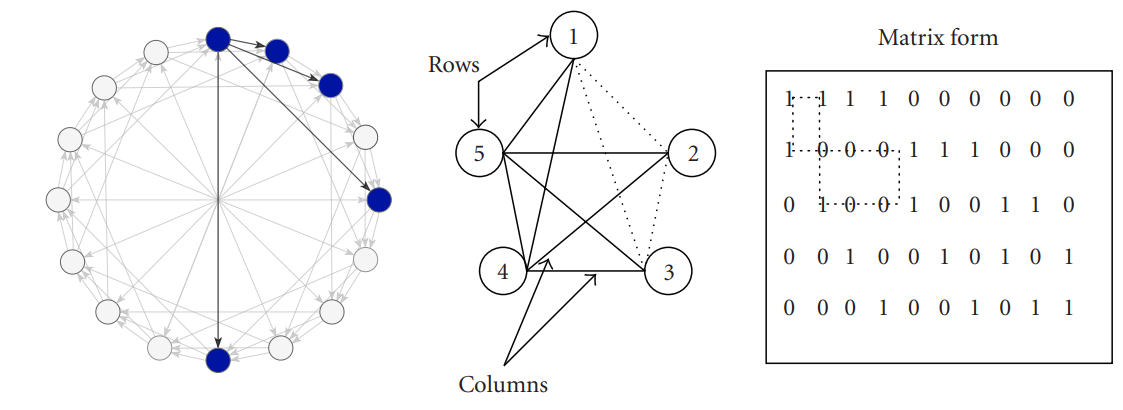} 
  \caption{Chord protocol weight 4 cage-graph. Distance graph and related LDPC code parity-check matrix}
  \label{Chord}
\end{figure}

Another state-of-the-art attention architecture from the long-range arena is presented in the article by ~\cite{Kha22}, which is based on the P2P Chord protocol described in \cite{Stoica2001}, \ref{Chord}, left. 
ChordMixer utilizes Cage graphs as distance graphs to design its attention mechanism, as shown in the research paper by ~\cite{Ma07}. Using Cage graphs allows ChordMixer to construct the attention mechanism in a way that is equivalent to the parity-check matrix of LDPC codes, \ref{Chord}. 

 \begin{figure}
\centering
\includegraphics[width=6.in]{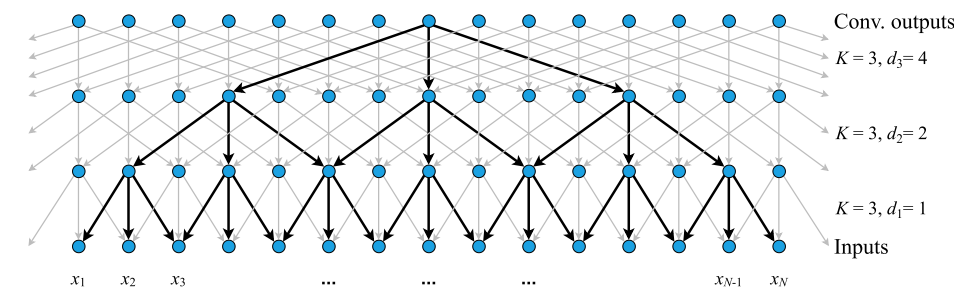} 
  \caption{The architecture of symmetric dilated convolutions. Computation graph of regular weight 3 codes}
  \label{CDIL}
\end{figure}

Lastly, we consider the dilated Convolutional model ("CDIL") from the article by ~\cite{Lei23}. This is a convolutional code that uses the 2P Chord protocol as its basis and is considered a column weight 3 convolutional code, as shown in the Fig. \ref{CDIL}. The trapping set of such a code was fully theoretically described by ~\cite{Vasic09}, proposed an effective decoding of such a code using \cite{Xi21} .

In conclusion, graph models (structured multi/hyper graph) serve as a valuable tool for characterizing the nonlinear correlations observed in data generated by dynamic systems. An essential factor to consider is the topology of the generalized channel responsible for generating the data. This includes analyzing the dynamic system's topology, identifying intersecting cycles within the graph forming code and pseudo-code words, and evaluating the operating temperature of the system, represented by the code rate. The code rate describes the ratio of the useful signal to noise or interference, acting as an entropy measure. In scenarios with high levels of noise, the symmetrical nature of the system becomes crucial, and the quality of data recovery relies on the linear size of Trapping sets, which are codewords with a large weight (Covariance/Density Evolution, "Waterfall" regime). As the temperature decreases, we transition to a mode that involves low probability events, where the thick ends of the distribution become significant. In this regime, codewords and pseudo-code words with low weight play a critical role (Importance Sampling, "Error-floor" regime). It is important to ensure that the topology of neural networks aligns with the topology and temperature of the generalized channel. This includes considering the nested system of code rates present in the graph. By establishing such correspondence, we can effectively leverage graph models in studying and understanding the intricate dynamics of complex systems. Understanding its properties such as code distance, weight spectrum enumerator, trapping sets enumerating, and density evolution threshold from  error-correcting coding methods can help in designing and optimizing (loss surface landscape) of these models.

\newpage

\section{Exploring the Connection Between Information Geometry and Hyperbolic Toric Topology of ISING model}

 The origins of information geometry can be traced back to the work of C. R. Rao, who was a pioneering statistician and mathematician. In the 1940s, Rao introduced the concept of the Fisher information matrix, which measures the amount of information that an observable random variable carries about an unknown parameter. In the 1960s, Rao extended this idea by showing that the Fisher information matrix can be interpreted as a Riemannian metric on a manifold of probability distributions. This allowed him to apply differential geometry techniques to the study of statistical models and establish a geometric framework for statistical inference. Rao's work laid the foundation for subsequent developments in the field of information geometry, which has since grown to encompass a wide range of topics in mathematics, statistics, and computer science. Nowadays, information geometry is a thriving research area with numerous applications in machine learning, signal processing, neuroscience, and quantum information theory, among others, \cite{Rao45}.  Shun'ichi Amari has made significant contributions to the modern theory of information geometry. In particular, Amari introduced the idea of using a Riemannian metric to describe the geometry of probability distributions and has developed many important results in this area, including the Amari-Chentsov tensor and the alpha-divergence family of divergence measures. He has also made significant contributions to the study of neural networks and deep learning, applying ideas from information geometry to develop new algorithms for training and optimization, \cite{Rao83,Rao00}].  In 2018, the journal Information Geometry was established, which is devoted to the field of information geometry. The journal publishes research articles, review papers, and book reviews on all aspects of information geometry, including its mathematical foundations, applications, and computational methods, \cite{Egu18, Ra18}. It serves as an important forum for researchers working in the field to share their latest results and exchange ideas. Frédéric Barbaresco, Frienk Nielsen, Nihat Au have made significant contributions to the theoretical development and application of information geometry to signal processing, 
\cite{Kha19, Fra18,FraBar21,NielBo10}. Frank Nielsen has developed various tools and techniques for geometrical data analysis, including distance functions, divergence measures, and optimization algorithms, to uncover the underlying geometric properties of data sets and probability distributions. Nielsen's work in information geometry has been influential in several areas, including machine learning, pattern recognition, and statistical analysis. His research has contributed to understanding the geometrical properties of statistical models, developing efficient algorithms for data analysis, and improving various machine learning techniques.  Barbaresco introduced the concept of information-theoretic sensor management, which involves using ideas from information theory to optimize the selection and deployment of sensors in a signal processing system. Barbaresco's work has had important implications for a wide range of applications, including radar and sonar systems, wireless sensor networks, and autonomous vehicles. His research has helped to establish a powerful theoretical framework for analyzing and designing signal processing systems based on principles of information theory. 

Xavier Pennec's article \cite{Pe18} is a  contribution to the field of geometric data analysis. The article focuses on the use of barycenter-based methods for analyzing high-dimensional data that lies on or near a manifold, which is a curved surface in a higher-dimensional space. The article introduces a new methodology called barycentric subspace analysis (BSA), which involves computing the barycenter of a set of data points and then projecting the data onto a subspace that is centered around the barycenter. This approach has several advantages over traditional methods, such as principal component analysis (PCA), which can be unstable when applied to data on manifolds. One key feature of BSA is that it is invariant to the choice of coordinate system used to represent the data. This is important when working with data on manifolds, where different coordinates can lead to different representations of the same underlying structure.

To demonstrate the relationship between the parity-check matrix representation of Quasi-cyclic LDPC codes and the ground states of the Ising model under Toric Hyperbolic Topology, we introduce the concept of information geometry. Our focus is on using the center of mass barycenter method to establish this connection. In particular, we explore the Torical topology version of this method under the Ising model, which leads us to the Quasi-Cyclic representation of LDPC codes. 
We use information geometry methods to analyze the ground states of the ISING model, and discover that this states are represented by a toric (hyperbolic) topology that can be described using Quasi-Cyclic (QC) and Quasi-Cyclic Low-Density Parity-Check (QC-LDPC) codes. Although the center of mass and information geometry seem unrelated at first glance, there is actually a deep connection between the two concepts. In particular, the center of mass can be viewed as a special case of the so-called barycenter, which is a concept from information geometry. The barycenter is a generalization of the center of mass that applies to probability distributions rather than physical objects. It describes the location of the weighted average of a set of probability distributions, where each distribution is assigned a weight based on its importance.

By using information geometry, it is possible to define the barycenter of a set of probability distributions in a rigorous mathematical way. This allows us to analyze the behavior of complex systems involving multiple distributions and optimize their performance. Projecting the data onto a subspace centered around the barycenter, it is possible to obtain a new set of coordinates that are more suitable for statistical analysis and optimization.

One example of the application of the barycenter to the Klein bottle topology of image data is in the study of graph embeddings. Graph embeddings are a technique used to map graphs onto surfaces, such as the Klein bottle, and are useful for studying their properties and relationships, \cite{Gu19}.

By using the barycenter to identify the symmetries and other geometric properties of the Klein bottle, it is possible to develop more efficient algorithms for graph embeddings and other applications in computer science and data analysis.

\subsection{QC codes and QC-LDPC codes as Hyperbolic Toric Topology Based Ground State Solution of Ising Model}

\textbf{Theorem 1: The structured (block) parity-check matrix of QC and QC-LDPC codes, under (Hyperbolic) Toric Topology represent the ground state solutions of the Ising Model (Markov Random Field).}

The Ising model is a mathematical model used to describe the behavior of ferromagnetic materials, such as iron and nickel. It consists of a lattice of spins that can be in either an "up" or "down" state, with interactions between adjacent spins influencing their orientation. 
 The Ising model has since been studied extensively in various fields, including computer science, image processing, and computer vision. In particular, it has been shown to have strong connections to Markov Random Fields (MRFs), which are widely used in image processing and computer vision as a probabilistic modeling framework for analyzing complex visual data.

MRFs are based on the concept of a graphical model, which represents a complex system as a collection of variables and their dependencies using a graph structure. In an MRF, the variables correspond to pixels or regions of an image, and the edges of the graph encode the relationships between them. By using the Ising model as a building block, MRFs can efficiently capture the spatial dependencies and correlations in image data, making them useful for tasks such as segmentation, denoising, and object recognition. In recent years, deep learning techniques have also been applied to MRFs, resulting in powerful new models such as Conditional Random Fields (CRFs) and Deep Belief Networks (DBNs). These models leverage the strengths of both graphical models and deep learning to achieve state-of-the-art results in many computer vision tasks.

The Edward-Anderson Hamiltonian Ising model is an important concept in statistical physics that was first proposed by S.F. Edwards and P.W. Anderson in 1975, \cite{EdAn75}. The model describes a physical system in which magnetic spins on a lattice interact with one another, resulting in a phase transition from a disordered to an ordered state as temperature decreases.

Marc Mézard, Nicolas Sourlas, Gérard Ben Arous, Andrea Montanari   made significant contributions to the understanding of the Edward-Anderson Hamiltonian Ising model and its phase transition properties. In particular, they developed a rigorous mathematical framework for studying the spin glasses using methods from statistical physics, probability theory, and computer science, \cite{LiHuTay18,Ge96,Me84,Sour89,Sour94,Sourlas07,Monta09}.

Numerous optimization problems, including those in computer vision (Quantum  Computer Vision, \cite{Yu22}), can be converted into a Quadratic Unconstrained Binary Optimization (QUBO) form, \cite{Ble23}. However, QUBO problems are typically NP-complete, which implies that finding solutions through classical means necessitates exploring an exponentially growing solution space as problem size increases. In contrast, quantum computing holds the promise of exponentially faster computation due to the superposition nature of qubits. The exponentially expanding Hilbert space of a quantum system naturally accommodates the solution space of combinatorial optimization problems, offering potential advantages over classical machines in solving such problems. The Quantum Approximate Optimization Algorithm (QAOA) is specifically designed to address QUBO problems by utilizing a quantum circuit to find approximate solutions. The goal is to overcome the inherent difficulty of approximation in classical computation by harnessing the capabilities of QAOA. QAOA is a quantum algorithm designed to solve combinatorial optimization problems. It utilizes a quantum circuit to find approximate solutions by iteratively applying a sequence of parameterized quantum gates. Given an optimization problem with a cost function to minimize, the QAOA constructs a time-dependent Hamiltonian that encodes the problem's objective function. This Hamiltonian is represented as a sum of Pauli operators acting on qubits. The algorithm then prepares an initial state on the qubits and applies alternating sequences of two types of gates: the problem-dependent mixer Hamiltonian and the driver Hamiltonian.

The mixer Hamiltonian evolves the state towards a superposition of different computational basis states, allowing exploration of different candidate solutions. The driver Hamiltonian serves to rotate the state closer to the optimal solution by implementing a unitary transformation. The final step involves measuring the qubits in the computational basis to obtain a classical bit string representing a potential solution. By running iterations of the QAOA with varying parameters and measuring the corresponding bit strings, one can determine approximate solutions for the original optimization problem. While not guaranteed to find the global optimum, the QAOA offers a promising approach to tackle hard optimization problems using quantum computers.

Fermionic QAOA is a well-suited approach for solving combinatorial optimization problems that involve constraints. It leverages the concept of fermion particle number preservation to naturally enforce these constraints, making it ideal for problem instances with specific limitations. The Fermionic QAOA employs a driver Hamiltonian based on fermions to handle optimization problems that incorporate constraints. This article provides valuable insights and analysis specifically focused on the Fermionic QAOA algorithm. In our future article, we will explore the application of codes on graph-based representations for bosonic systems, \cite{Pont23}.

 QUBO problems fall under the NP-complete class, guaranteeing that any NP-complete problem can be efficiently transformed into a QUBO problem. This mapping of Karp's 21 NP-complete problems to QUBO is extensively discussed in paper \cite{Lu23}. In addition to MaxCut, other relevant optimization problems such as Graph Coloring   \cite{Ta20}, Number Partitioning, and Quadratic Knapsack (\cite{Glo19}) have been successfully formulated as QUBO problems.

In a QUBO problem, the unknown vector $\mathbf{x} = (x_1,\ldots,x_n)$ consists of decision variables taking discrete binary values, i.e., $\mathbf{x} \in \{0,1\}^n$. The problem is defined by a square symmetric matrix $\mathbf{Q} \in \mathbb{R}^{n \times n}$. The objective is to find the optimal vector $\mathbf{x}^*$ that minimizes the cost function:
\begin{equation} 
C(\mathbf{x}) = \mathbf{x}^T \mathbf{Q} \mathbf{x} = \sum_{i, j=1}^{n} Q_{ij}x_i x_j
\end{equation} 

QUBO problems can also be framed as maximization problems by inverting the sign of the cost function. It is important to note that QUBO problems do not have any constraints on the variables $\mathbf{x}$.

QUBO instances are closely related to Ising models. They can be mapped to each other with a one-to-one correspondence, where the QUBO variables $\mathbf{x} \in \{0,1\}^n$ are replaced by Ising variables $\mathbf{z} \in \{-1,1\}^n$, with $z_i = 2x_i - 1$ for $i=1,\ldots,n$. The Ising Hamiltonian, dependent on $\mathbf{z}$, is equivalent to the QUBO cost function, with a constant term irrelevant for optimization. The relationship between QUBO and Ising models is explained in more detail in paper \cite{Glo19}. Inspired by the adiabatic theorem, annealing methods are commonly employed to find the ground state of physical systems. Similarly, solutions to Ising problems often utilize annealing techniques, \cite{Mo22}. As QUBO problems can be formulated as Ising models, they can be effectively solved through quantum annealing using adiabatic quantum computing, \cite{Da19}.

\begin{figure*}[t]
	\begin{center}
            \includegraphics[width=3.in]{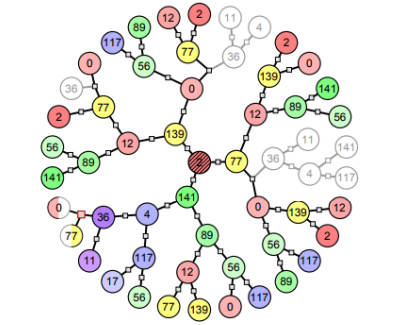}
	\end{center}
	\caption {Covering Tree of QC-LDPC Tanner Code
[155,64,20] for p=4 layers, 4 decoding iteration \cite{Ta001}}
	\label{fig:images00007}
\end{figure*}

The Quantum Approximate Optimization Algorithm (QAOA) is a versatile algorithm developed to tackle combinatorial optimization problems. Its effectiveness improves with the increase in the number of layers, denoted as $p$, \cite{Fa22}. When applied to the Sherrington-Kirkpatrick Ising model, the QAOA exhibits similarities to the back-propagation optimization technique used in training deep neural networks (DNNs). This similarity gives rise to a shared challenge involving Trapping sets pseudo-codewords, which bears resemblance to the Belief Propagation (BP) soft decoding process. Furthermore, the depth of the QAOA, represented by the parameter $p$, corresponds to the number of iterations executed in the Wiberg decoding tree for BP decoding (in~Fig.~\ref{fig:images00007}), \cite{Wibb96}. The value of $p$ aligns with the number of spins connected by the parity-check matrix and affects the number of iterations in the Belief Propagation decoding algorithm or Mézard's Replica symmetry breaking method (Cavity method), \cite{Me87}. The paper \cite{Bai17}  demonstrates the analysis of the free energy in the bipartite spherical Sherrington-Kirkpatrick model. The findings reveal that the free energy tends to follow a Gaussian distribution when the temperature is above the critical temperature. Conversely, when the temperature falls below the critical point, the free energy tends to converge to the GOE Tracy-Widom distribution. This behavior is well-established through density evolution, particularly in more precious terms of covariance evolution(Ornstein-Uhlenbeck stochastic process, \cite{Am09,Sti16}), observed in the waterfall region. Additionally, importance sampling is utilized in the error floor region for codes on the graph for phase transaction when the temperature falls below the critical point \cite{
Cole0Wi,Cole06,Cole08,CheSte07,CheSte07_2,CCSV09,CheSte11,VasCNP09,KaBa12,RaDeVa20,AbDeDiRy10,
ToBa14,BuSi14,VeSuDra18,PaShiChu12,KiMyJe15,KaBa20,Faba21,RoYt09}. Yedidia at al. significantly expanded Belief propagation analysis of Pearl \cite{Pe88},  through their paper \cite{Yedi00,Yedi05}. This publication delved into the understanding of belief propagation and its broader implications. 
Forney's normal graph model, \cite{For01}  included an insightful anticipation of the development of polar codes by clustering cycles in the Reed-Muller code model, and create normal graph model for multivariate functions of statistical physics systems, while Vontobel-Smarandache extensively researched the sublinear size of trapping sets, Bethe permanents, Bethe energy, and bounds on pseudocodewords, \cite{VoKo03,
Voko11,Sm11}. 

\textbf{Code on the graph research(MacKay, Luby, Wibberg decoding tree, Yedidia, Forney, Vontobel, Smarandache and other) revealed an important observation regarding the concavity of the Bethe entropy function. Generally, the Bethe entropy function is not concave. However, when considering codes on the graph and appropriately parameterizing (embedding) it (under Toric Hyperbolic and Spherical Topology, which we show below), the Bethe entropy function exhibits concavity, \cite{Voko11,MiSe19}. Complexity and error if Bethe energy minimization depend from Trapping set enumerators (codewords - TS(a,0) bounded by permanent \cite{Sma12} and pseudocodewords - TS(a,b)  betha permanents, graph cover/stochastic matrix property/multivariate function optimization surface/stable solutions of a linear dependent local subsystem of equations \cite{CheTan05}) and estimate by Covariance Evolution (linear size Trapping Sets, high temperature, low SNR) and Importance Sampling (sublinear size Trapping Sets, Low temperature, high SNR). }

\newpage

Presented herein is a detailed exposition that offers evidence supporting the claim that the ground state solutions of the Ising Model are accurately represented by the parity-check matrix of  QC-LDPC codes within the realm of (Hyperbolic) Toric Topology.

Consider Edward-Anderson Hamiltonian Ising (spin) model:
\begin{equation} 
 H_{EA}=-\sum_{i=1,\dots ,n}{\sum_{a=1,\dots ,m}{C_{ij}J_{ij} \sigma_i\sigma_j}}, 
\end{equation} 
where $C_{ij}~$- connectivity matrix, the element of which are 1 if two spin interact and 0 otherwise, $J_{ij}$- is weigh interaction power between spins , $\sigma_i$- are Ising spins, $n$ -number of column (variable nodes), $m$ - number of rows (check nodes).  The $J_{ij}$s give the strength of the two-spin interaction and are usually taken as independent random variables with a known probability distribution, where $J_0$ denote mean, $\triangle J^2$ the variance of distribution.
 $H_{ECC}$ could be rewrites to
  infinite-range model:
\begin{equation} 
H_{ECC}  =  -\sum_p{\sum_{i_{1,\dots ,i_p}=1,\dots ,n}{C^{(p)}_{i_{1,\dots ,i_p}}J_{i_{1,\dots ,i_p}}\sigma_{i_1}\dots \sigma_{i_p}}},
\end{equation} 
where $J^{(p)}_0,\triangle \ J^2_{(p)}$ mean and variance of $J_{ij}$.

Let's $a_i$  be $n$ bits of codeword and $\sigma_i=2a_i-1$ the spins associated with them. The coded message, that is,  the input to the channel, corresponds  to the matrix element $J^0_{i_{1,\dots ,i_p}}=\sigma_{i_1}\dots \sigma_{i_p}$. Decoding and estimation of marginal corresponding to finding the ground state of Hamiltonian $H_{EA_{\inf}}$. The $Js$ from channel equal to the $J^0$ plus noise. Code parity-check matrix is specified by the connective matrices $C$. 

For one value of p in $H_{ECC}$ and coordinate number $z_i=\sum_{j_{2,\dots ,j_p}}{C_{i,j_{2,\dots ,j_p}}}=z$ is independent of $i$, rate of the code equal $R={p} / {z}$. If $p\to \infty $ we get random energy model, if $\ p=2$ we get Sherrington-Kirkpatrick Spin Glasses for which each spin interact with all other spins. For decreasing compexity of ground state estimation in Sum-Product we can assume that each variable node has a tree “neighborhood. Such model of short range correlation at spin glass model used in Replica symmetry breaking method (Cavity method). 
Since the code which has the ability to correct error contain cycles, the residual curvature of the resulting universal covering tree contain curvatures.

The Ising Model is an example of Markov Random Fields, which is closely related to vision and image processing problems. In this subsection, we demonstrate the causes of symmetry, specifically the quasi-cyclic symmetry that arises in Markov random fields. As a result, such types of symmetries will occur in ML data manifolds. Quasi-cyclic LDPC automorphisms appear under (quasi)toric boundary conditions of the Ising model. Such conditions with hyperbolic curvature on the surface allow for sparse structured equilibrium states (Hamiltonian ground state) of Ising systems.

According to the Ising model, each element has a spin value equal to +1 or -1. Each of the options for the arrangement of particles and spins is associated with the energy
\begin{equation} 
E\left(\sigma \right)=\sum _{i,j}C_{ij} J_{ij} \sigma _{i} \sigma _{j}  ,
\end{equation} 
where $C_{ij} $ is the coupling matrix element (parity-check matrix) and $\sigma _{i} ,\sigma _{j} $ is the interaction energy between spins.

\begin{figure*}[t]
	\begin{center}
            \includegraphics[width=3.in]{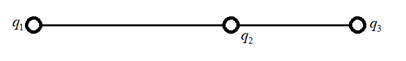}
	\end{center}
	\caption {Example of 3 particles under one line (1D) interaction system}
	\label{fig:images00}
\end{figure*}

Let's consider the interaction of particles (for convenience of presentation, we assume they have charges of the same sign) 1, 2, and 3 in the one-dimensional case using the Ising model, as shown in Figure \ref{fig:images00}.

In the Ising model, we can represent the interaction energy between spins as a function of the distance between them, where closer spins have a stronger interaction energy. This leads to the more general description of $J_{ij}$ being inversely proportional to the distance between the spins:

\begin{equation} 
J_{ij} \mathtt{\sim} \frac{1}{r_{ij}}. 
\end{equation} 

Therefore, in the Ising model, we can relate the interaction energy between particles to the coupling matrix element between spins under spin-fermion  model.

Charges 1 and 3 have the same sign and create an electric (magnetic) field. The energy of the field created by charge $i$ at point $j$ is directly proportional to the magnitude of the charge and inversely proportional to the square of the distance between the points: 

\begin{equation} 
E_i = k\frac{Q_i}{r_{ij}^2},
\end{equation} 

where $k$ is a constant of proportionality and $Q_i$ is the charge of particle $i$.

\subsection{Quasi-cyclic as ground states of ISING model, ring of polynomial, one row of circulants}

We define values $q_{1}, q_{3}, r_{12}^2=R_1$, and $r_{32}^2=R_3$ such that the electric field created by charge 1 at point 2 balances the field created at point 2 by charge 3. To determine the possible states of the system in which the magnitude of the field created by neighboring charges at the desired point reaches a minimum, we introduce the concept of a circulant matrix denoted by $I_n$. This is a unit matrix of size $e\times e$, subjected to cyclic value shift.

Since the distances between particles are rational numbers, they can be represented as proportional integers. The size of the circulant, which we consider as the maximum possible analogue of the distance between charges 1 and 3, is introduced as:
\begin{equation} 
e=\sum\left(R_{1},R_{3}\right). 
\end{equation} 
This allows us to represent the interactions between the particles using the elements of the circulant matrix.

From a physical point of view, the charge values are multiples of an elementary charge, which means they can be represented as an integer (discretely quantized) number of elementary charges. Thus, integer ratios of possible charges to the square of the distance, not exceeding the size of the circulant, will be equal to the cyclic shift of the circulant:
\begin{equation} 
\frac{q_i}{R_i}=n, 0\leq n<e.
\end{equation} 
Here, $n$ is an integer representing the number of elementary charges and $e$ is the size of the circulant. This allows us to represent the charges and their interactions using the elements of the circulant matrix, making it possible to use tools from algebraic graph theory to analyze the system.

Thus, the minimum of the field energy at the desired point forms a set of pairs of circulants of the form:
\begin{equation} 
\begin{array}{l} {\frac{q_{1}}{R_{1}}=n_{1}; \frac{q_{3}}{R_{3}}=n_{3},} \\ {n_1+n_3=e-1.} \end{array}
\end{equation} 
Here, $n_1$ and $n_3$ are integers representing the number of elementary charges in particles 1 and 3, respectively.

The matrix describing the interaction of electric fields 1 and 3 at point 2, where the fields at point 2 are balanced, can be represented as:
\begin{equation} 
H=\left[I_{n_{1}}|0| I_{n_{3}}\right].
\end{equation} 
In the case of three particles, the zero circulant can be removed since it does not contribute to the interaction energy between particles.

\begin{figure*}[t]
	\begin{center}
            \includegraphics[width=6.in]{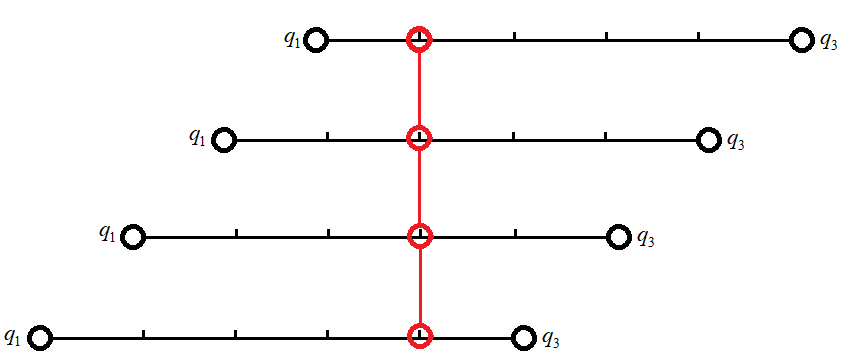}
	\end{center}
	\caption {Example of Quasi-Cyclic Automorphism described equilibrium (ground) state of the 3 particles system}
	\label{fig:images2}
\end{figure*}

Let's consider a simple example where $e=\sum\left(R_1, R_3\right)=5$, which corresponds to a circulant size of 5. This is illustrated in Figure \ref{fig:images2}.

\begin{figure*}[t]
	\begin{center}
            \includegraphics[width=\textwidth]{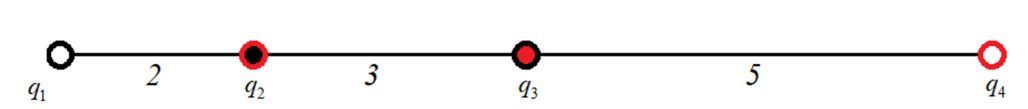}
	\end{center}
	\caption {Particle configuration after interaction screening }
	\label{fig:images001}
\end{figure*}

In this example, we have a possible set of circulants that delivers the minimum field energy at point 2, in the form:
\begin{equation} 
\begin{array}{l} {\left[I_{0}| I_{5}\right]} \\ {\left[I_{1}| I_{4}\right]} \\ {\left[I_{2}| I_{3}\right]} \\ {\left[I_{3}| I_{2}\right]} \\ {\left[I_{4}| I_{1}\right]} \\ {\left[I_{5}| I_{0}\right]} \end{array}.
\end{equation} 
However, the circulant with zero shift (the identity matrix) is the equilibrium (ground) state of the system, which corresponds to an infinite number of charge configurations. Let us single out from the set those states that can be used to describe the final system:
\begin{equation} 
\begin{array}{l} {\left(\begin{array}{l} {\left[I_{0}| I_{5}\right]} \\ {\left[I_{1}| I_{4}\right]} \\ {\left[I_{2}| I_{3}\right]} \\ {\left[I_{3}| I_{2}\right]} \\ {\left[I_{4}| I_{1}\right]} \\ {\left[I_{5}| I_{0}\right]} \end{array}\right)\Rightarrow \left(\begin{array}{l} {\left[I_{1}| I_{4}\right]} \\ {\left[I_{2}| I_{3}\right]} \\ {\left[I_{3}| I_{2}\right]} \\ {\left[I_{4}| I_{1}\right]} \end{array}\right)} \end{array}.
\end{equation} 
This reduced set contains the states that can be used to describe the final system.

Proposition 1$\dagger$: For a system of four particles in one dimension described by the Ising model, with charges having the same sign and creating an electric (magnetic) field, and distances between particles given by $r_{12}^{2}=a$, $r_{23}^{2}=b$, and $r_{34}^{2}=a+b$, the possible states of the system that result in the lowest magnitude of the field energy at a desired point correspond to a set of pairs of circulants. Due to the interaction screening (localization) of particles, particle 1 does not affect particle 3, and particle 4 does not affect particle 2. The matrices describing the effect of particles 1, 3 on particle 2 and particles 2, 4 on particle 3 are defined by circulant matrices of different sizes, which can be adjusted to the same size using the least common multiple of the distances between particles. The final parity-check matrix has a circulant size equal to the adjusted size and contains circulants with shifts values of 16, 15, 24, and 25. Such system desribed by Quasi-cyclic parity check matrix $H=\left[I_{n_{1}}|0| I_{n_{3}}\right]$.

\subsection{LCM as law of different sizes QC ground states combining (renormalization group), from several row of different size of circulants to one row representation}

\begin{figure*}[t]
	\begin{center}
            \includegraphics[width=5.in]{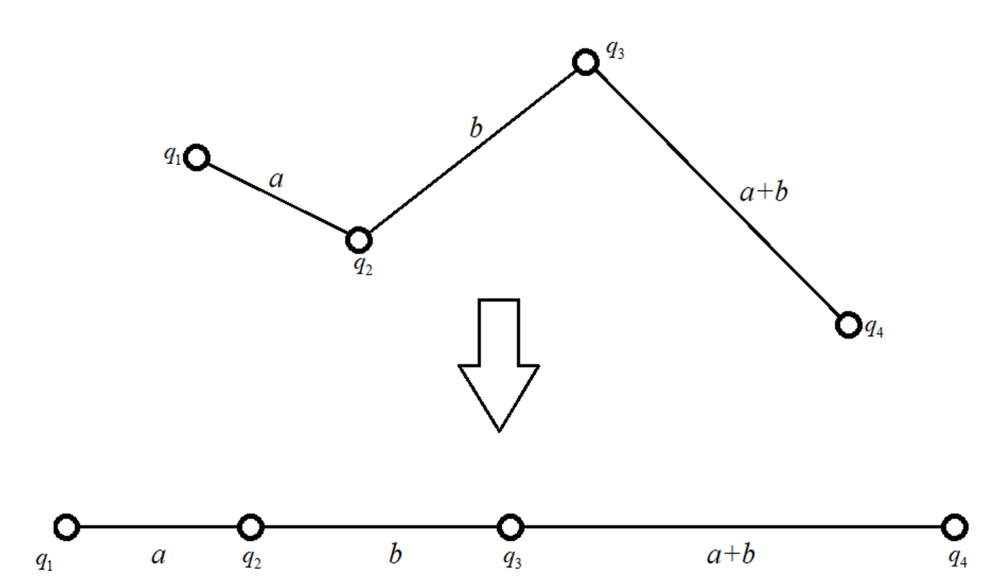}
	\end{center}
	\caption {Projection of a two-dimensional problem onto a one-dimensional case}
	\label{fig:images0001}
\end{figure*}

Let's consider the problem of four particles located on a straight line, which is the projection of the two-dimensional Ising problem. This is illustrated in Figure \ref{fig:images0001}. Let the squares of distances between particles be:
\begin{equation} 
\begin{array}{l} {r_{12}^2=a} \\ {r_{23}^2=b} \\ {r_{34}^2=a+b}\end{array}
\end{equation} 
Here, $a$ and $b$ represent the squares of the distances between the particles.

Due to the interaction screening (localization) of particles, particle 1 does not affect particle 3, and particle 4 does not affect particle 2. 

Let's find solutions for two problems: the equilibrium/ground state of particles 1-3 and the equilibrium/ground state of particles 2-4.

For the 1-3 system, the matrix describing the interaction between charges can be represented as:
\begin{equation} 
\begin{array}{cccc} {I_{a}^{a+b}} & {0} & {I_{b}^{a+b}} & {0} \\ {0} & {I_{b}^{a+2b}} & {0} & {I_{a+b}^{a+2b}} \end{array}
\end{equation} 
This matrix consists of four circulants. The first circulant represents the interaction between particles 1 and 2, the second circulant represents the interaction between particles 2 and 3, the third circulant represents the interaction between particles 3 and 4, and the fourth circulant represents the interaction between particles 1 and 4. Here, $I_p^o$ denotes a circulant of size $o$ with a cyclic shift of $p$.

To bring the circulants to the same size, we can use the following relationships:
\begin{equation} 
\begin{array}{l} {c=\text{LCM}\left(a+b,a+2b\right)} \\ {\overline{a}=\frac{a\cdot c}{a+b}; \overline{b_1}=\frac{b\cdot c}{a+b}} \\ {\overline{b_2}=\frac{b\cdot c}{a+2b}; \overline{a+b}=\frac{\left(a+b\right)\cdot c}{a+2b}} \end{array}
\end{equation} 
Here, $\text{LCM}(x,y)$ denotes the least common multiple of $x$ and $y$. The bars over $a$, $b_1$, $b_2$, and $a+b$ denote the adjusted values.

Using these relationships, we can obtain a system of interaction of four particles in quasi-cyclic notation:
\begin{equation} 
\begin{array}{cccc} {I_{\overline{a}}^{c}} & {I_{\overline{b_1}}^{c}} & {I_{\overline{b_2}}^{c}} & {I_{\overline{a+b}}^{c}} \end{array}
\end{equation} 
This matrix consists of four circulants of the same size $c$, which corresponds to the minimum field energy at points 1-3 and 2-4.

Let's consider an example with the following distances:
\begin{equation} 
\begin{array}{l} {r_{12}^{2} =2} \\ {r_{23}^{2} =3} \\ {r_{34}^{2} =5} \end{array}
\end{equation}

Describing the effect of particles 1, 3 on particle 2 and particles 2, 4 on particle 3, we can obtain the following matrix:

\[\begin{array}{cccc} {I_{2}^{5}} & {0} & {I_{3}^{5}} & {0} \\ {0} & {I_{3}^{8}} & {0} & {I_{5}^{8}} \end{array}\]

This matrix consists of four circulants with sizes 5 and 8. The first circulant represents the interaction between particles 1 and 2, the second circulant represents the interaction between particles 2 and 3, the third circulant represents the interaction between particles 3 and 4, and the fourth circulant represents the interaction between particles 1 and 4.

Note that the circulants have different sizes due to the need for adjustment to ensure that they all have the same size. In this case, the adjusted sizes are 5 and 8. This matrix contains the information about the possible charge configurations that result in the minimum field energy at points 2-3.

To bring the description to circulants of the same size, we can use the following relationships:
\begin{equation} 
\begin{array}{l} {c=\text{LCM}(5,8)=40} \\ {\overline{2}=\frac{2\cdot c}{2+3}=16; \overline{3}=\frac{3\cdot c}{2+3}=24} \\ {\overline{5}=\frac{5\cdot c}{3+5}=15; \overline{2+3}=\frac{(2+3)\cdot c}{3+5}=25}\end{array}
\end{equation} 
Using these relationships, we can obtain a parity-check matrix with circulant size 40 and cyclic shift values of 16, 15, 24, and 25:
\begin{equation} 
\begin{array}{cccc} {I_{16}^{40}} & {I_{15}^{40}} & {I_{24}^{40}} & {I_{25}^{40}} \end{array}
\end{equation} 
This matrix contains the information about the possible charge configurations that result in the minimum field energy at point 2-3.

Using the above result, we can formulate Proposition 2$\dagger$:

Proposition 2 $\dagger$ For four particles located on a straight line, which is the projection of the two-dimensional Ising problem Exist Quasi-cyclic parity-check matrix which represent ground states of Ising model.

The Proposition states that for a system of four particles located on a straight line in the projection of a two-dimensional Ising problem, two circulant matrices can be obtained to find the equilibrium/ground state of particles 1-3 and particles 2-4. The interaction between charges is described using matrices consisting of four circulants, with each circulant representing the interaction between two adjacent particles. The circulants can be adjusted to the same size using relationships involving the least common multiple of the distances between particles. The result is a system of interaction in quasi-cyclic notation with circulants of the same size containing information about the possible charge configurations resulting in the minimum field energy at points 1-3 and 2-4. This theorem provides a method for finding the equilibrium/ground state of charges in a system of four particles on a straight line.

In other words, if we have a system of three particles in one dimension with charges $q_1$ and $q_3$ creating an electric or magnetic field, we can find the set of pairs of circulants that correspond to the lowest magnitude of the field energy at a desired point. The matrix describing the interaction of fields 1 and 3 at the desired point will be given by circulant matrices of size $e\times e$ subjected to cyclic value shift. However, only certain charge configurations from the set of all possible states will correspond to the final equilibrium state of the system.

The ground state of the system can be represented by a quasi-cyclic parity-check matrix. The matrix will consist of circulants of the same size, which are subjected to cyclic value shift. The parity-check matrix describes the interaction between charges and contains the information about the possible charge configurations that result in the minimum field energy at the desired point.

To find the possible charge configurations that result in the equilibrium state of the system with the lowest magnitude of the field energy at a desired point, we can use the same approach as in the one-dimensional case. We need to construct a parity-check matrix that describes the interaction between charges.

We can then use these distances to calculate the interaction between charges using Coulomb's law or the equivalent equation for magnetic fields,  \cite{Spa04}. Using the results obtained in the one-dimensional case, we can construct a quasi-cyclic parity-check matrix that describes the interaction between charges in the two-dimensional case.

Note that the size of the circulants in the parity-check matrix will depend on the distances between particles and may not all be the same. Therefore, we may need to adjust the circulant sizes to ensure that they are all the same. We can use the least common multiple of the circulant sizes to obtain a common size for all circulants in the matrix.

In summary, we can use the same approach as in the one-dimensional case to find the possible charge configurations that result in the equilibrium state of the system with the lowest magnitude of the field energy at a desired point. We can construct a parity-check matrix that describes the interaction between charges in the two-dimensional case using the distances between particles and the results obtained in the one-dimensional case.

\subsection{Interaction of four particles in two-dimensional space without projecting the particles onto one axis}

\begin{figure*}[t]
	\begin{center}
            \includegraphics[width=7.in]{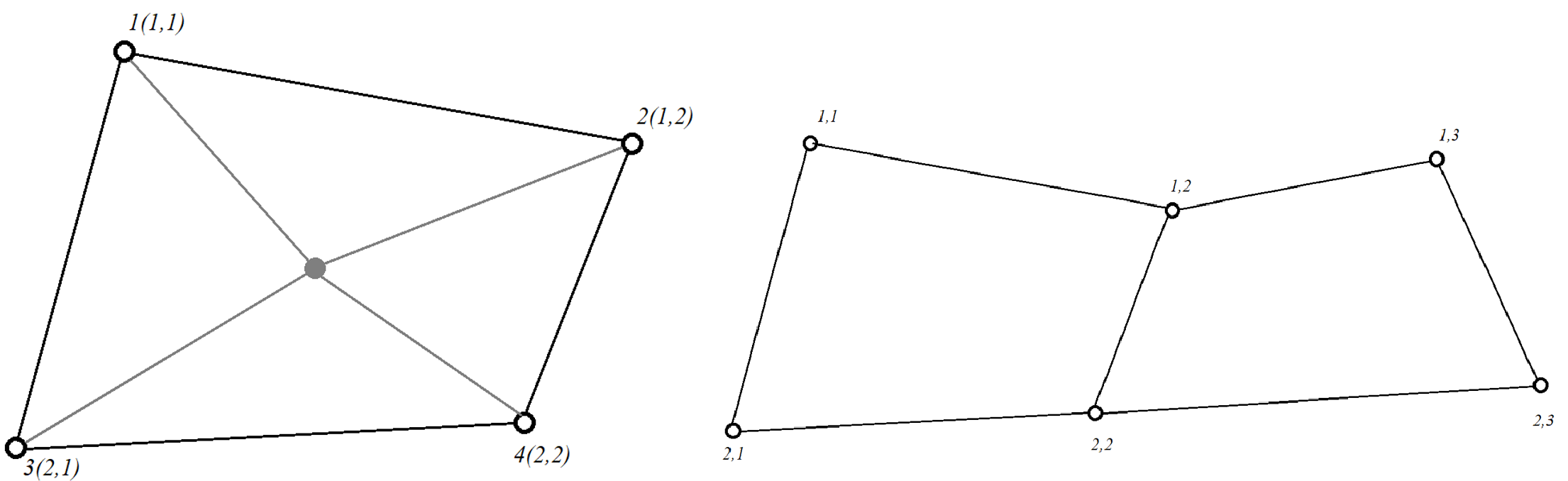}
	\end{center}
	\caption {(left) Interaction of 4 particles in 2d space. (right) Interaction of 6 particles in 2d space}
	\label{fig:images0011111}
\end{figure*}

Consider now the problem of the interaction of four particles in two-dimensional space, without projecting the particles onto one axis (in~Fig.~\ref{fig:images0011111}) left).

In the case of four particles in two-dimensional space, we can construct a parity-check matrix with two circulant rows: one for interaction along the X-axis and one for interaction along the Y-axis. The order of each circulant will be determined by the sum of the projections of distances between particles onto the respective axis. We can define these values as:
\begin{equation} 
\begin{array}{l} {L_{x} =\sum _{i=1}^{4}L_{x_{i} x^{*} }  ;L_{x_{i} x^{*} } =L_{x_{i} } } \\ {L_{y} =\sum _{i=1}^{4}L_{y_{i} y^{*} } ;L_{y_{i} y^{*} } =L_{y_{i} }  } \end{array}
\end{equation} 
We can then define an interaction matrix as:
\begin{equation} 
H^{j} =\left(\begin{array}{cccc} {I_{L_{x1}^{j} }^{L_{x}^{j} } } & {I_{L_{x2}^{j} }^{L_{x}^{j} } } & {I_{L_{x3}^{j} }^{L_{x}^{j} } } & {I_{L_{x4}^{j} }^{L_{x}^{j} } } \\ {I_{L_{y1}^{j} }^{L_{y}^{j} } } & {I_{L_{y2}^{j} }^{L_{y}^{j} } } & {I_{L_{y3}^{j} }^{L_{y}^{j} } } & {I_{L_{y4}^{j} }^{L_{y}^{j} } } \end{array}\right)
\end{equation} 
However, we may need to adjust the sizes of the circulants in the interaction matrix to ensure that they are all the same. We can do this by finding the least common multiple of $L_{x}^{j}$ and $L_{y}^{j}$ and scaling the sizes of the circulants accordingly:

\[\begin{array}{l} {L^{j} =LCM\left(L_{x}^{j} ,L_{y}^{j} \right)} \\ {\overline{L_{xi}^{j} }=L_{xi}^{j} \cdot \frac{L^{j} }{L_{x}^{j} } ;\overline{L_{yi}^{j} }=L_{yi}^{j} \cdot \frac{L^{j} }{L_{y}^{j} } } \end{array}\]

The final form of the interaction matrix will be:
\begin{equation} 
H^{j} =\left(\begin{array}{cccc} {I_{\overline{L_{x1}^{j} }}^{L_{}^{j} } } & {I_{\overline{L_{x2}^{j} }}^{L_{}^{j} } } & {I_{\overline{L_{x3}^{j} }}^{L_{}^{j} } } & {I_{\overline{L_{x4}^{j} }}^{L_{}^{j} } } \\ {I_{\overline{L_{y1}^{j} }}^{L_{}^{j} } } & {I_{\overline{L_{y2}^{j} }}^{L_{}^{j} } } & {I_{\overline{L_{y3}^{j} }}^{L_{}^{j} } } & {I_{\overline{L_{y4}^{j} }}^{L_{}^{j} } } \end{array}\right)
\end{equation} 
This matrix describes the interaction between charges in two-dimensional space and contains the information about the possible charge configurations that result in the lowest magnitude of the field energy at a desired point.

\subsection{2D ISING ground states QC representation under Toroidal Topology}

Let's consider a situation involving the interaction of six particles in two-dimensional space, where a pair of particles is involved in both j and j+1 cells (as shown in Fig. \ref{fig:images0011111}). We can define the parity-check matrices for these two adjacent cells as:
\begin{equation} 
  H^{j} = \begin{pmatrix} I_{\overline{L_{x1}^{j}}}^{L_{}^{j}} & I_{\overline{L_{x2}^{j}}}^{L_{}^{j}} & I_{\overline{L_{x3}^{j}}}^{L_{}^{j}} & I_{\overline{L_{x4}^{j}}}^{L_{}^{j}} \\ I_{\overline{L_{y1}^{j}}}^{L_{}^{j}} & I_{\overline{L_{y2}^{j}}}^{L_{}^{j}} & I_{\overline{L_{y3}^{j}}}^{L_{}^{j}} & I_{\overline{L_{y4}^{j}}}^{L_{}^{j}} \end{pmatrix},  H^{j+1} = \begin{pmatrix} I_{\overline{L_{x2}^{j}}}^{L_{}^{j}} & I_{\overline{L_{x2}^{j}}}^{L_{}^{j+1}} & I_{\overline{L_{x4}^{j}}}^{L_{}^{j}} & I_{\overline{L_{x4}^{j}}}^{L_{}^{j+1}} \\ I_{\overline{L_{y2}^{j}}}^{L_{}^{j}} & I_{\overline{L_{y2}^{j+1}}}^{L_{}^{j+1}} & I_{\overline{L_{y4}^{j}}}^{L_{}^{j}} & I_{\overline{L_{y4}^{j+1}}}^{L_{}^{j+1}} \end{pmatrix}
  \end{equation} 

The columns of each matrix represent the contributions of particles when listed with row priority. We can extend this approach to form quadrangular or triangular cells in four directions using sets of particles. The space configuration where the particles are located is toroidal, allowing for circular bypass of the configuration both "from bottom to top" and "from left to right" from a physical perspective on the boundary conditions.

Theorem 3: For a system of interacting particles in a two-dimensional space described by the Ising model, with charges having the same sign and creating an electric (magnetic) field, and distances between particles given by the sum of projections on the X and Y axes, the interaction matrix will have two circulant rows: interaction along the X axis and along the Y axis. The matrix describing the interaction of particles can be defined as a pair of circulant matrices that corresponds to a set of pairs of circulants. To bring the orders of the circulants to a common value, the least common multiple of the distances is used to adjust the sizes. When considering a situation describing the interaction of six particles in two dimensions, a pair of particles will be involved in both j and j + 1 cells. The configuration of the space in which the particles are located is a toroidal space.

\subsection{Circulant of weight 2 case, MET QC-LDPC Codes, multigraph/hypergraph Ising model ground states solution}

Let us now consider the situation in which two particles are located at an equal distance along the X axis from the center of equilibrium of the system. In such a situation, we get in one of the columns of the matrix a null circulant and a circulant of weight 2 at column 3, row 2:
\begin{equation} 
H^{j} =\left(\begin{array}{cccc} {I_{\overline{L_{x1}^{j} }}^{L_{}^{j} } } & {I_{\overline{L_{x2}^{j} }}^{L_{}^{j} } } & {I_{\overline{L_{x1}^{j} }}^{L_{}^{j} } } & {I_{\overline{L_{x4}^{j} }}^{L_{}^{j} } } \\ {I_{\overline{L_{y1}^{j} }}^{L_{}^{j} } } & {I_{\overline{L_{y2}^{j} }}^{L_{}^{j} } } & {I_{\overline{L_{y3}^{j} }}^{L_{}^{j} } } & {I_{\overline{L_{y4}^{j} }}^{L_{}^{j} } } \end{array}\right)\sim \left(\begin{array}{cccc} {I_{\overline{L_{x1}^{j} }}^{L_{}^{j} } } & {I_{\overline{L_{x2}^{j} }}^{L_{}^{j} } } & {0} & {I_{\overline{L_{x4}^{j} }}^{L_{}^{j} } } \\ {I_{\overline{L_{y1}^{j} }}^{L_{}^{j} } } & {I_{\overline{L_{y2}^{j} }}^{L_{}^{j} } } & {I_{\overline{L_{y1}^{j} }}^{L_{}^{j} }  } +{I_{\overline{L_{y3}^{j} }}^{L_{}^{j} }  }  & {I_{\overline{L_{y4}^{j} }}^{L_{}^{j} } } \end{array}\right)
\end{equation}


Similarly, if the positions of the particles coincide, we obtain a triangular cell
\begin{equation} 
H^{j} =\left(\begin{array}{cccc} {I_{\overline{L_{x1}^{j} }}^{L_{}^{j} } } & {I_{\overline{L_{x1}^{j} }}^{L_{}^{j} } } & {I_{\overline{L_{x3}^{j} }}^{L_{}^{j} } } & {I_{\overline{L_{x4}^{j} }}^{L_{}^{j} } } \\ {I_{\overline{L_{y1}^{j} }}^{L_{}^{j} } } & {I_{\overline{L_{y1}^{j} }}^{L_{}^{j} } } & {I_{\overline{L_{y3}^{j} }}^{L_{}^{j} } } & {I_{\overline{L_{y4}^{j} }}^{L_{}^{j} } } \end{array}\right)\sim \left(\begin{array}{cccc} {I_{\overline{L_{x1}^{j} }}^{L_{}^{j} } } & {0} & {I_{\overline{L_{x1}^{j} }}^{L_{}^{j} } } & {I_{\overline{L_{x4}^{j} }}^{L_{}^{j} } } \\ {I_{\overline{L_{y1}^{j} }}^{L_{}^{j} } } & {0} & {I_{\overline{L_{y3}^{j} }}^{L_{}^{j} } } & {I_{\overline{L_{y4}^{j} }}^{L_{}^{j} } } \end{array}\right)
\end{equation}

For the parallelogram configuration, we obtain multi-edge type LDPC codes. These codes can be represented using a hypergraph or a multi-graph notation as described in the section on Hyperbolic Toric Topology Based Ground State Solution of Ising Model: QC-LDPC Codes.

In the hypergraph representation, the rows of the parity-check matrix correspond to hyperedges, and the columns correspond to vertices. Each vertex is connected to multiple hyperedges, which leads to the term "multi-edge" in the name of these codes. 

In the multi-graph representation, the parity-check matrix is represented as a collection of edges connecting vertices (circulant of weight more than 1). The multi-graph representation is more intuitive from a graphical perspective, as it directly shows the connections between the variables and constraints.

\subsection{Abelization Cases and the Connection Between Ising Models and Number Geometry Lattice}

\begin{figure*}[t]
	\begin{center}
            \includegraphics[width=3.in]{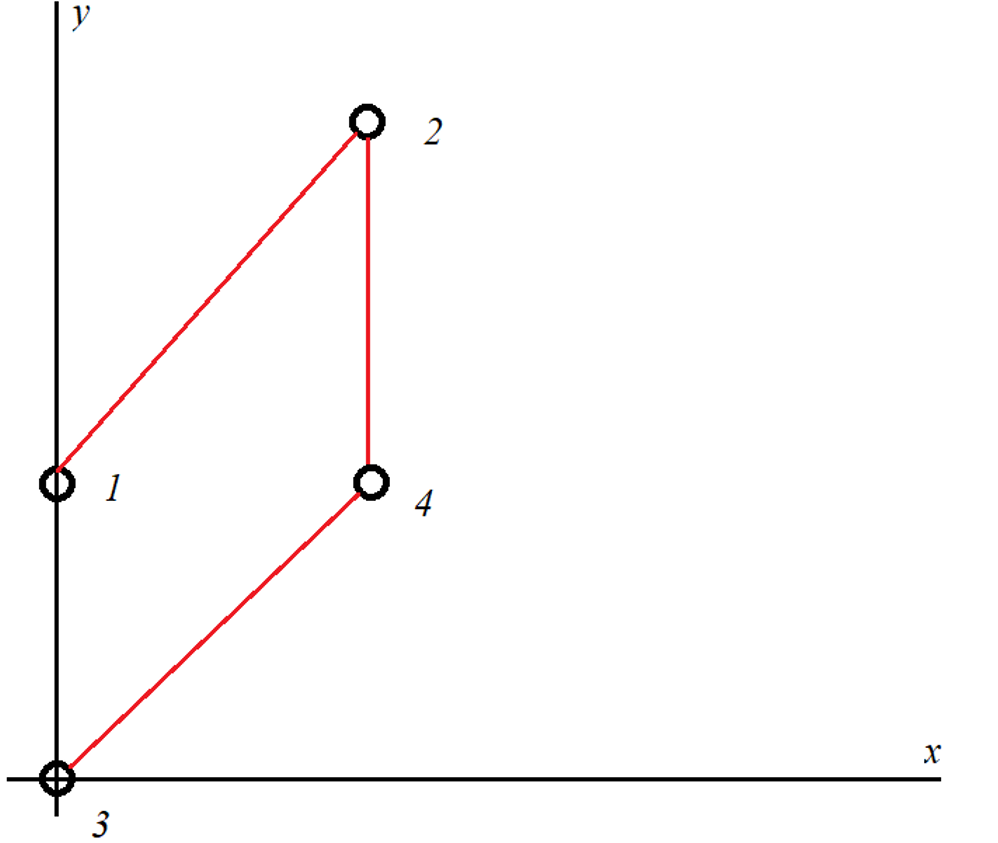}
	\end{center}
	\caption {Diamond-shaped cell configuration}
	\label{fig:images0013}
\end{figure*}

The parallelogram configuration with a ground state forming an abelian subgroup is an interesting case. In this situation, the lattice formed by the spins can be analyzed using lattice theory (Hessian, quadratic form), and it is closely connected to the square lattice (in~Fig.~\ref{fig:images0013}).

In the abelization process, the spins are grouped into clusters such that they satisfy commutation relations, which allows for the representation of the ground state as an abelian group. The resulting lattice formed by these clusters has properties similar to those of the square lattice, including translational symmetry and nearest-neighbor interactions. It allow to use Number Geometry methods for solution of such problem.

The connection between the lattice formed by the spin clusters and the square lattice allows for the use of tools and techniques from lattice theory in the analysis of the codes obtained from the parallelogram configuration. This includes the use of Fourier transforms.

Overall, the abelian subgroup formed by the ground state of the parallelogram configuration allows for a deeper understanding of the properties and behavior of DNN based on such quasi-cyclic codes.

The Hessian is a quadratic form, which is a polynomial of second degree from $n$ variables. It is defined on the tangent space and is invariant to linear transformations. 

The general form of the Hessian can be expressed as:
\begin{equation} 
f(x)=\sum_{i=1}^n q_{ii}x_i^2 + 2 \sum_{1 \leq i < j \leq n} q_{ij} x_i x_j = q_{11}x_1^2 + q_{22}x_2^2 + \dots + q_{nn}x_n^2 + 2q_{12}x_{1}x_{2} + \dots + 2q_{n-1,n}x_{n-1}x_{n},
\end{equation} 
where $q_{ij}$ are the coefficients of the quadratic form.

A quadratic form is positive definite if for any nontrivial column vector $x$, the inequality $f(x) > 0$ holds. This means that the quadratic form always evaluates to a positive value, except when the input vector is the zero vector.

In optimization and machine learning, the Hessian plays an important role in determining the behavior of algorithms such as gradient descent and Newton's method. Specifically, positive definiteness of the Hessian ensures convergence to a unique minimum point, while indefinite or negative definiteness can lead to convergence to saddle points or local maxima.  A lattice is a discrete Abelian subgroup defined under $\mathbb{R}^n$.

Given a linearly independent basis $B = \{b_1, b_2, \dots, b_n\}$ defined under $\mathbb{R}^m$, the lattice points can be represented as a linear integer combination of the basis vectors:
\begin{equation} 
 L(B) = \left\{\sum_{i=1}^n x_i b_i : (x_1, x_2, \dots, x_n) \in \mathbb{Z}^n\right\}, 
\end{equation} 
where $m$ and $n$ are the dimension and rank of the lattice, respectively. The set of coefficients $(x_1, x_2, \dots, x_n)$ is called the representation of a given point in the lattice with respect to the basis $B$.

More over due to Quasi-Cyclic automorphism (circulant form polynomial ring)  of Ising model ground state such lattice can be ideal lattice. 
An ideal lattice is a special type of lattice where the basis vectors have a certain property known as "orthogonality". In other words, the basis vectors are chosen in such a way that they are perpendicular to each other. This ensures that the volume spanned by the basis vectors is minimized, which leads to more efficient computations in high-dimensional spaces.

\begin{figure*}[t]
	\begin{center}
            \includegraphics[width=5.in]{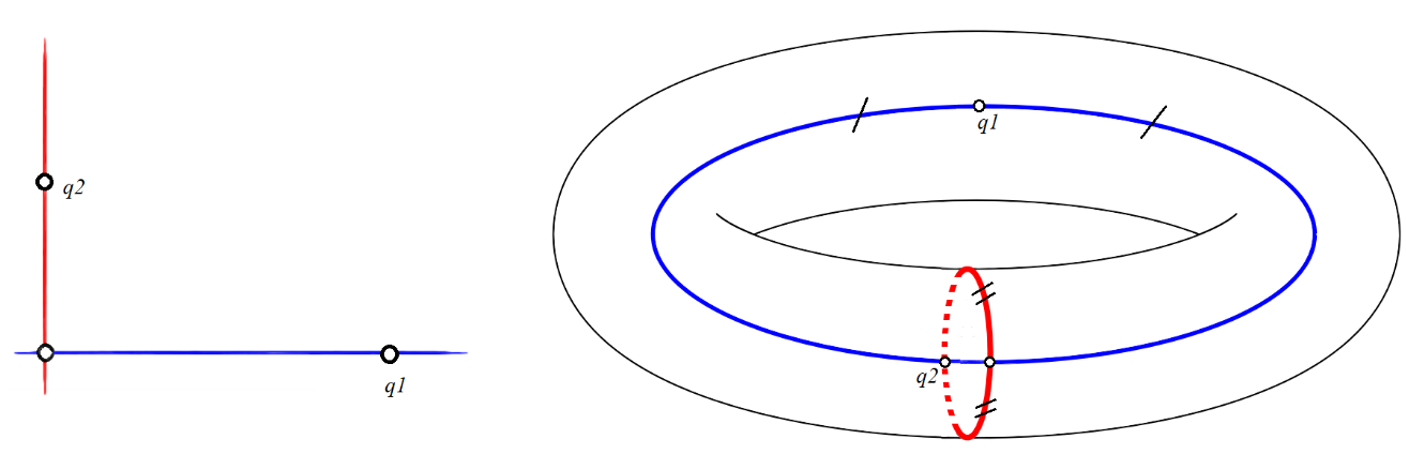}
	\end{center}
	\caption {Lattice Shortest Basis (left) and it Toric topology (right)}
	\label{fig:images0014}
\end{figure*}

\begin{figure*}[t]
	\begin{center}
            \includegraphics[width=5.in]{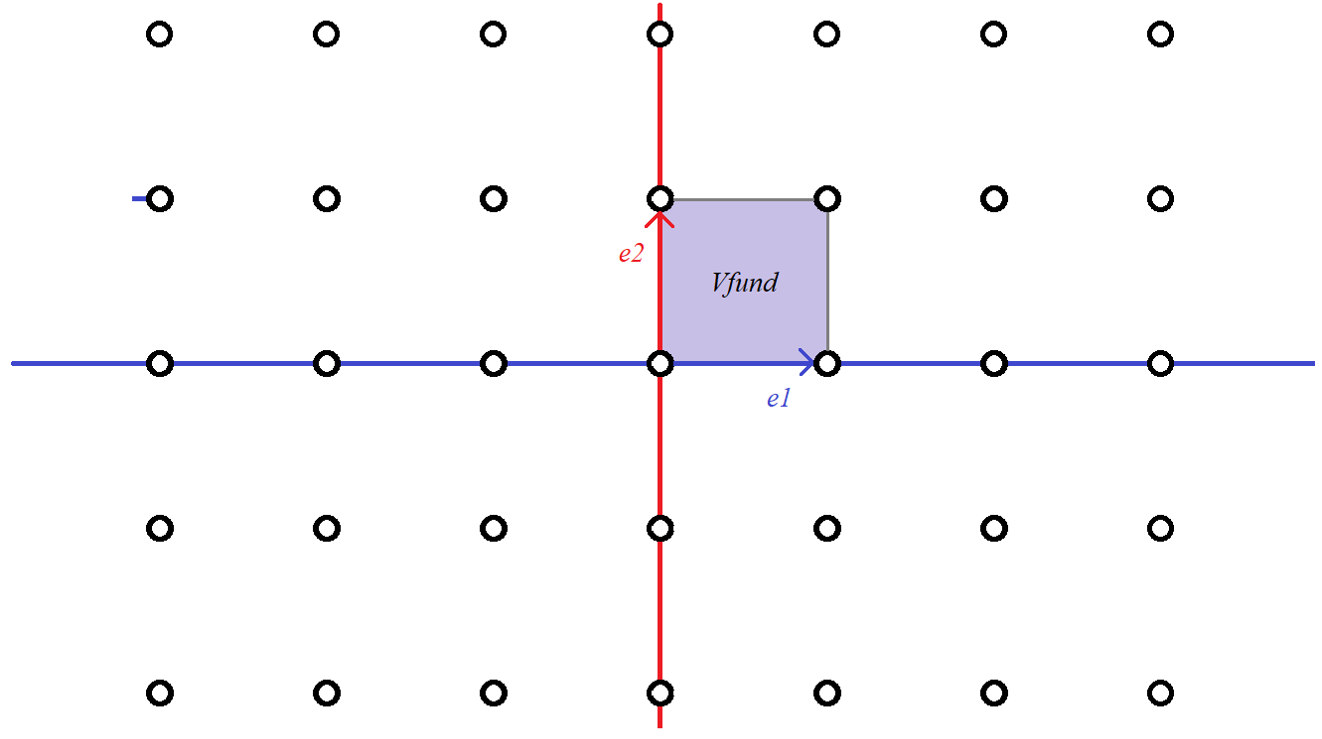}
	\end{center}
	\caption {Abelization (lattice) ground state solution of Diamond-shaped cell under Toric topology}
	\label{fig:images0015}
\end{figure*}

The lattice is said to be ideal if it satisfies the following conditions:

1. $b_i \cdot b_j = 0$ for all $i \neq j$.
2. $|b_i| = |b_j|$ for all $i,j$.
3. The determinant of the Gram matrix $G = \big[b_i \cdot b_j\big]$ is minimized among all lattices with basis vectors satisfying conditions 1 and 2.

In other words to define an ideal lattice, we need to specify three additional properties in addition to linear independence of its basis vectors:
\begin{enumerate}
\item Orthogonality: The basis vectors of an ideal lattice must be orthogonal, which means that their dot products are zero.

\item  Equal length: All basis vectors of an ideal lattice must have the same length.

\item  Minimal volume: An ideal lattice must have the minimal possible volume within its class of lattices with respect to the above two conditions. This means that among all lattices satisfying the first two conditions, the ideal lattice has the smallest determinant of its Gram matrix.

\end{enumerate}

Ideal lattices also have the property that they are invariant under certain transformations, such as rotations or reflections.  In the context of hyperbolic toric topology for self-attention transformer architecture optimization under the Ising model, the use of an ideal lattice guarantees positive definiteness of the Hessian, which is essential for ensuring convergence to a unique global minimum.

If the basis vectors of a lattice are linearly independent, then the lattice is said to be non-degenerate. In this case, the lattice may or may not be positive definite. If the Hessian associated with the optimization problem defined by the lattice is positive definite, then the lattice will also be positive definite.

However, if the basis vectors are not linearly independent, then the lattice is said to be degenerate. In this case, the lattice cannot be positive definite since there exists a linear combination of the basis vectors that gives zero.

\textbf{Therefore, for hyperbolic toric topology under the Ising model for self-attention transformer architecture optimization, it is important to choose an ideal lattice with linearly independent basis vectors that satisfy certain additional constraints. This ensures that the lattice is positive definite and guarantees convergence to a unique global minimum.}

Positive definiteness of the Hessian still improve optimization even under stochastic gradient descent (SGD) backpropagation training. Although SGD does not explicitly compute or use the Hessian matrix, it can still benefit from its positive definiteness.

The reason for this is that positive definiteness of the Hessian indicates that all eigenvalues are positive. This implies that the curvature of the optimization surface is well-behaved and that there are no saddle points or other undesirable stationary points that can cause convergence issues. As a result, SGD can converge faster and more reliably when the Hessian is positive definite.

Furthermore, recent research has shown that incorporating second-order information, such as the Hessian, can improve the performance of SGD even further. For example, techniques such as Hessian-free optimization and natural gradient descent explicitly use the Hessian to accelerate convergence and improve generalization.  Although SGD does not rely on the Hessian in the same way as Newton's method or other second-order optimization algorithms, positive definiteness of the Hessian can still improve optimization and accelerate convergence under SGD backpropagation training.

For example, A diamond-shaped cell lattice (~Fig.~\ref{fig:images0013}) which we get from QC-LDPC codes is a type of Bravais lattice that has four basis vectors pointing towards the corners of a rhombus. The lattice is generated by translating these basis vectors along all possible integer combinations.

To determine if the diamond-shaped cell lattice is ideal, we need to check if it satisfies the three conditions for ideality.

Therefore, the diamond-shaped cell lattice is not an ideal lattice in general. However, it is possible to modify the lattice by rotating or scaling its basis vectors to obtain an ideal lattice with similar properties.

Lattice basis reduction can be used to get as close as possible to orthogonality and minimal volume for a given lattice. The lattice basis reduction problem involves finding a new set of basis vectors that span the same lattice but are closer to being orthogonal and have smaller length - solution of the Shortest Basis Problem.

By applying lattice basis reduction to the diamond-shaped cell lattice or any other lattice, we can obtain a new set of basis vectors that are closer to being orthogonal and have smaller length, thus achieving a more efficient representation of the lattice, ~Fig.~\ref{fig:images0014}. This can improve the performance of optimization algorithms that rely on the lattice structure, such as hyperbolic toric topology for bayesian inference or DNN architectures, ex. self-attention transformer architecture optimization under the Ising model.

The Shortest Basis Problem, also known as the $\Delta$-Short Basis Problem or ${\rm S}BP_{\Delta}(m)$, refers to the task of finding a lattice basis that satisfies certain criteria. Specifically, given a full-rank lattice basis $B$ and a real value $\Delta > 1$, our goal is to find another lattice basis $B'=\{ b_{1}^{'} ,b_{2}^{'} ,..., b_{m}^{'} \}$ such that it has the same lattice as $B$ ($L(B)=L(B')$) and the product of the norms of the new basis vectors is minimized subject to the constraint that it is no larger than $\Delta$ times the product of the norms of the dual basis vectors, denoted as $b_i^{\bot}$. 

In mathematical notation, this can be expressed as:
\begin{equation} 
\prod_{i=1}^m\|b_i'\| \leq \Delta \cdot \prod_{i=1}^m\|b_i^\bot\|,
\end{equation} 
where $\|\cdot\|$ denotes the Euclidean norm.

The Shortest Basis Problem has important applications in various fields, including cryptography and coding theory, where it is used for constructing efficient lattice-based cryptographic schemes, designing high-performance error-correcting codes, and solving integer programming problems. The problem is computationally challenging, and many approximation algorithms have been developed to solve it efficiently, \cite{KZ90, DeRy71}.

The lattice basis $B=\{ b_{1} ,b_{2} ,...,b_{n} \}$ is associated with a quadratic form $QF_B(x_1, x_2, \dots, x_n) = \sum_{1\leq i,j\leq n}\langle b_i, b_j\rangle x_ix_j$. When the input variables $x_1, x_2, \dots, x_n$ are integers, $QF_B$ takes on values corresponding to the lengths of the vectors in the lattice generated by $B$, denoted as $L(B)$: specifically, $QF_B(x_1, x_2, \dots, x_n) = \|\sum_{i=1}^nx_ib_i\|^2$.

In other words, the quadratic form $QF_B$ measures the square of the length of any vector in the lattice $L(B)$ that can be expressed as a linear combination of the basis vectors $b_1, b_2, \dots, b_n$ with integer coefficients. The values of the quadratic form for each such vector correspond to the square of its Euclidean norm, which in turn is equal to the sum of the squares of its components.

The relationship between the lattice basis and the quadratic form has many applications in mathematics and computer science, including the analysis of error-correcting codes, optimization problems, and cryptographic schemes based on lattices.

The quadratic form $QF_B$ can be expressed as a function of the column vector $x=(x_1, x_2, \dots, x_n)^T$ using an $S$-matrix:
\begin{equation} 
f(x) = x^T S x, 
\end{equation} 
where $S$ is a matrix defined in terms of the basis vectors $b_1, b_2, \dots, b_n$ and their inner products. Specifically, the $(i,j)$-th entry of $S$ is given by $\frac{\langle b_i, b_j\rangle}{2}$ if $i \neq j$, or $|| b_i|| ^2$ if $i=j$:

\begin{equation} 
S =
\begin{pmatrix}
|| b_1|| ^2 & \frac{\langle b_1, b_2\rangle}{2} & \cdots & \frac{\langle b_1, b_n\rangle}{2} \\
\frac{\langle b_2, b_1\rangle}{2} & || b_2|| ^2 & \cdots & \frac{\langle b_2, b_n\rangle}{2} \\
\vdots & \vdots & \ddots & \vdots \\
\frac{\langle b_n, b_1\rangle}{2} & \frac{\langle b_n, b_2\rangle}{2} & \cdots & || b_n|| ^2 
\end{pmatrix}.
\end{equation} 

The $S$-matrix is symmetric and positive semidefinite, which makes it useful for analyzing the properties of lattices and quadratic forms. It plays an important role in lattice-based cryptography, coding theory, and optimization problems.

Two quadratic forms $f(x)$ and $f'(x)$: $f(x)=\sum _{i=1}^{n}q_{ii} x_{i}^{2}  +2\sum _{1\le i<j\le n}q_{ij} x_{i} x_{j}  ,q_{ij}$ and $f'(x)=\sum _{i=1}^{n}q'_{ii} x_{i}^{2}  +2\sum _{1\le i<j\le n}q'_{ij} x_{i} x_{j}  ,q'_{ij}$ congruent (equivalent), if there exists such a unimodular matrix $U\in GL_{n} (Z)$, $[q_{ij} ]_{1\le i,j\le n} =U^{T} [q'_{ij} ]_{1\le i,j\le n} U$.
Search of transform matrix $U$ for $QF_{B}$ which have smallest non-diagonal values in $S$, equivalent to choice Hessian grid (Korobov optimal grid) solving  ${\rm S}BP_{\Delta }$-problem or Topology Alpha-complex calculation,  \cite{Kor04}.

Two quadratic forms $f(x)$ and $f'(x)$ are said to be congruent or equivalent if there exists a unimodular matrix $U \in GL_n(\mathbb{Z})$ such that the coefficients of $f$ and $f'$ satisfy the relation: $[q_{ij}]_{1\le i,j\le n} = U^T [q'_{ij}]_{1\le i,j\le n} U$. 

In other words, two quadratic forms are equivalent if they become identical after a linear transformation using an integer-valued invertible matrix. 

Finding a transformation matrix $U$ for the quadratic form represented by the lattice basis $B$, such that the corresponding $S$-matrix has minimal non-diagonal values is equivalent to solving the ${\rm S}BP_\Delta$ problem, constructing a Hessian grid (Korobov optimal grid, \cite{Kor04}), or calculating the topology alpha-complex, for example under Toroidal grid \cite{Sc22}.

\begin{equation} 
\begin{array}{l} {H^{j} =\left(\begin{array}{cccc} {I_{\overline{L_{x1}^{j} }}^{L_{x}^{j} } } & {I_{\overline{L_{x1}^{j} }}^{L_{}^{j} } } & {I_{\overline{L_{x1}^{j} }}^{L_{x}^{j} } } & {I_{\overline{L_{x1}^{j} }}^{L_{x}^{j} } } \\ {0} & {I_{\overline{L_{y2}^{j} }}^{L_{y}^{j} } } & {0} & {I_{\overline{L_{y2} }}^{L_{y}^{j} } } \end{array}\right)\sim \left(\begin{array}{cccc} {I_{\overline{L_{x1}^{j} }}^{L_{x}^{j} } } & {0} & {I_{\overline{L_{x1}^{j} }}^{L_{x}^{j} } } & {0} \\ {0} & {I_{\overline{L_{y2}^{j} }}^{L_{y}^{j} } } & {0} & {I_{\overline{L_{y2} }}^{L_{y}^{j} } } \end{array}\right)\sim } \\ {\sim \left(\begin{array}{cc} {I_{\overline{L_{x1}^{j} }}^{L_{x}^{j} } } & {0} \\ {0} & {I_{\overline{L_{y2}^{j} }}^{L_{y}^{j} } } \end{array}\right)} \end{array} 
\end{equation}

 Need to mention that minimal volume of a lattice and the code distance close related. Code distance improvement could be result of improving EMD spectrum, which improve Trapping sets distribution contain TS(a,0), which a codeword of Hamming weight a.

In coding theory, a code is a set of vectors in an ambient space that satisfies certain constraints. The distance of a code is defined as the minimum Hamming distance between any two distinct code words, where the Hamming distance measures the number of different components between two vectors. The code distance is a measure of how well the code can correct errors or detect them.

The minimal volume of a lattice is closely related to the code distance of a linear code generated by the basis vectors of the lattice. Specifically, if we take the basis vectors of a lattice and use them to generate a linear code, then the code distance of the resulting code is proportional to the inverse square root of the determinant of the Gram matrix of the lattice.

This relationship is known as Minkowski's theorem and provides a powerful tool for analyzing the performance of linear codes in terms of their lattice structure. By optimizing the lattice structure to minimize the volume of the lattice, we can obtain a more efficient representation of the code that can achieve larger code distances and better error-correction capabilities.

Therefore, in hyperbolic toric topology for self-attention transformer architecture optimization under the Ising model, the use of an ideal lattice with minimal volume can lead to improved performance of the optimization algorithm by improving the code distance of the linear code generated by the lattice basis vectors.

The configuration described above corresponds to a diamond-shaped cell that can be employed to close hyperbolic configurations in non-hyperbolic spaces, such as toroidal space. Although it may seem that a matrix consisting of two circulants placed on the diagonal has no physical interpretation, in toroidal space, due to the closedness of geodesic lines, we have a system of two particles that balance the field at the intersection point of the geodesic lines.

This configuration is useful in solving geometric problems in spaces with non-trivial topologies, such as toroidal spaces, which arise naturally in physical and computational contexts. By closing hyperbolic configurations using the diamond-shaped cell, we can define a lattice structure that allows us to study the properties of the space using techniques from lattice theory.

The concept of balancing the field at the intersection points of geodesic lines is important for understanding the behavior of physical systems in toroidal space. It can help us to analyze the distribution of energy, matter, or other physical quantities in the space and make predictions about their dynamics. (in~Fig.~\ref{fig:images0014}):

A set of particles reduced to the same circulant order, represented by the matrix equation:
\begin{equation} 
\left(\begin{array}{cc} {I_{\overline{L_{x1}^{j} }}^{L_{x}^{j} } } & {0} \\ {0} & {I_{\overline{L_{y2}^{j} }}^{L_{y}^{j} } } \end{array}\right)\sim \left(\begin{array}{cc} {I_{S_{j+1,k} }^{N} } & {0} \\ {0} & {I_{S_{j,k+1} }^{N} } \end{array}\right), 
\end{equation} 
form a system of particles located at the nodes of a regular orthogonal uniform grid (as shown in Fig.~\ref{fig:images0015}). The volume of an elementary parallelepiped built on mesh edges remains invariant under coordinate transformations. With sufficiently fine space quantization, the ratio of the shift of the circulant to the size of the circulant remains constant for each circulant, regardless of the change in the dimension of the space.

This configuration has important applications in various fields, including computational physics, quantum mechanics, and signal processing. By representing physical systems as ensembles of particles arranged on regular grids, we can study their behavior using techniques from lattice theory and statistical mechanics. The regularity and invariance of the grid structure simplify many mathematical and numerical calculations and allow us to make predictions about the dynamics of the system.

In summary, the configuration of particles on a regular orthogonal uniform grid is a powerful tool for analyzing physical and mathematical systems in spaces with non-trivial topologies. It provides a framework for studying the properties of lattices, circulants, and other mathematical objects that arise naturally in many areas of science and engineering.

Based on the analysis presented, we can formulate the following proposition:

\textbf{Proposition 2: In a two-dimensional Ising model with charges of the same sign creating an electric or magnetic field and distances between particles given by the sum of projections on the X and Y axes, with sufficiently detailed space quantization, the ratio of the shift of the circulant to the size of the circulant remains constant for each circulant,  regardless of the change in the dimension of the space. }

Such relation related to period of torus per each dimensions.

\subsection{Ising models ground states configuration from famous QC-LDPC codes: Tanner codes and etc}
\begin{figure*}[t]
	\begin{center}
            \includegraphics[width=4.in]{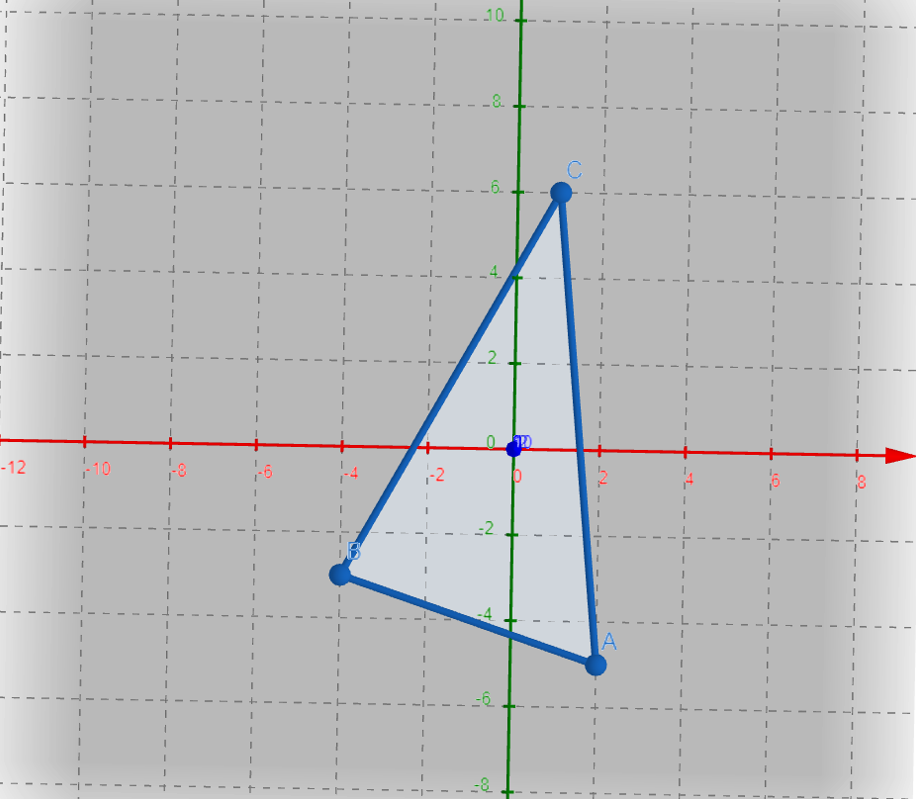}
	\end{center}
	\caption {Interaction of three particles on a plane. Example 2 from article \cite{Ta04}}
	\label{fig:images0016}
\end{figure*}

In the paper \cite{Ta04} , quasi-cyclic block and convolution LDPC codes based on circulant matrices are studied. We can construct their geometric images using the proposed hypothesis. The elementary codes from this paper, which are based on 2x3, 3x4, and 3x5 circulants, can be described as a set of particles in Euclidean space.

For example, consider the code given by the parity-check matrix:
\begin{equation} 
H = \begin{pmatrix}
I_1^7 & I_2^7 & I_4^7 \\
I_6^7 & I_5^7 & I_3^7
\end{pmatrix},
 \end{equation} 
which represents the interaction of three particles on a plane (as shown in Fig.~\ref{fig:images0016}). Each row of the parity-check matrix corresponds to a particle, and each column corresponds to a check node. The identity matrices $I_i^7$ represent the connection between the particle and the check nodes, which are arranged in a circulant pattern.

The position of the center of mass of a triangular plate with particles at the vertices is close to the position of the origin. An analogue in two-dimensional toric space for such a configuration is an isosceles right triangle with vertices equidistant from the origin.

\begin{figure*}[t]
	\begin{center}
            \includegraphics[width=5.in]{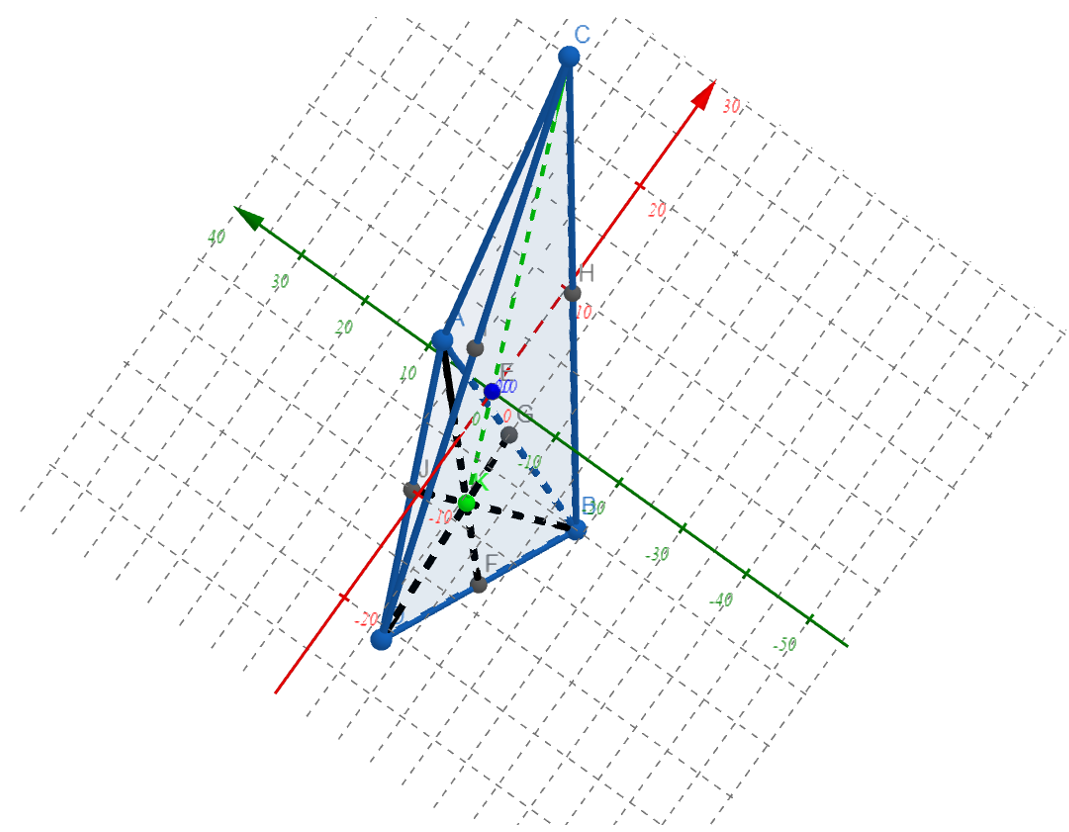}
	\end{center}
	\caption {Interaction of 4 particles in space. Example 3 from article LDPC Block and Convolutional Codes Based on Circulant
Matrices \cite{Ta04}}
	\label{fig:images0017}
\end{figure*}

 A QC code with parity-check matrix (Example 3 from article LDPC Block and Convolutional Codes Based on Circulant Matrices \cite{Ta04}) 
 \begin{equation} 
H=\left(\begin{array}{cccc} {I_{1}^{26} } & {I_{5}^{26} } & {I_{25}^{26} } & {I_{21}^{26} } \\ {\begin{array}{c} {I_{9}^{26} } \\ {I_{3}^{26} } \end{array}} & {\begin{array}{c} {I_{19}^{26} } \\ {I_{15}^{26} } \end{array}} & {\begin{array}{c} {I_{17}^{26} } \\ {I_{23}^{26} } \end{array}} & {\begin{array}{c} {I_{11}^{26} } \\ {I_{11}^{26} } \end{array}} \end{array}\right)
\end{equation} 
can be represented as a system of 4 particles in space  (in~Fig.~\ref{fig:images0017}).


\begin{figure*}[t]
	\begin{center}
            \includegraphics[width=\textwidth]{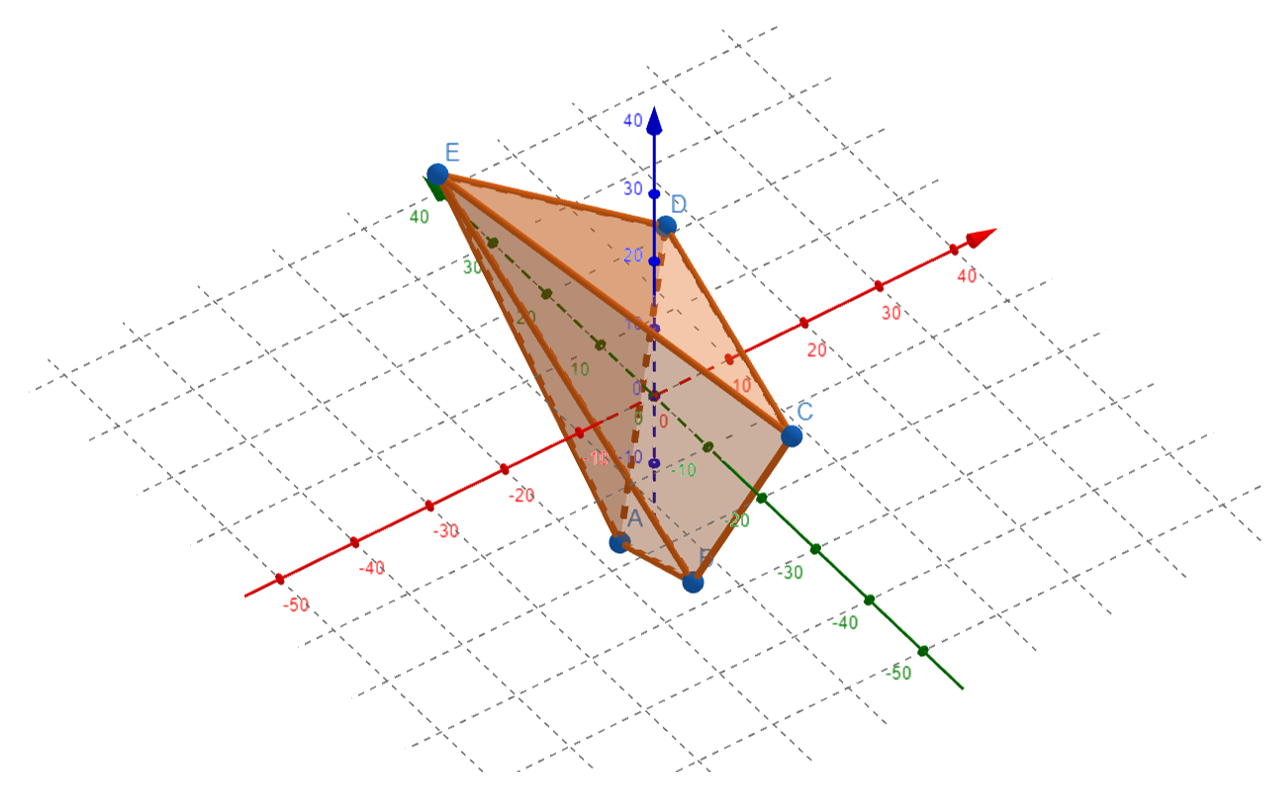}
	\end{center}
	\caption {Interaction of five particles in  space }
	\label{fig:images0019}
\end{figure*}

Four particles form a tetrahedron with the center of mass coinciding with the origin. In Toric three-dimensional space, this configuration corresponds to four points of the cube, forming a trihedral angle, equidistant from the origin. Parity-check matrix 
\begin{equation} 
H=\left(\begin{array}{cccc} {I_{1}^{31} } & {I_{2}^{31} } & {I_{4}^{31} } & {\begin{array}{cc} {I_{8}^{31} } & {I_{16}^{31} } \end{array}} \\ {\begin{array}{c} {I_{5}^{31} } \\ {I_{25}^{31} } \end{array}} & {\begin{array}{c} {I_{10}^{31} } \\ {I_{19}^{31} } \end{array}} & {\begin{array}{c} {I_{20}^{31} } \\ {I_{7}^{31} } \end{array}} & {\begin{array}{c} {\begin{array}{cc} {I_{9}^{31} } & {I_{18}^{31} } \end{array}} \\ {\begin{array}{cc} {I_{14}^{31} } & {I_{28}^{31} } \end{array}} \end{array}} \end{array}\right) 
\end{equation} 

describes the interaction of five particles in space. The particle configuration that defines the code is shown in~Fig.~\ref{fig:images0019}. 


\begin{figure*}[t]
	\begin{center}
            \includegraphics[width=5.in]{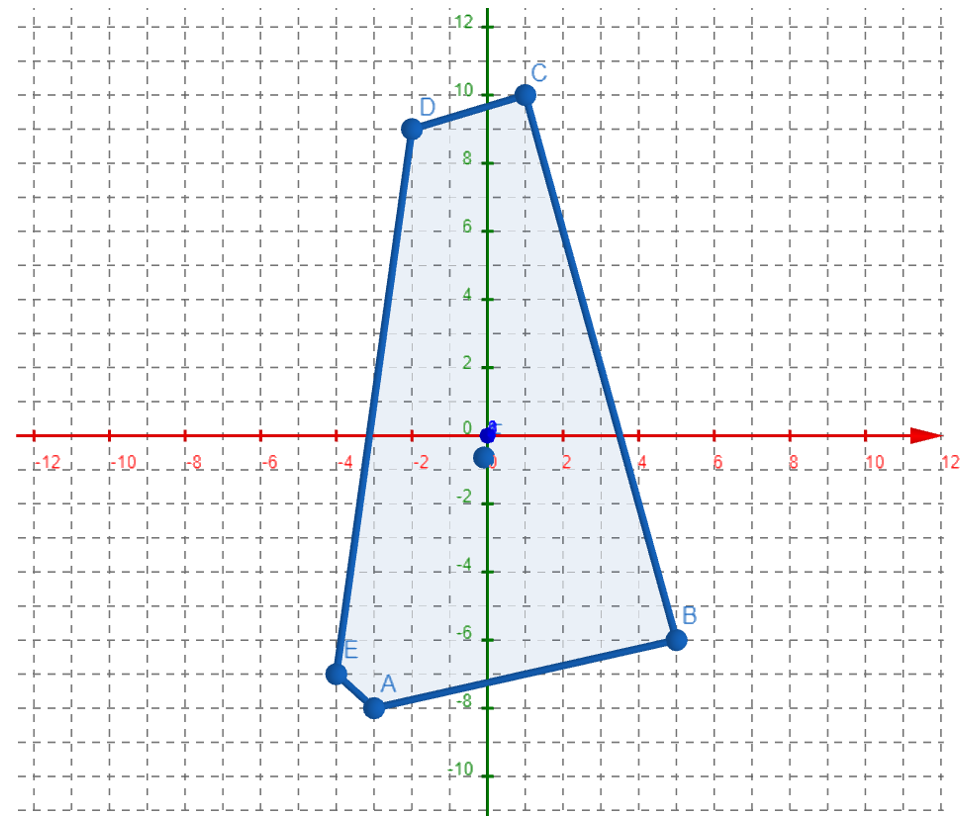}
	\end{center}
	\caption {Interaction of five particles in a plane}
	\label{fig:images0020}
\end{figure*}

\begin{figure*}[t]
	\begin{center}
            \includegraphics[width=3.in]{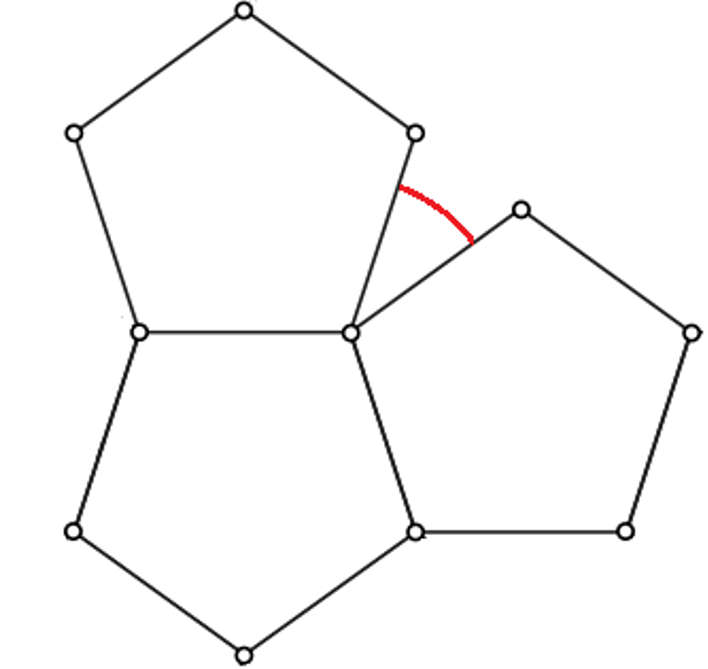}
	\end{center}
	\caption {Lack of angle when trying to tessellate a plane with regular pentagons}
	\label{fig:images0021}
\end{figure*}

Consider the following parity-check matrix

\begin{equation}  H=\left(\begin{array}{ccccc} {I_{10}^{11} } & {I_{9}^{11} } & {I_{8}^{11} } & {I_{7}^{11} } & {I_{6}^{11} } \\ {I_{1}^{11} } & {I_{2}^{11} } & {I_{3}^{11} } & {I_{4}^{11} } & {I_{5}^{11} } \end{array}\right)
\end{equation}

describes the interaction of five particles in a plane. A polygon with charges at the vertices is shown in (in~Fig.~\ref{fig:images0020}). The cell center is close to the origin. It is impossible to tile a Cartesian plane (in~Fig.~\ref{fig:images0021})  or a two-dimensional torus with regular pentagons (Pentagonal tiling problem). Such a tiling occurs on hyperbolic surfaces or, as a central projection of an inscribed dodecahedron, on a sphere. A cell formed by five particles is a hyperbolic cell. The use of such cells (either alone or in combination with quadrangular cells) makes it possible to construct Low Density Quasi-Cyclic parity-check matrices (QC-LDPC codes) corresponding to states with minimum energy (ground state).

\textbf{Theorem 3: QC-LDPC codes  corresponding to states with minimum energy (ground state, equilibrium under Toric Hyperbolic Topology) can be constructed using hyperbolic cells formed by particles. These cells cannot tile a Cartesian plane or a two-dimensional torus with regular pentagons due to the Pentagonal tiling problem, but they can occur on hyperbolic surfaces or as a central projection of an inscribed dodecahedron on a sphere. The use of hyperbolic cells (either alone or in combination with quadrangular cells) provides a way to construct efficient binary and non-binary QC-LDPC codes for embedding  of Deep Neural Network (Quantum System) models such as (dilated) convolutional neural networks (binary Sherrington-Kirkpatrick/non-binary Potts models),  (self-attention) transformers and other deep/shallow neural network.}

By incorporating these codes into the computation of the self-attention mechanism in the Transformer-based architectures of neural networks, the computational efficiency and accuracy of the model can be improved. Specifically, the use of QC-LDPC codes can reduce the amount of computation required during the self-attention process, leading to faster training and inference times, as well as improved accuracy in downstream tasks. Therefore, incorporating QC-LDPC codes into the architecture of deep neural networks is a promising direction for optimizing their performance.

By representing the quasi-cyclic codes as ensembles of particles on a regular grid, we can study their properties using techniques from lattice theory and statistical mechanics. The geometry of the particles and their interactions reflects the structure of the circulant matrices used to construct the parity-check matrix. This approach provides a new perspective on the design and analysis of neural network architecture using LDPC code parity-check matrix. 

In summary, the geometric interpretation of quasi-cyclic codes based on circulant matrices offers a novel way to visualize and understand their structure and behavior. By representing these codes as ensembles of particles on a regular grid, we can apply concepts from lattice theory and statistical mechanics to study their properties and improve their performance.

The geometric interpretation of toric lattices and codes has important applications in various fields, including deep neural networks (DNNs). Another application is to use the geometry of toric lattices to design efficient DNN architectures. In particular, the use of toric grids can simplify the structure of the DNN and reduce the computational complexity of training and inference. By exploiting the regularity and symmetries of the toric grid, we can develop more efficient algorithms for computing convolutions, pooling, and other operations common in DNNs.

Furthermore, the use of toric lattices and codes can provide a new perspective on the behavior and performance of DNNs. By analyzing the geometry and topology of the toric lattice associated with a DNN, we can gain insights into its properties and limitations. For example, the connectivity patterns of the toric lattice can influence the learning capacity and generalization ability of the DNN.

Summary proposed method can be considered as a form of decurvature of the data manifold. Specifically, when we embed high-dimensional data into a lower-dimensional subspace, we are effectively reducing the curvature or complexity of the underlying data manifold. In high-dimensional spaces, data manifolds can exhibit complex structures that can make it difficult to analyze or visualize the data. By embedding the data into a lower-dimensional subspace, we can obtain a simplified representation of the manifold that captures its essential features while reducing its curvature and complexity.This process can be thought of as a form of nonlinear dimensionality reduction, where we aim to preserve the essential structure of the data while reducing its dimensionality and curvature. Nonlinear embedding techniques, such as manifold learning or kernel methods, can be used to construct an embedding that preserves the essential geometric properties of the data manifold.The resulting lower-dimensional representation can then be used for further analysis or processing, such as clustering, classification, or visualization. The effectiveness of the embedding depends on the choice of embedding method and the criteria used to measure the preservation of the essential features of the original data. Overall, compressing the data by embedding it into another space provides a powerful tool for analyzing large-scale datasets with complex structural properties, and can enable more accurate and robust solutions to problems in machine learning, data analysis, and signal processing by decurving the underlying data manifold.

Overall, the geometric interpretation of toric lattices and codes offers a promising avenue for designing more efficient, robust, and interpretable DNNs. By leveraging the mathematical tools and insights from lattice theory, coding theory, and statistical mechanics, we can develop new approaches for addressing the challenges and opportunities in DNN research and applications.

\newpage

\section{Connection between Quasi-Cyclic Codes with Complex Automorphism and Cyclic Codes (BCH, RS) to the Ground State of Atomic Models in the Periodic System of Mendeleev on Spherical Topology and Loss Surfaces of Multilayer Networks Data Manifold}

\textbf{\underline{Rule of Existence}: A Quantum Systems, Chemical elements, Data Manifolds exists if there is a code on the graph parity-check matrix (non-linear projection of graph codes) that reaches the capacity of the corresponding Quantum, Data channels at a finite length. If QC ground states allow it to counteract channel entropy.} 

\textbf{Particular consequences of the Rule of Existence (Fundamental Postulate of survival)}: QC-LDPC codes parity-check matrix (with mixed automorphism, circulant sizes) under Spherical Topology, corresponds to the energy levels of chemical elements from Mendeleev's table and Loss Surfaces of Multilayer Networks of Data Manifold.

Proposal: The Graph model parity-check matrix and it multidimensional non-linear product (like product code, Toric code) can be used to MAP/ML estimation the phase diagram of complex dynamic system.

Proposal: The pressure (and it derivatives: gravitas 'weight', etc) is physical equivalent of decoder complexity, measure of system locality.

\textbf{Observation}: The problems associated with modern ML  seem to be futile. Most of research and current solutions are focused on "determining the orbit of an electron cloud based on a predefined set of states" (some samples from Data Manifold), rather than understanding if an electron can exist with fixed parameters (such as energy) and considering factors like topology, calculation accuracy, energy level (temperature, SNR). This reliance on additional properties or data is incapable of achieving the intended goal of efficient and high precious ML. This common approach approach, known as "AI, ML, DNN, ...", appears to be an exercise in senselessness and fundamental misunderstanding.

The loss surfaces of multilayer networks (DNN) can be described through the representation of Trapping Sets in graph models using Spherical mixed automorphism QC-LDPC codes. The connection between loss surfaces and the spherical Ising model is extensively discussed in the paper cited as \cite{Choro14}. In this study, we will demonstrate the relationship between spherical topology and mixed automorphism QC-LDPC codes on the graph model. 

\begin{figure*}[t]
	\begin{center}
            \includegraphics[width=4.in]{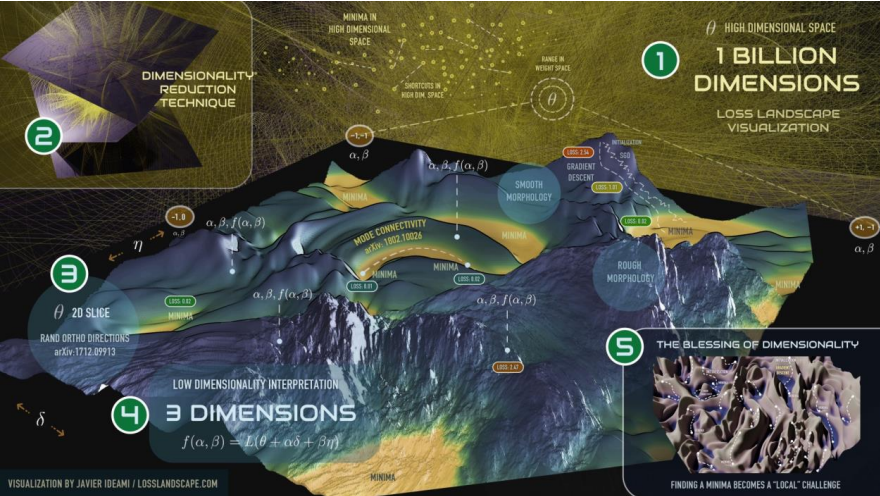}
	\end{center}
	\caption {Example of loss landscape visualization of codewords TS(a,0) and pseudo-codewords (TS(a,b)) generated by cycles under defined non linear dimension reduction metric, \cite{LiHuTay18} }
	\label{fig:images00211111}
\end{figure*}

Short cycles in a DNN can have an impact on the loss function landscapes. These short cycles refer to connections or paths in the network that form loops of relatively few layers. The presence of short cycles can introduce challenges in optimizing the DNN's loss function. One potential effect is the creation of local optima or flat regions in the loss landscape. This can hinder the convergence of optimization algorithms, as they may get trapped in these suboptimal solutions instead of finding the global minimum. Additionally, short cycles can lead to gradient vanishing or exploding problems during backpropagation, where gradients either diminish or grow rapidly as they propagate through the network. This can result in difficulties in training the DNN effectively and could slow down or prevent convergence altogether. Moreover, short cycles can introduce non-smoothness or discontinuities in the loss function landscape, making it harder to navigate and optimize. They can cause abrupt changes in the loss surface, which may affect the stability and reliability of the optimization process. Loss function landscapes emerge around codewords TS(a,0) and pseudo-codewords (TS(a,b)) generated by cycles, (in~Fig.~\ref{fig:images00211111}).  Indeed, physical pressure or decoding complexity can be utilized to decurvature short cycles in the context of DNNs, but this comes with certain trade-offs. One approach is to impose physical pressure on the optimization process by modifying the loss function or introducing regularization terms. By increasing the energy cost associated with short cycles, the optimization algorithm is encouraged to prioritize more spatially local solutions over globally optimal ones. This can help in decurving the cycles and promoting better convergence towards desired solutions. However, it's important to carefully balance the strength of the pressure to avoid unintended consequences such as getting stuck in local optima or sacrificing overall model performance, \cite{Li18}. 

By promoting spatial locality through matching metrics, the optimization algorithm aims to ensure that neighboring nodes or components in the network have similar characteristics or behavior. This can lead to desirable global curvature or topology, where the overall structure of the network aligns with the intended design principles. However, it's important to note that defining and incorporating appropriate matching metrics can be a challenging task. The choice of metrics depends on the specific problem domain, network architecture, and optimization objectives. Additionally, determining the optimal balance between prioritizing spatial locality and achieving the desired global curvature requires careful consideration.

Decurvature or matching metric approaches may face challenges when dealing with higher-dimensional manifolds beyond three dimensions. In such cases, the Thurston geometrization conjecture (Perelman proved), which provides a classification of three-dimensional manifolds into eight geometric types, does not directly apply, \cite{Per02, Per03, Per031,Be2010} .

The task of performing topology data analysis in higher dimensions becomes significantly more complex. The inherent intricacies and properties of high-dimensional spaces make it challenging to define appropriate matching metrics that capture the desired topological features accurately.

In higher-dimensional spaces, the complexity of topological structures increases exponentially, leading to difficulties in characterizing and analyzing the topology using traditional techniques. This includes challenges in identifying critical points, classifying topological features, and measuring distances or similarities between high-dimensional objects.

To address these challenges, researchers have developed specialized techniques in computational topology and algebraic topology to study higher-dimensional data and extract meaningful topological information from complex datasets. These techniques often involve sophisticated algorithms for simplicial complexes, persistent homology, or other topological descriptors that can handle high-dimensional data effectively.

It's important to note that topology data analysis in higher dimensions remains an active area of research, and new methods continue to emerge. While decurvature or direct application of the Thurston geometrization conjecture may not be applicable, alternative approaches rooted in advanced computational and algebraic topology offer promising avenues for exploring and understanding the topology of higher-dimensional manifolds.

 Consider the projective QC-LDPC code with mixed automorphism structures given in the paper \cite{Fos01}:

\begin{equation} \label{1.1)} 
H=\left(\begin{array}{c} {\begin{array}{ccccc} {I_{17} } & {I_{17} } & {I_{17} } & {I_{17} } & {I_{17} } \end{array}} \\ {H_{1} } \\ {H_{2} } \\ {H_{3} } \\ {H_{4} } \end{array}\right)=\left(\begin{array}{c} {\begin{array}{ccccc} {I_{17} } & {I_{17} } & {I_{17} } & {I_{17} } & {I_{17} } \end{array}} \\ {C_{85}^{0} +C_{85}^{24} +C_{85}^{40} +C_{85}^{71} +C_{85}^{84} } \\ {C_{85}^{1} +C_{85}^{49} +C_{85}^{58} +C_{85}^{81} +C_{85}^{84} } \\ {C_{85}^{3} +C_{85}^{14} +C_{85}^{32} +C_{85}^{78} +C_{85}^{84} } \\ {C_{85}^{16} +C_{85}^{33} +C_{85}^{50} +C_{85}^{67} +C_{85}^{84} } \end{array}\right) 
\end{equation}

where $H_{i} $ -- circulant size 85, weight 5 (according to the number of circulant terms of weight 1). For the convenience of geometric analogies, we take the first 3 rows of the matrix H:

\begin{equation} \label{1.2)} 
H=\left(\begin{array}{c} {\begin{array}{ccccc} {I_{17} } & {I_{17} } & {I_{17} } & {I_{17} } & {I_{17} } \end{array}} \\ {H_{1} } \\ {H_{2} } \end{array}\right)=\left(\begin{array}{c} {\begin{array}{ccccc} {I_{17} } & {I_{17} } & {I_{17} } & {I_{17} } & {I_{17} } \end{array}} \\ {C_{85}^{0} +C_{85}^{24} +C_{85}^{40} +C_{85}^{71} +C_{85}^{84} } \\ {C_{85}^{1} +C_{85}^{49} +C_{85}^{58} +C_{85}^{81} +C_{85}^{84} } \end{array}\right) 
\end{equation}

Within the framework of the model under consideration, this matrix can be obtained by collapsing the matrix along one of the coordinates:
\begin{equation} \label{1.3)} 
\left(\begin{array}{c} {\begin{array}{ccccc} {C_{85}^{k} } & {C_{85}^{k} } & {C_{85}^{k} } & {C_{85}^{k} } & {C_{85}^{k} } \end{array}} \\ {\begin{array}{ccccc} {C_{85}^{0} } & {C_{85}^{24} } & {C_{85}^{40} } & {C_{85}^{71} } & {C_{85}^{84} } \end{array}} \\ {\begin{array}{ccccc} {C_{85}^{1} } & {C_{85}^{49} } & {C_{85}^{58} } & {C_{85}^{81} } & {C_{85}^{84} } \end{array}} \end{array}\right)\to \left(\begin{array}{c} {\begin{array}{ccccc} {I_{17} } & {I_{17} } & {I_{17} } & {I_{17} } & {I_{17} } \end{array}} \\ {C_{85}^{0} +C_{85}^{24} +C_{85}^{40} +C_{85}^{71} +C_{85}^{84} } \\ {C_{85}^{1} +C_{85}^{49} +C_{85}^{58} +C_{85}^{81} +C_{85}^{84} } \end{array}\right) 
\end{equation} 

If we introduce a spherical coordinate system $\left(r,\varphi ,\theta \right)$ in this matrix, then the case under consideration is 5 particles, indistinguishable from each other, remote at the same distance from the center (let's call this the generalized average position of electrons in the cloud on such and such an orbital).

Let us now consider, within the framework of the proposed model, the electron clouds of the carbon atom. The charge of carbon is 6: 2 electrons in the first energy level and 4 electrons in the second. Let us form a code matrix based on the proposed considerations. Let's write it in general form:
\begin{equation} \label{1.4)} 
\left\{\begin{array}{c} {r} \\ {\varphi } \\ {\theta } \end{array}\right\}\left(\begin{array}{cccccc} {C_{k}^{r_{1} } } & {C_{k}^{r_{1} } } & {C_{k}^{r_{2} } } & {C_{k}^{r_{2} } } & {C_{k}^{r_{2} } } & {C_{k}^{r_{2} } } \\ {C_{k}^{\varphi _{1} } } & {C_{k}^{\varphi _{2} } } & {C_{k}^{\varphi _{3} } } & {C_{k}^{\varphi _{4} } } & {C_{k}^{\varphi _{5} } } & {C_{k}^{\varphi _{6} } } \\ {C_{k}^{\theta _{1} } } & {C_{k}^{\theta _{2} } } & {C_{k}^{\theta _{3} } } & {C_{k}^{\theta _{4} } } & {C_{k}^{\theta _{5} } } & {C_{k}^{\theta _{6} } } \end{array}\right) 
\end{equation}

To determine the numerical characteristics of the circulants of the matrix, we set some conditions. First, to collapse the matrix along the radii, it is necessary to observe the multiplicity of the order of the circulant to the radii and their sum:
\begin{equation} \label{1.5)} 
\begin{array}{l} {k\mod{r_{1}} =0} \\ {k \mod{r_{2} =0}} 
\\ {k \mod{\left(r_{1} +r_{2} \right)
}=0} \end{array} 
\end{equation}

in the generalized case, we can write as

\begin{equation} \label{1.6)} 
\begin{array}{l} {k \mod{r_{i} =0}} \\ {k \mod{\sum _{i=1}^{N}r_{i}  =0} } \end{array} 
\end{equation}

or, in the most general case (for example, when the radii are prime numbers)
\begin{equation} \label{1.7)} 
\begin{array}{l} {k \mod{\left(a\left(\sum _{i=1}^{N}r_{i}  \right)\prod _{i=1}^{n}r_{i}  \right)}} \\ {} \\ {a,r_{i} \in {\mathbb N}} \end{array} 
\end{equation}

In the case under consideration, the following circulant size and matrix shifts satisfy the criteria:
\begin{equation} \label{1.8)} 
\left\{\begin{array}{c} {r} \\ {\varphi } \\ {\theta } \end{array}\right\}\left(\begin{array}{cccccc} {C_{48}^{24} } & {C_{48}^{24} } & {C_{48}^{36} } & {C_{48}^{36} } & {C_{48}^{36} } & {C_{48}^{36} } \\ {C_{48}^{\varphi _{1} } } & {C_{48}^{\varphi _{2} } } & {C_{48}^{\varphi _{3} } } & {C_{48}^{\varphi _{4} } } & {C_{48}^{\varphi _{5} } } & {C_{48}^{\varphi _{6} } } \\ {C_{48}^{\theta _{1} } } & {C_{48}^{\theta _{2} } } & {C_{48}^{\theta _{3} } } & {C_{48}^{\theta _{4} } } & {C_{48}^{\theta _{5} } } & {C_{48}^{\theta _{6} } } \end{array}\right) 
\end{equation}

To determine the values of  $\varphi $, $\theta $, we use the criterion for the presence of a cycle of length 6 (\textbf{call it cycle based gauge)} in the quasi-cyclic check matrix, \cite{Fossorier04}. Based on the electron cloud associated with an atom of a chemical element, we can refer this equation as the Schrödinger-Heisenberg-Bohr-Fossorier electron cloud Gauge (SHBF Cycle Gauge):

\begin{equation} \label{1.9)} 
\left(\sum _{i=1}^{N-1}\Delta _{ji,ji+1} \left(l_{i} \right) \right) \mod {k}=0 
\end{equation}

that is, the sum of shifts of the circulant along the row (column) is a multiple of the dimension of the circulant. One of the matrices that meet this criterion will be:
\begin{equation} \label{1.10)} 
\left\{\begin{array}{c} {r} \\ {\varphi } \\ {\theta } \end{array}\right\}\left(\begin{array}{cccccc} {C_{48}^{24} } & {C_{48}^{24} } & {C_{48}^{36} } & {C_{48}^{36} } & {C_{48}^{36} } & {C_{48}^{36} } \\ {C_{48}^{1} } & {C_{48}^{7} } & {C_{48}^{13} } & {C_{48}^{19} } & {C_{48}^{25} } & {C_{48}^{31} } \\ {C_{48}^{23} } & {C_{48}^{17} } & {C_{48}^{47} } & {C_{48}^{41} } & {C_{48}^{35} } & {C_{48}^{29} } \end{array}\right) 
\end{equation}

It is important that the condition along the row is satisfied when the shift of all circulants of the row changes at a time by a multiple of
\begin{equation} \label{1.11)} 
S=k/N 
\end{equation}

Now let's collapse the matrix along the radii:
\begin{equation} \label{1.12)} 
H=\left(\begin{array}{c} {\begin{array}{cccccc} {C_{8}^{0} } & {C_{8}^{0} } & {C_{8}^{2} } & {C_{8}^{2} } & {C_{8}^{2} } & {C_{8}^{2} } \end{array}} \\ {C_{48}^{1} +C_{48}^{7} +C_{48}^{13} +C_{48}^{19} +C_{48}^{25} +C_{48}^{31} } \\ {C_{48}^{23} +C_{48}^{17} +C_{48}^{47} +C_{48}^{41} +C_{48}^{35} +C_{48}^{29} } \end{array}\right)  \to \left(\begin{array}{c} {\begin{array}{cccccc} {I_{8} } & {I_{8} } & {I_{8} } & {I_{8}} & {I_{8}} & {I_{8} } \end{array}} \\ {C_{48}^{1} +C_{48}^{7} +C_{48}^{13} +C_{48}^{19} +C_{48}^{25} +C_{48}^{31} } \\ {C_{48}^{23} +C_{48}^{17} +C_{48}^{47} +C_{48}^{41} +C_{48}^{35} +C_{48}^{29} } \end{array}\right)
\end{equation}

The first row, on which the collapse was carried out, contains 6 circulants of size 8 and weight 1. 2 shift 0 circulants correspond to the first energy level, 4 shift 2 circulants correspond to the second energy level:
\begin{equation} \label{1.13)} 
1s^{2} ;2s^{2} 2p^{2}  
\end{equation}

\begin{figure*}[t]
	\begin{center}
            \includegraphics[width=3.5in]{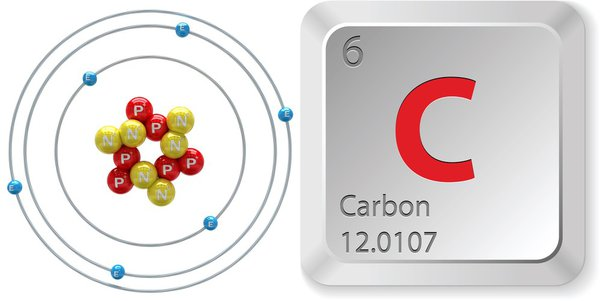}
	\end{center}
	\caption {The electronic configuration of Carbon}
	\label{fig:Carbon}
\end{figure*}

Lines 2 and 3, respectively, are circulants of weight 6 and size 48. The parity-check matrix of the quasi-cyclic LDPC code from the work of Fossorier-Shu Lin \cite{Fos01}, example 2 automorphism structures corresponds electron clouds (ground state) to the element Carbon under Spherical Topology and SHBF Cycle Gauge, and related TS, Fig. \ref{fig:Carbon}. 

\begin{figure*}[t]
	\begin{center}
            \includegraphics[width=3.5in]{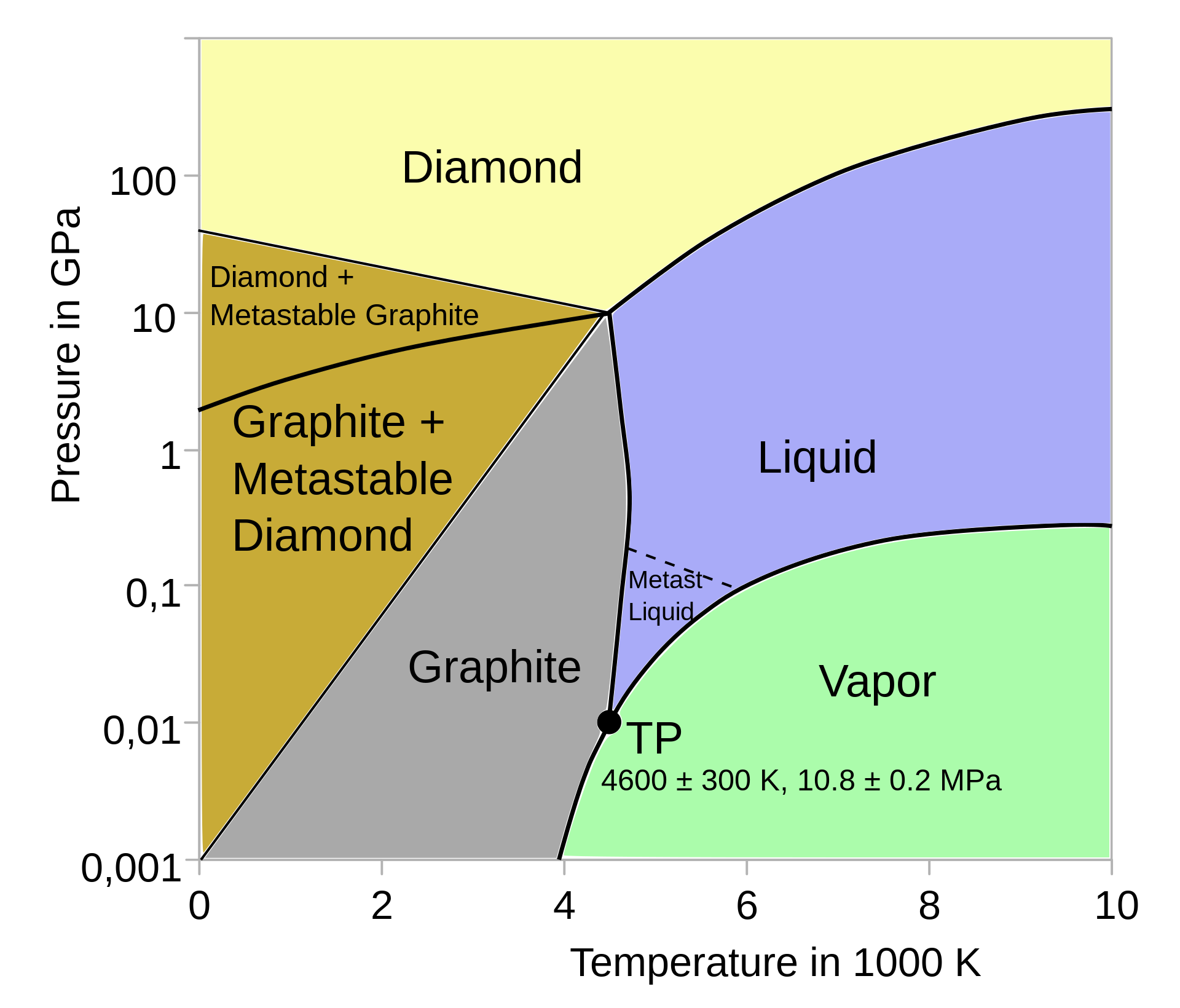}
	\end{center}
	\caption {Theoretically predicted phase diagram of carbon, from paper \cite{Ege2009} and it relation to error-correction curve of Fossorier-Shu Lin mixed automorphism QC-LDPC code, \cite{Fos01}  }
	\label{fig:Carbon2}
\end{figure*}

\textbf{Proposition: In this case, the carbon state diagram can be interpreted as a transition to the states of matter (non-convergence, waterfall, error-floor areas on BER to SNR figure) depending on temperature (signal-to-noise ratio), pressure (complexity of the decoder breaking cycles in the graph, Trapping sets, density of cycles in the graph after decoder). By increasing the complexity of the decoder, we can break cycles in the graph by actually varying the \underline{pressure (physical equivalent of decoder complexity, measure of system locality).} Temperature is the signal-to-noise ratio, the number of cycles in the graph after the decoder has run, a function of the complexity of the decoder and the topology of the cycles in the graph - pressure, } \ref{fig:Carbon2} .

\newpage

\subsection{ MET QC-LDPC Graph Model for pumped up Fermions showing the behavior peculiar to Bosons }

\begin{figure*}[t]
	\begin{center}
            \includegraphics[width=3.5in]{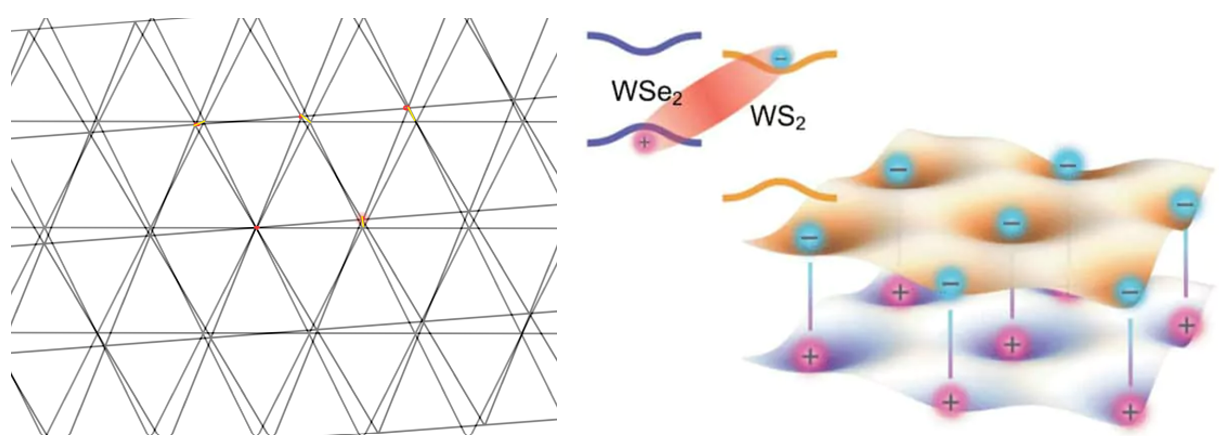}
	\end{center}
	\caption {Fermions are held in shifted lattices by a certain angle  (left),  regular triangular shape made of tungsten disulfide and wolfram diselenide from \cite{Xio23}  }
	\label{fig:FermBosLat}
\end{figure*}

The paper \cite{Xio23} considers a  experiment in which two lattices of regular triangular shape made of tungsten disulfide and wolfram diselenide having negatively charged electrons and positively charged electron holes in the nodes, respectively. Fermions are held in lattice nodes by strong interaction. When the lattices  
(Fig. \ref{fig:FermBosLat}) are shifted by a certain angle $ \alpha $ relative to the stationary center of rotation and the photon energy is pumped, fermions enter into electromagnetic interaction, forming dipoles-excitons. Due to the dipole nature, excitons are dielectric insulators, they are held in a strict lattice by a strong interaction.

\begin{figure*}[t]
	\begin{center}
            \includegraphics[width=3.5in]{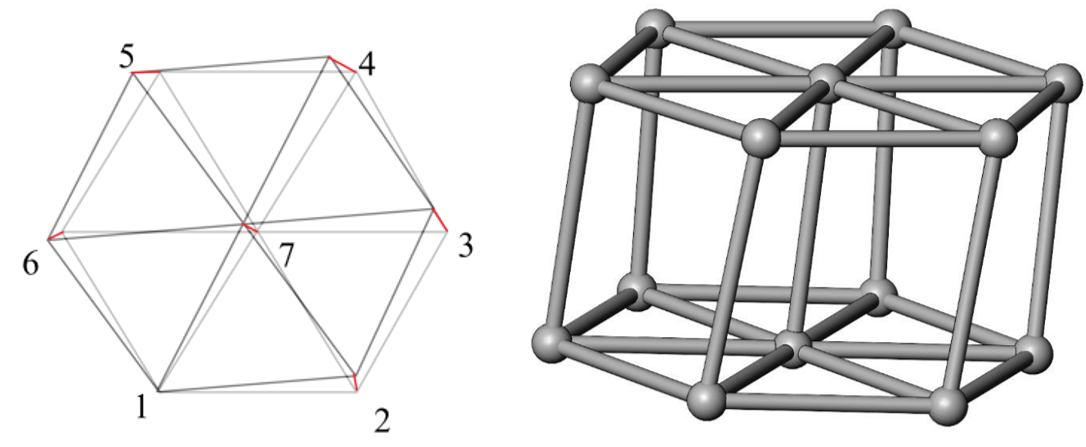}
	\end{center}
	\caption { The 7 dipoles hexagonal prism   }
	\label{fig:HexPrism}
\end{figure*}

Let's consider one hexagonal prism including 7 dipoles (Fig. \ref{fig:HexPrism}) , one of which is the center of rotation of the grid. By introducing a rectangular Cartesian coordinate system centered at the center of the grid rotation, according to the proposed hypothesis, we can determine a matrix describing the effect of grid displacement on the interaction of particles in a fixed center.

Let R be the step of the triangular grid, $L_{ij}$-- the distance between two nodes, then

\begin{equation} \label{1.21)} 
\begin{array}{l} {\Delta R_{jx} =L_{1jx} -L_{1jx} \cos \alpha } \\ {\Delta R_{jy} =L_{1jy} -L_{1jy} \sin \alpha } \end{array} 
\end{equation}

Let's introduce quantization:
\begin{equation} \label{1.22)} 
{dR=\left\lfloor \frac{R}{q} \right\rfloor } ,  {qdR=\overline{R}} , {\overline{\Delta R}=\left\lfloor \frac{\Delta R}{dR} \right\rfloor } 
\end{equation}

Then we can describe the interaction matrix as follows:
\begin{equation} \label{1.23)} 
H=\left(\begin{array}{ccccccc} {C_{0}^{q} } & {C_{\overline{\Delta R_{2x} }}^{q} } & {C_{\overline{\Delta R_{3x} }}^{q} } & {C_{\overline{\Delta R_{4x} }}^{q} } & {C_{\overline{\Delta R_{5x} }}^{q} } & {C_{\overline{\Delta R_{6x} }}^{q} } & {C_{\overline{\Delta R_{7x} }}^{q} } \\ {C_{0}^{q} } & {C_{\overline{\Delta R_{3y} }}^{q} } & {C_{\overline{\Delta R_{3y} }}^{q} } & {C_{\overline{\Delta R_{4y} }}^{q} } & {C_{\overline{\Delta R_{5y} }}^{q} } & {C_{\overline{\Delta R_{6y} }}^{q} } & {C_{\overline{\Delta R_{7y} }}^{q} } \end{array}\right) 
\end{equation}

At a sufficient distance from the center of rotation, the bevel of the dipoles can exceed the value at which the connection between the i-th electrons and the holes of the corresponding layers is preserved. In this case, the influence passes to neighboring nodes, while changing the direction of the circulant shift.

In addition, 3 main directions can be distinguished on such a grid and a local non-orthogonal system of three coordinates L1-3 can be introduced. Coordinates vary from 0 to 1 (exceeding 1 when the grid is beveled), \cite{Sege76}.

The local coordinate system is connected by perpendiculars, pubescent from a point to a side, each of the coordinates expresses the ratio of the area of the triangle constructed on the corresponding side and point to the area of the original triangle Fig. \ref{fig:SegLoc}, than, the coordinate of point B in the L-system will have the value:
\begin{equation} \label{1.24)} 
\begin{array}{l} {B_{L} =\left(L_{1} ;L_{2} ;L_{3} \right)=\left(\frac{S_{\Delta Bjk} }{S_{\Delta ijk} } ;\frac{S_{\Delta Bki} }{S_{\Delta ijk} } ;\frac{S_{\Delta Bij} }{S_{\Delta ijk} } \right)=} \\ {=\left(\frac{h_{B1} }{h_{1} } ;\frac{h_{B2}^{} }{h_{2} } ;\frac{h_{B3} }{h_{3} } \right)} \end{array} 
\end{equation}

\begin{figure*}[t]
	\begin{center}
            \includegraphics[width=3.5in]{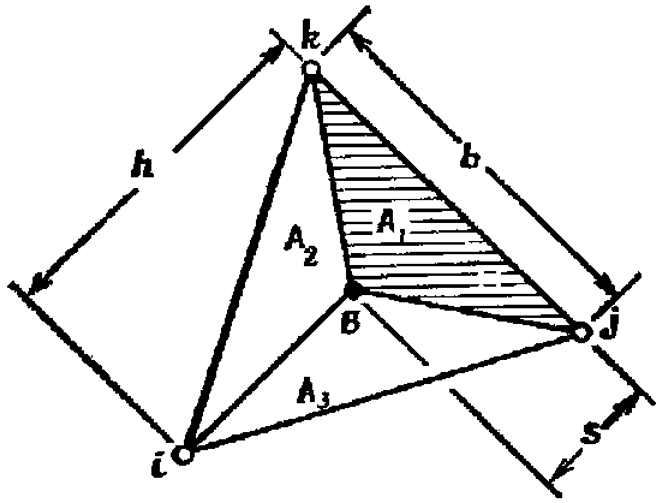}
	\end{center}
	\caption { Finite Element Analysis local coordinate system Grid from book \cite{Sege76}  }
	\label{fig:SegLoc}
\end{figure*}

Each displaced lattice node participates in the formation of six grid cells and can be represented, according to the proposed hypothesis, by six sets of columns of three circulants describing the shift of the node relative to the local coordinates of each of the 6 cells. By the properties of additivity, the columns describing the interaction in each cell will be the sum of three columns describing the shifts of each of the cell nodes relative to the local coordinate system.

\begin{figure*}[t]
	\begin{center}
            \includegraphics[width=3.5in]{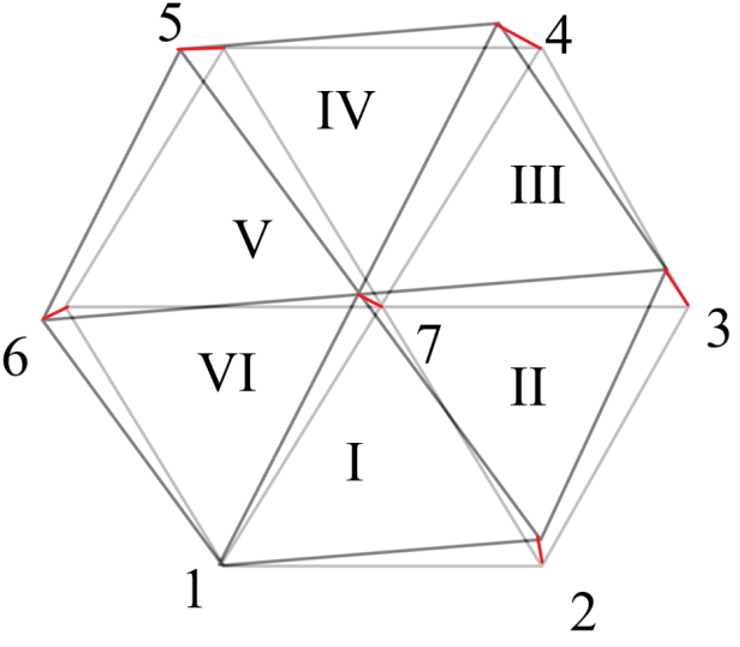}
	\end{center}
	\caption { Grid discretization of interaction  }
	\label{fig:FBcells}
\end{figure*}

We introduce the discretization similarly (eq. \ref{1.1)}, \ref{1.2)}). Let $\Delta L_{ij}^{k,m} $ - relative discretized change of the local coordinate, indexes $ij$ show which unbiased side the height is omitted, index $k$ shows from which offset node the height is omitted, index $m$ shows which cell is in consideration. We will use Roman numerals to number the cells (Fig. \ref{fig:FBcells}).

The cells will be traversed counterclockwise from the smallest index of the element, $C^{\Delta L_{ij}^{k,m} } $- shift circulant $\Delta L_{ij}^{k,m} $, order $LCM\left(\Delta L_{ij}^{k,m} \right)$. Let's construct an interaction matrix for element IV:

\begin{equation} \label{1.25)} 
H^{IV} =\left(\begin{array}{ccc} {C^{\Delta L_{47}^{3,IV} } } & {C^{\Delta L_{47}^{4,IV} } } & {C^{\Delta L_{47}^{7,IV} } } \\ {C^{\Delta L_{73}^{3,IV} } } & {C^{\Delta L_{73}^{4,IV} } } & {C^{\Delta L_{73}^{7,IV} } } \\ {C^{\Delta L_{34}^{3,IV} } } & {C^{\Delta L_{34}^{4,IV} } } & {C^{\Delta L_{34}^{7,IV} } } \end{array}\right) 
\end{equation}

Then, by the properties of additivity, the local interaction matrix for node 3 will consist of the sum (modulo 2) of matrices taking into account the column and row indexes for elements III, IV. We get MET QC-LDPC code, \ref{proto}:
\begin{equation} \label{1.26)} 
H_{3} =H^{III} \oplus H^{IV} =\left(\begin{array}{l} {\begin{array}{cccc} {C^{\Delta L_{47}^{2,III} } } & {C^{\Delta L_{47}^{3,IV} } +C^{\Delta L_{47}^{3,III} } } & {C^{\Delta L_{47}^{4,IV} } } & {C^{\Delta L_{47}^{7,IV} } +C^{\Delta L_{47}^{7,III} } } \\ {C^{\Delta L_{73}^{2,III} } } & {C^{\Delta L_{73}^{3,IV} } +C^{\Delta L_{73}^{3,III} } } & {C^{\Delta L_{73}^{4,IV} } } & {C^{\Delta L_{73}^{7,IV} } +C^{\Delta L_{73}^{7,III} } } \\ {0} & {C^{\Delta L_{34}^{3,IV} } } & {C^{\Delta L_{34}^{4,IV} } } & {C^{\Delta L_{34}^{7,IV} } } \end{array}} \\ {\begin{array}{cccc} {C^{\Delta L_{24}^{2,III} } \, \, \, \, \, \, \, \, } & {C^{\Delta L_{24}^{3,III} } \, \, \, \, \, \, \, \, \, \, } & {0\, \, \, \, \, \, \, \, \, \, \, \, \, \, \, \, \, } & {C^{\Delta L_{24}^{7,III} } } \end{array}} \end{array}\right) 
\end{equation}

In the case of uniformity of the grid and sufficiently precious discretization, \textbf{due to the cyclical shift of the circulant, the sums of circulants related to one node, with an even number of terms, turn to zero}. With insufficient sampling, the density of the interaction matrix will increase (sparsity decrease).

\textbf{Important observation 1:  The parity of the number (even) of elements in the cycle with sufficient grid resolution (code length, number of nodes in the graph) leads to sparsity. The symmetry of the connection (subgraphs, submatrix, subtensors) equivalent to codewords (Trapping set (a,0) by definition, \cite{USA23}, \ref{Trapping sets}) leads to sparsity.} 

\textbf{Important observation 2: The size of even degree cycles, ..., equivalent to codewords (TS(a,0)) doesn't need to be large to get MAP/ML estimation (inference) capacity bound, \cite{ChanUrRic01}.}

\textbf{Summary: proposed method allows to describe quantum systems using dual to Tensor Network Multi-edge Type QC-LDPC graph model: fermion (AntiSymmetrical wave function) and Boson (more simple because Symmetrical wave function) quantum systems}.

\subsection{ Exploring the Relationship Between Nishimori Temperature (Signal to Noise Ration) and Bethe Free Energy under Codes on the Graph  }

In the research paper \cite{Lore21}," authored by Lorenzo Dall'Amico et al., a comprehensive investigation is carried out to unveil the intriguing connection between the Nishimori temperature and the Bethe free energy within the context of a random Erdős-Rényi graph.

Let us begin by introducing the Erdős-Rényi graph, denoted as $\zeta (\nu, \varepsilon)$. Here, $(\nu, \varepsilon)$ signifies the set of vertices and edges, respectively. A matrix $J$ is associated with this graph, which encapsulates its structural characteristics. The non-zero matrix elements are generated based on the following rule:

\begin{equation} \label{4.1)} 
P(x) = p(|x|)e^{\beta_N x}
\end{equation}

Here, $p(|x|)$ represents a non-negative function, and $\beta_N > 0$ is subsequently identified as the Nishimori temperature. In the realm of statistical physics, the Nishimori temperature finds significant utility in the Ising model, \cite{Nishimori80,Nishimori81,Nishimori86}. In this model, the vector $s=\{-1, 1\}^n$ is a random vector following the Boltzmann distribution:

\begin{equation} \label{4.2)} 
\mu(s) = \frac{e^{-\beta H_j(s)}}{Z_{j,\beta}}
\end{equation}

Here, $\beta$ takes on positive values, and $Z_{j,\beta}$ serves as the normalization coefficient, while $H_j(s)=-s^TJs$.

The crucial point of interest arises when $\beta = \beta_N$. This signifies a scenario where the system temperature aligns precisely with the Nishimori temperature. The average value of a specific variable $E$ across multiple realizations of matrix $J$ can be determined by averaging Equation \eqref{4.2)} over $\beta$.

\begin{figure*}[t]
	\begin{center}
            \includegraphics[width=4.1in]{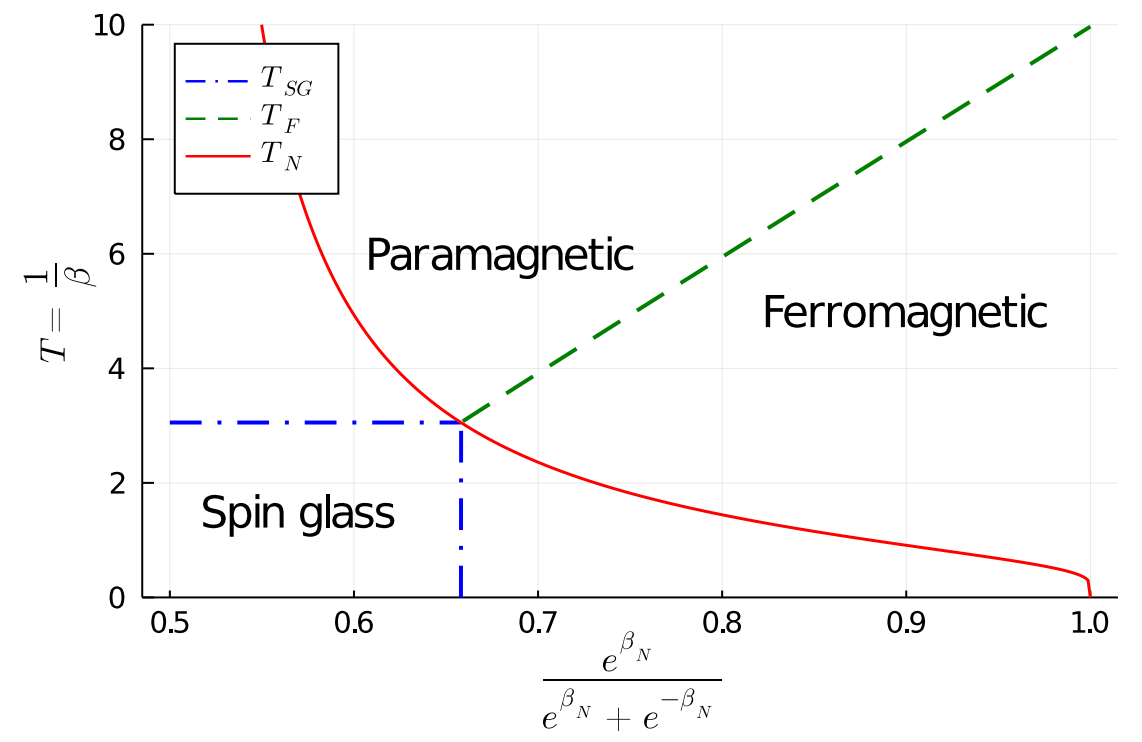}
	\end{center}
	\caption {Phase diagram of spin glasses depicted in terms of the Nishimori temperature, $\beta$} 
	\label{fig:SegLoc}
\end{figure*}

Reformulation of Equation \eqref{4.1)} and condition on $p_0$.

Let's revisit the relationship expressed in Equation \eqref{4.1)} and rewrite it in the following form:

\begin{equation} \label{4.3)} 
\begin{array}{l}
\forall (ij)\in \varepsilon, i<j \\
J_{ij} \sim P \\
P(x) = p_{0} (|x|)e^{\beta _{N} x}
\end{array}
\end{equation}
Here, we require that the function $p_{0} (|x|)$ satisfies the normalization condition:

\begin{equation} \label{4.4)} 
\begin{array}{l}
\int p_{0} (|x|)e^{\beta _{N} x} dx = 1 \\
J_{ij} = 0 \Leftrightarrow (ij)\notin \varepsilon
\end{array}
\end{equation}

The matrix $J$ and the vector $s=\{-1,1\}^{n}$ together define the Hamiltonian of the Ising model as:

\begin{equation} \label{4.5)} 
H_{J} (s) = -\sum _{(ij)\in \varepsilon} J_{ij} s_{i} s_{j} = -s^{T} Js
\end{equation}

Notably, the Nishimori temperature is determined independently of the specific graph being considered. In contrast, $p_{0}$ and the Nishimori temperature are intricately related through the normalization condition.

\subsection{Relationships between the Nishimori Temperature and the Bethe-Hessian Matrix}

Definition 1. (Weighted Non-backtracking Matrix)

Consider a graph $\zeta(\nu, \varepsilon)$ and a function $f: \varepsilon \to \mathbb{R}$ such that for all $e \in \varepsilon, f(e) = \omega_{e}$, where $\omega_{e}$ represents the weight corresponding to edge $e$. The matrix $B$ (weighted non-backtracking matrix) is defined for the edges of the graph as:

\begin{equation} \label{4.6)} 
\begin{array}{l}
B_{(ij),(kl)} = \delta_{jk} (1 - \delta_{il}) \omega_{kl} \\
B \in \mathbb{R}^{2|\varepsilon| \times 2|\varepsilon|}
\end{array}
\end{equation}

Definition 2. (Bethe-Hessian)

Consider a graph $\zeta(\nu, \varepsilon)$ and a function $f: \varepsilon \to \mathbb{R}$ such that for all $e \in \varepsilon, f(e) = \omega_{e}$, and $x \in \mathbb{C} \setminus \{\pm \omega_{ij}\}_{(ij) \in \varepsilon}$. The Bethe-Hessian is defined as:

\begin{equation} \label{4.7)} 
\begin{array}{l}
H_{ij}(x) = \left(1 + \sum_{k \in \partial i} \frac{\omega_{ik}^{2}}{x^{2} - \omega_{ik}^{2}}\right) \delta_{ij} - \frac{x\omega_{ij}}{x^{2} - \omega_{ij}^{2}} \\
H(x) \in \mathbb{C}^{n \times n}
\end{array}
\end{equation}

It is noted that the eigenvalues of matrices $B$ and $H$ are strongly related, as shown in:

\begin{equation} \label{4.8)} 
\det [xI_{2|\varepsilon|} - B] = \det [H(x)]\prod_{(ij) \in \varepsilon} (x^{2} - \omega_{ij}^{2})
\end{equation}

This implies that for any $x$ from the spectrum of $B$, $\det [H(x)] = 0$.

Now, let's establish a connection between the Bethe-Hessian and the Nishimori temperature. For this purpose, we consider the Bethe approximation $\widetilde{F}_{J\beta}(q)$ of the free energy $F_{J\beta}$ for a certain set of parameters $q$:

\begin{equation} \label{4.9)} 
\begin{array}{l}
F_{J\beta} = \sum_{s} \mu(s) \left(\beta H_{J}(s) + \ln \mu(s)\right) \\
\widetilde{F}_{J\beta}(q) = \sum_{s} p_{q}(s) \left(\beta H_{J}(s) + \ln p_{q}(s)\right)
\end{array}
\end{equation}

The function $F_{J\beta}$ represents the moment-generating function of the Boltzmann distribution \eqref{4.2)}, which, in general, cannot be exactly calculated.

The variational free energy $\widetilde{F}_{J\beta}$ serves as an approximation to it. By using a parameterized family of distributions $p_{q}(s)$, the optimization of free energy $q$ minimizes the Kullback-Leibler divergence to achieve the best possible approximation of the free energy.

The distribution $p_{q}$, averaged and covariantly parameterized, is defined as $p_{q} = p_{m,\chi}$ as follows:

\begin{equation} \label{4.10)} 
\begin{array}{l}
p_{m,\chi} = \prod_{(ij) \in \varepsilon} \frac{1 + m_{i}s_{i} + m_{j}s_{j} + \chi_{ij}s_{i}s_{j}}{4} \prod_{i=1;n} \left[\frac{1 + m_{i}s_{i}}{2}\right]^{1-d_{i}}
\end{array}
\end{equation}

Here, $m_{i}, \chi_{ij}$ represent averages over $s_{i}, s_{ij}$ by distribution, and $d_{i} = |\{j: (ij) \in \varepsilon\}|$ represents the weight of the $i$-th node.

By combining expressions \eqref{4.10)} and \eqref{4.11)}, it is observed that $\left. \widetilde{F}_{J\beta}^{Bethe}(m,\chi)\right|_{m=0} = 0$, implying that the paramagnetic point always provides the minimum energy according to Bethe.

\begin{equation} \label{4.11)} 
\left. \frac{\partial^{2} \widetilde{F}_{J\beta}^{Bethe}}{\partial m_{i} \partial m_{j}} \right|_{m=0} = \delta_{ij} \left(1 + \sum_{k \in \partial i} \frac{\chi_{ik}^{2}}{1 - \chi_{ik}^{2}}\right) - \frac{\chi_{ik}}{1 - \chi_{ik}^{2}}
\end{equation}

Now, by calculating the Bethe energy gradient with respect to $\chi$ and assuming $\chi_{ij} = \tanh(\beta J_{ij})$, we obtain an expression for the Hessian approximation:

\begin{equation} \label{4.12)} 
\left(H_{\beta, J}\right) = \delta_{ij} \left(1 + \sum_{k \in \partial i} \frac{\tanh^{2}(\beta J_{ik})}{1 - \tanh^{2}(\beta J_{ik})}\right) - \frac{\tanh(\beta J_{ik})}{1 - \tanh^{2}(\beta J_{ik})}
\end{equation}

Based on the relationship between Bethe energy and the Hessian, and the fact that the paramagnetic point always results in the minimum Bethe energy, as demonstrated earlier, we can deduce that the Nishimori temperature approximation corresponding to the paramagnetic point is defined as the maximum temperature at which the minimum coinciding eigenvalues of the Hessian occur (or in the special case of no coinciding minimal eigenvalues, the temperature corresponding to the minimal eigenvalue):

\begin{equation} \label{4.13)} 
\widehat{\beta}_{N} = \max_{\beta} (\beta: \gamma_{\min}(H_{\beta , J}) = 0),
\end{equation}

where $\gamma_{\min}$ corresponds to the smallest eigenvalue.

\subsection{ Bethe Permanent and Bethe Energy  }

The connection between formulas \eqref{4.11)} and \eqref{4.12)} reveals that the Bethe Hessian can also be computed through the second mixed derivative concerning independent directions of the Bethe energy. Assuming the matrix of the interaction graph is toric, we can establish an analogous difference for the second mixed derivative of the Bethe energy at a node. Let's consider the difference analogue of the mixed derivative for a certain function on a regular uniform two-dimensional grid using general terminology:

\begin{equation} \label{5.1)} 
\begin{array}{l}
\frac{\partial^{2} U}{\partial x_{i} \partial x_{j}} = \frac{\partial}{\partial x_{i}}\left(\frac{\partial U}{\partial x_{j}}\right) \\
\frac{\partial U}{\partial x_{j}} \sim \frac{U_{i,j+1} - U_{i,j-1}}{2h_{j}} \\
\frac{\partial^{2} U}{\partial x_{i} \partial x_{j}} \sim \frac{\left(U_{i+1,j+1} - U_{i+1,j-1}\right) - \left(U_{i-1,j+1} - U_{i-1,j-1}\right)}{4h_{j}h_{i}}
\end{array}
\end{equation}

By substituting the Bethe energy for nodal values, we obtain the Bethe-Hessian of the matrix corresponding to formula \ref{4.12)}. To correctly calculate the nodal Bethe energy, we introduce the concept of the Bethe permanent of a matrix.

For a non-negative matrix $W [n \times n]$, the matrix permanent is defined as, \cite{Von10,Smarandache13,Von20,Von23}:

\begin{equation} \label{5.2)} 
\sum_{\pi \in S_{n}} \prod_{i=1}^{n} W_{i\pi(i)}
\end{equation}

Here, $S_{n}$ represents the symmetric group over the set {1, ..., n}, and it can be seen as the set of all permutations of the matrix columns.

Now, let's define the matrix $W$ as some function over $S_{n}$:

\begin{equation} 
\begin{array}{l}
\text{perm}(W) = \sum_{\pi \in S_{n}} f(\pi, W) \\
f(\pi, W) = \prod_{i=1}^{n} W_{i\pi(i)}
\end{array}
\end{equation}

Since the value of the function is non-negative everywhere, we consider it as a density function over the space of all permutations.

If we envision permutations as perfect matches or connections between two sets of n objects, A and B, we can define sets of connections, X and Y, such that:

\[X = \{x_{1}, ..., x_{n}\}, Y = \{y_{1}, ..., y_{n}\}, x_{i}, y_{j} \in \{1, ..., n\} \text{ for all } i, j.\]

We define functions $\phi$ and $\psi$ as follows:

\begin{equation} \label{5.2)} 
\begin{array}{l}
\phi(x_{i}) = \sqrt{W_{ix_{i}}} , \phi(y_{j}) = \sqrt{W_{y_{j}j}} \\
\psi(x_{i}, y_{j}) = I(\neg (j = x_{i} \oplus i = y_{j}))
\end{array}
\end{equation}

Here, we use square roots because equal potentials will be multiplied in the final ratio. The logical function $I$ represents the negation of the XOR operation, returning 1 when the logical operands match. The result of the function is 0 when the pair of operands does not form a true permutation. In other words, if object i from A is associated with object j from B, which, in turn, is associated only with object i from A and no other (and vice versa), we detect a perfect match, and the function takes the value 1. Using this data curation, we introduce an analogue of the density function:

\begin{equation}  
\hat{f}(X,Y) = \prod_{i,j} \psi(x_{i}, y_{j}) \prod_{k} \phi(x_{k})\phi(y_{k})
\end{equation}

Thus, we can represent the matrix permanent as:

\begin{equation}  
\text{perm}(W) = \sum_{X,Y} \hat{f}(X,Y)
\end{equation}

To work in terms of probabilities, we introduce normalization:

\begin{equation} \label{5.3)} 
P(X,Y) = \frac{1}{Z(W)} \prod_{i,j} \psi(x_{i}, y_{j}) \prod_{k} \phi(x_{k})\phi(y_{k})
\end{equation}

Here, $Z(W)$ represents the normalization value, which is the sum of function values $\hat{f}(X,Y)$ over all possible pairs $(X,Y)$. We will refer to $P(X,Y)$ as the distribution function of the matrix permanent $W$.

\subsection{Free Bethe Energy}

To approximate the distribution function, we will employ the Bethe free energy. The Bethe free energy of our distribution, considering plausible states $b$:

\begin{equation} \label{5.4)} 
\begin{array}{l}
F_{Bethe} = -\sum_{i,j}\sum_{x_{i},y_{j}} b(x_{i},y_{j}) \ln \psi(x_{i},y_{j})\phi(x_{i})\phi(y_{j}) + \\
\hspace{1cm} + \sum_{i,j}\sum_{x_{i},y_{j}} b(x_{i},y_{j}) \ln b(x_{i},y_{j}) - \\
\hspace{1cm} - (n-1)\sum_{i}\sum_{x_{i}} b(x_{i}) \ln b(x_{i}) - \\
\hspace{1cm} - (n-1)\sum_{j}\sum_{y_{j}} b(y_{j}) \ln b(y_{j})
\end{array}
\end{equation}

Here, the likelihood states $b$ represent a set of pseudo-marginal distribution values that are locally consistent with each other but not necessarily consistent in the global distribution. In the bipartite graph under consideration, the pseudo-marginal values, in addition to being non-negative, are linearly locally consistent:

\begin{equation} 
\begin{array}{l}
\sum_{y_{j}} b(x_{i},y_{j}) = b(x_{i}), \sum_{x_{i}} b(x_{i},y_{j}) = b(y_{j}), \forall i,j, \\
\sum_{x_{i}y_{j}} b(x_{i},y_{j}) = 1
\end{array}
\end{equation} 

The class of true limit values is a subclass of pseudo-marginal values.

To approximate the permanent, we will use the expression, \cite{Yed03,Von10,Smarandache13}:

\begin{equation} \label{5.5)} 
\text{perm}(W) \approx \exp\left(-\min_{b} F_{Bethe}(b)\right)
\end{equation} 

Equation (2.5) is of interest to us primarily to determine the Bethe permanent and will be used in subsequent subsection. In the context of previous subsection, we are interested in methods to directly calculate the argument of a function, specifically the minimum of the Bethe energy. Let's define the necessary relations for its calculation.

\subsection{Belief Propagation}

An original algorithm for local minimization of Bethe free energy is Belief Propagation (Sum-Product). Belief Propagation is a message-passing algorithm that alternatively updates values at nodes by exchanging messages between variables. Let's assume that \(m_{x_l}(y_j)\) is the message from \(x_i\) to \(y_j\), then the likelihoods are defined via messages as:

\begin{equation} \label{5.6)} 
\begin{array}{l}
b\left(x_i, y_j\right) \sim \psi\left(x_i, y_j\right)\phi\left(x_i\right)\phi\left(y_j\right)\prod_{k\neq j}m_{y_k}\left(x_i\right)\prod_{l\neq i}m_{x_l}\left(y_j\right) \\
b\left(x_i\right) \sim \phi\left(x_i\right)\prod_{k\neq j}m_{y_k}\left(x_i\right); \quad b\left(y_j\right) \sim \phi\left(y_j\right)\prod_{l\neq i}m_{x_l}\left(y_j\right)
\end{array}
\end{equation}

At each iteration, messages are updated according to the rule:

\begin{equation} \label{5.7)} 
m_{x_i}^\text{new}\left(y_j\right) = \sum_{x_i} \left[\phi\left(x_i\right)\psi\left(x_i, y_j\right)\prod_{k\neq j}m_{y_k}\left(x_i\right)\right]
\end{equation}

Finally, the messages are attenuated using a scaling factor to provide smoother optimization in logarithmic space:

\begin{equation} \label{5.8)} 
\ln m_{x_i}\left(y_j\right) \leftarrow \ln m_{x_i}\left(y_j\right) + \alpha\left[\ln m_{x_i}^\text{new}\left(y_j\right) - \ln m_{x_i}\left(y_j\right)\right]
\end{equation}

\subsection{Belief Propagation Algorithm Low Temperature Approximation}

The sum-product implementation of the BP algorithm is computationally difficult (polynomial). We will use the normalized min-sum implementation (linear, Jacobian approximation for high SNR, low temperature). We reduce all possible messages to two options: when the message between the variables does not match and when they match:

\begin{equation} \label{5.8)} 
\begin{array}{l}
y_j \neq i: \\
m_{x_i y_j}^\text{not} = \sum_{x_i \neq j} \phi(x_j) m_{y_{x_i} x_i}^\text{match} \prod_{k\neq j, k\neq x_i} m_{y_k x_i}^\text{not}
\end{array}
\end{equation}

\begin{equation} \label{5.10)} 
\begin{array}{l}
y_j = i: \\
m_{x_i y_j}^\text{match} = \phi(x_i = j) \prod_{k\neq j} m_{y_k x_i}^\text{not}
\end{array}
\end{equation}

Normalized by \(m_{y_k x_i}^\text{not}\), we get:

\begin{equation} \label{5.11)} 
\begin{array}{l}
m_{x_i y_j}^\text{not} = 1 \\
m_{x_i y_j}^\text{match} = \frac{\phi(x_i = j)}{\sum_{k\neq j}^{} \phi(x_i = k) m_{y_k x_i}^\text{match}}
\end{array}
\end{equation}

Let's define a fast message update rule based on the fact that only one value needs to be updated between each pair of variables.

\begin{equation} \label{5.12)} 
m_{x_i y_j}^{} \leftarrow \frac{1}{Z} \phi(x_i = j)/\sum_{k\neq j}^{} \phi(x_i = k) m_{y_k x_i}^{}
\end{equation}

We can rewrite the formulas to update the values using the new message values:

\begin{equation} \label{5.13)} 
\begin{array}{l}
b(x_i = j, y_j = i) = \frac{1}{Z_{ij}} \phi(x_i) \phi(y_j) \\
b(x_i \neq j, y_j \neq i) = \frac{1}{Z_{ij}} \phi(x_i) \phi(y_j) m_{y_{x_i} x_i}^{} m_{x_{y_j} y_j}^{} \\
b(x_i) = \frac{1}{Z} \phi(x_i) m_{y_{x_i} x_i}^{} \\
b(x_i) = \frac{1}{Z} \phi(y_j) m_{x_{y_j} y_j}^{}
\end{array}
\end{equation}

Applying relation \eqref{4.1)}, based on the toroidality of the matrix, we can determine the Bethe-Hessian by the Bethe energy at each of the nodes by the method of algebraic additions, calculating the Bethe energy for each \(i,j\) element from the relations for the permanent of the matrix, from which the \(i\) row and \(j\) column are deleted. In addition to optimizing the likelihood function, there is also a combinatorial method for calculating the Bethe permanent. However, relation \eqref{5.5)} makes it possible to pass from the value of the permanent to the value of the energy.

\subsection{Combinatorial Computation of Bethe Energy} 

In order to introduce a combinatorial definition of the Bethe permanent, introducing new notations, let us turn to the basic properties of permanent calculus:

Let \(H = (H_{ij}) \sim [m \times m]\) be a square matrix. Then we define the matrix permanent \(H\) as:

\begin{equation} \label{6.1)} 
perm(H) = \sum_{\sigma \in S_{m}} \prod_{i \in [m]} H_{i \sigma(i)}
\end{equation}

where \(S_{m}\) over the set \([m]\) is the set of all possible permutations.

Relation \eqref{6.1)} differs from the definition of the determinant of a matrix only in the absence of sign alternation and in that it can be extended to non-square matrices.

So, for a matrix of the form \(H = \begin{pmatrix} 1 & 1 \\ 1 & 1 \end{pmatrix}\), the permanent is defined as \(perm(H) = 1 \cdot 1 + 1 \cdot 1 = 2\).

Approximations of the matrix permanent are the M-permanent Beta and the Beta permanent. They are defined for a P-extended matrix. For the matrix discussed above, the P-extended matrix will be:

\begin{equation} \label{6.2)} 
H^{P_{M} \uparrow} = \begin{pmatrix} P_{11} & P_{12} \\ P_{21} & P_{22} \end{pmatrix}
\end{equation}

where \(P_{ij} \sim [M \times M]\), \(P_{ij} \in \psi_{m, M}\) - permutation matrices from \(\psi_{m, M}\) - the set of all permutation matrices of size \([M \times M]\).

Let us describe the properties of permutation matrices:

- If the order of the rows in a unit matrix is changed, the resulting matrix is called a permutation matrix. In other words, a square matrix, in each row and in each column of which only one element is non-zero and equal to one, is called a permutation matrix.

- The following properties of the permutation matrix are easily verified by direct calculation.

    - Left multiplication of a permutation matrix by a rectangular matrix A results in a permutation of the rows of matrix A.
    
    - Right multiplication of a permutation matrix by a rectangular matrix A results in a permutation of the columns of matrix A.

    - Let, for example, the fifth row of the permutation matrix be a row of the form \((0, 1, 0, 0, ..., 0)\). Then the result of multiplying this row by the columns of a rectangular matrix \(A = (a_{ij})\) is the second line \((a_{21}, a_{22}, a_{23}, ...)\) matrix A, which is located in the position of the fifth row of the resulting matrix.

Thus, if in the i-th row of the permutation matrix P the unit is located in the j-th column, then multiplying the matrix P on the left by the matrix A leads to moving the j-th row of the matrix A to the position of the i-th row.

Similarly, if in the i-th column of the permutation matrix P the unit is located in the j-th row, then multiplying the matrix P on the right by the matrix A leads to moving the j-th column of the matrix A to the position of the i-th column.

- If a permutation matrix P is obtained from an identity matrix E by swapping two rows (or two columns), then such a matrix is called an elementary permutation matrix.

- When multiplying an elementary permutation matrix by a matrix A on the left, the corresponding rows of the matrix A are permuted.

- Right multiplication of an elementary permutation matrix by matrix A results in permutation of the corresponding columns of matrix A.

Property 1:
\begin{equation} 
PP^{T} = P^{T}P = I
\end{equation} 

Where \(P^{T}\) - transposed permutation matrix; \(I\) - identity matrix.

Property 1 is true because:

\begin{equation} \label{6.3)} 
 (PP^{T})_{ij} = \sum_{k=1}^{n} (P)_{ik} (P^{T})_{kj} = \delta_{ij}
\end{equation} 

where \(\delta_{ij}\) is the Kronecker delta symbol.

Theorem 1. The product of permutation matrices of the same order is a permutation matrix.

Theorem 2. A permutation matrix of the nth order can be represented as a product of (n - 1) elementary permutation matrices.

Theorem 3. The square of the elementary permutation matrix is the identity matrix.

Knowing this, we define the M-permanent of the Beta matrix \(H\), \cite{Von10,Smarandache13}:

\begin{equation} \label{6.4)} 
perm_{BM}(H) = \sqrt[M]{\left\langle perm\left(H^{P_{M} \uparrow}\right)\right\rangle_{\psi_{m,M}}}
\end{equation}
where the angle brackets mean the average over all elements of the set \(\psi\). This statement would seem to lead us to complex calculations: instead of one permanent of the 2 x 2 matrix, it is necessary to calculate \(M!\) permanent matrices of size \(2M \times 2M\).

Using property 1 and theorem 1, we transform the matrix \(H^{P_{M} \uparrow}\):

\begin{equation} \label{6.5)} 
\begin{array}{l}
H^{P_{M} \uparrow} = \begin{pmatrix} P_{11} & P_{12} \\ P_{21} & P_{22} \end{pmatrix} = \begin{pmatrix} P_{11}P_{11}^{T} & P_{12}P_{11}^{T} \\ P_{21}P_{21}^{T} & P_{22}P_{21}^{T} \end{pmatrix} = \\
= \begin{pmatrix} I & \widetilde{P_{12}} \\ I & \widetilde{P_{22}} \end{pmatrix} = \begin{pmatrix} II^{T} & \widetilde{P_{12}}\widetilde{P_{12}}^{T} \\ II^{T} & \widetilde{P_{22}}\widetilde{P_{12}}^{T} \end{pmatrix} = \begin{pmatrix} I & I \\ I & \widehat{P_{22}} \end{pmatrix}
\end{array}
\end{equation}

Now there is only one block-permutant of size \(M \times M\) in the matrix. The M-permanent Beta of such a matrix can be determined using the cyclic index of the symmetry group:

\begin{equation} \label{6.6)} 
\left\langle perm\left(H^{P_{M} \uparrow}\right)\right\rangle_{\psi_{m,M}} = Z(S_{M})
\end{equation}

where \(Z(S_{M})\) is the cyclic index of the symmetry group.

The cyclic index of the symmetry group can be defined in terms of Bell polynomials:

\begin{equation} \label{6.7)} 
Z(S_{M}) = \frac{B_{M}(0!a_{1}, 1!a_{2}, \ldots, (n-1)!a_{M})}{M!}
\end{equation}

where \(B_{M}(0!a_{1}, 1!a_{2}, \ldots, (n-1)!a_{M})\) is the Bell polynomial. The calculation comes down to applying the recursive relation:

\begin{equation} \label{6.8)} 
\begin{array}{l}
Z(S_{M}) = \frac{1}{M!} \sum_{l=1}^{M} C_{l-1}^{M-1} (l-1)!a_{l}(M-l)!Z(S_{M-1}) \\
Z(S_{0}) = 1
\end{array}
\end{equation}

where \(a_{l} = 2^{c}\) - Bell variables, \(c \sim [1;M]\) - the number of cycles formed by the considered substitution.

The value of the Bethe energy from the obtained values of the permanent from the cyclic index of the group can be restored using the relation \eqref{5.5)}.

\subsection{Relation of ground states of ISING models and  Topology data analysis}

Gunnar Carlsson is a mathematician who has made significant contributions to the field of applied topology, particularly in the area of topological data analysis. His work has focused on developing mathematical tools and techniques for analyzing complex data sets using methods from algebraic topology. One of Carlsson's major contributions has been the development of persistent homology, which provides a way to extract multi-scale topological features from data. This approach has been widely adopted in the field of topological data analysis and has led to numerous applications in areas such as biology, neuroscience, and social networks. Carlsson has also worked on developing algorithms for clustering and classification using topological methods. These approaches have shown promise in providing more robust and interpretable results compared to traditional machine learning techniques.

At paper \cite{Gu18} authors explores the use of topological data analysis (TDA) to study the structure of neural networks. The paper argues that TDA provides a powerful tool for understanding the behavior of these networks, particularly with respect to their ability to generalize from training data to new examples.

The paper begins by providing an overview of neural network topology, which refers to the arrangement of neurons and connections within the network. The authors argue that traditional methods for analyzing this topology, such as graph theory and clustering, are limited in their ability to capture the complexity and high dimensionality of neural networks.

To overcome these limitations, the authors propose using TDA, which involves applying mathematical tools from topology to analyze the structure of complex data sets. In particular, they focus on the use of persistent homology, a technique from TDA that measures the robustness and persistence of features in a data set.

The authors demonstrate the effectiveness of this approach by applying it to a variety of neural network architectures and datasets. They show that persistent homology can be used to identify important features of the network topology, including clusters of neurons and regions of high connectivity.

\subsection{Topology Data Analysis (TDA) Fundamentals}

Topology data analysis can be used as a tool to select the appropriate curvature architecture for a given problem

Curvature architecture refers to the choice of the functional form for the curvature component in a geometric model. The appropriate curvature architecture depends on the underlying topology of the data being analyzed.

Topology data analysis involves studying the global shape and structure of data sets using tools from algebraic topology. These tools can provide insights into the topological features of the data, such as connected components, holes, and voids.

By analyzing the topological structure of the data, we can determine the most suitable curvature architecture for the problem at hand. For example, if the data has a toroidal topology with nontrivial fundamental group, a hyperbolic curvature architecture may be more appropriate than a Euclidean one.

Therefore, incorporating topology data analysis into the process of selecting a curvature architecture can improve the performance of geometric models by ensuring that the chosen architecture is well-suited to the topological structure of the data being analyzed.

 Two continuous functions from one topological space to another are called homotopic (similar), if one can be "continuously deformed" into the other, such a deformation being called a homotopy between the two functions.

 Persistent homology is a method for computing topological features of a space at different spatial resolutions. Persistence describe range of function, critical points of function, by pairing they min and max (thresholds) values. More persistent features are detected over a wide range of spatial scales and are deemed more likely to represent true features of the underlying space rather than artifacts of sampling, noise, or particular choice of parameters.

\begin{figure}
\centering
\includegraphics[width=4.2in,height=3.5in]{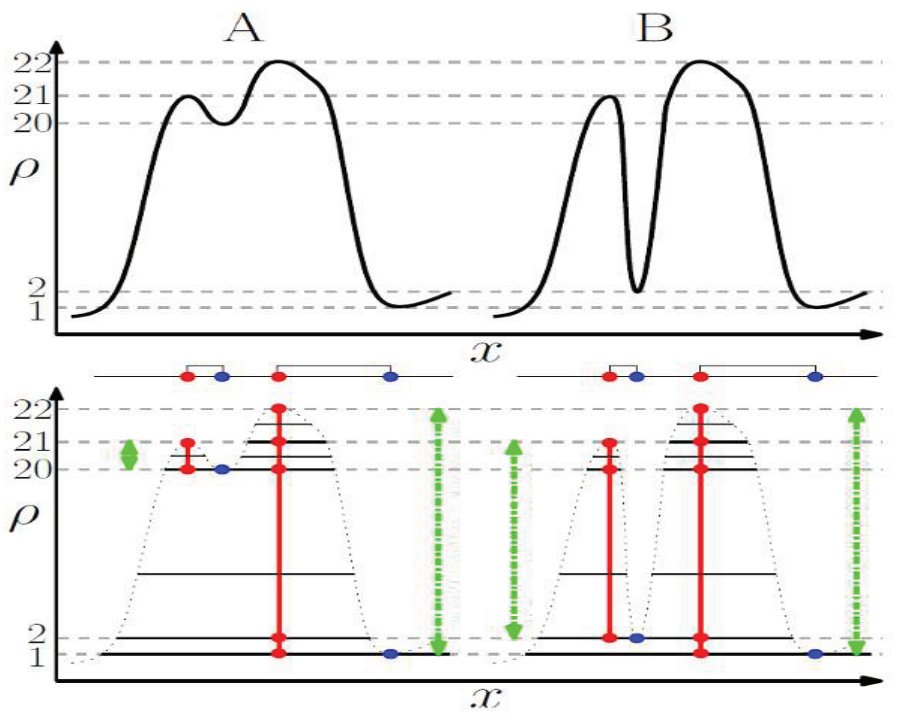} 
  \caption{Illustration of Homotopy between two functions}
  \label{Fig3}
\end{figure}

 Topology of its excursion sets (i.e. the set of points with value higher than a given threshold) evolves when the threshold is continuously and monotonically changing. Whenever the threshold crosses the value of a critical point, the topology of the excursion set change, Fig. \ref{Fig3}. This simple concept can be extended to higher dimensions and different area of application from automorphic function to automorphic bundles, for detail read \cite{Frenkel04,Knapp04,AscFeZu08,RobGhris12,Rob14,Ghris17}.

 Let ${\rm X}\subset {{\rm R}}^{{\rm n}}$ be a point cloud and fix $\alpha >0$.  The ball with radius$~~\alpha $ and center in ${\rm x}$ will be denoted by ${{\rm B}}_{{\rm a}}{\rm (x).}$

 A Cech complex ${\mathcal C}\left({\mathbf X},{\mathbf a}\right)$ is a simplicial complex with the vertex set ${\rm X,}$ such that ${{\rm x}}_{{{\rm i}}_0}{\rm ,\dots ,}{{\rm x}}_{{{\rm i}}_{{\rm n}}}$forms an ${\rm n-}$simplex in $\left({\mathbf X},{\mathbf a}\right),~$if and only if $\bigcap^{{\rm n}}_{{\rm k=0}}{{{\rm B}}_{{\rm a}}{\rm (}{{\rm x}}_{{{\rm i}}_{{\rm k}}}{\rm )}\ne {\rm 0.}}$ 

 For each ${\mathbf a}$ we have homology group ${{\rm H}}_{{\rm n}}{\left({\mathcal C}\left({\mathbf X},{\mathbf a}\right)\right)}_{{\rm a}}$-vector spaces (over fixed field F) and for any ${\alpha }_{{\rm 1}}{\rm <}{\alpha }_{{\rm 2}}$ there is a linear map: ${{\rm H}}_{{\rm n}}{\left({\mathcal C}\left({\mathbf X},{\mathbf a}\right)\right)}_{{{\mathbf a}}_{{\mathbf 1}}}\to {{\rm H}}_{{\rm n}}{\left({\mathcal C}\left({\mathbf X},{\mathbf a}\right)\right)}_{{{\mathbf a}}_{{\mathbf 2}}}.$

 Dimension of ${{\rm H}}_0{\left({\mathcal C}\left({\mathbf X},{\mathbf a}\right)\right)}_{{{\mathbf a}}_{{\mathbf 0}}}$ is a number of connected component ${\mathcal C}\left({\mathbf X},{\mathbf a}\right)$ or Betti number 0, ${{\rm Betti}}_0$. Dimension of ${{\rm H}}_{{\rm 1}}{\left({\mathcal C}\left({\mathbf X},{\mathbf a}\right)\right)}_{{{\mathbf a}}_{{\mathbf 1}}}$is a number of "1-dimensional holes" at ${\mathcal C}\left({\mathbf X},{\mathbf a}\right){\mathbf ,\ }{{\rm Betti}}_{{\rm 1}}$. By n-th persistent homology of K we will understand a collection of the vector spaces ${{\rm H}}_{{\rm n}}{\left({\mathcal C}\left({\mathbf X},{\mathbf a}\right)\right)}_{{\mathbf a}}$together the maps ${{\rm H}}_{{\rm n}}{\left({\mathcal C}\left({\mathbf X},{\mathbf a}\right)\right)}_{{{\mathbf a}}_{{\mathbf 1}}}\to {{\rm H}}_{{\rm n}}{\left({\mathcal C}\left({\mathbf X},{\mathbf a}\right)\right)}_{{{\mathbf a}}_{{\mathbf 2}}}.$ 
 
 Betti number describe topology properties of function and field spaces. A barcode is a visual of persistent homology as a graded module over F[x]. Homology could be estimated using any type of data which allow to estimate distance matrix and its statistical analogues - correlation matrix. To calculation persistent homology of low dimension data used Alpha complex. Alpha complex is equivalent to the Cech complex and much smaller if you do not bound the radii. Let ${\rm X}$ be a finite subset of ${{\mathbb R}}^{{\rm n}}$ and ${\rm x}\in $X.

 A Voronoi cell of ${\rm X}$ is a subset 
 ${\rm V} \left ({\rm X,x} \right )$ 
 of ${ \mathbb R}^{ \rm n}$ 
 given by  ${ \rm V} \left ({\rm X,x} \right ){ \rm =} \left \{ \rm v \in { \mathbb R}^{ \rm n} { \rm |} \forall { \rm x}^{ \rm '} \in { \rm X } ,  || {v-x} || \leq ||  {v-x'}|| \right \}. $

 Example of Voronoi cell represented on Fig. \ref{Fig444}.

 A Delanay complex ${\rm D(X)}$ is a simplicial complex with vertex set ${\rm X}$ such that ${{\rm x}}_{{{\rm i}}_0}{\rm ,\dots ,}{{\rm x}}_{{{\rm i}}_{{\rm n}}}$forms a ${\rm n-}$simplex in ${\rm D}\left({\rm X}\right)$, if and only if $\bigcap^{{\rm n}}_{{\rm k=0}}{{\rm V}\left({\rm X,}{{\rm x}}_{{{\rm i}}_{{\rm k}}}\right)\ne \emptyset }.$ For ${\rm x}\in {\rm X}$ and ${\rm a}>0$ let ${{\rm V}}_{{\rm a}}\left({\rm X,x}\right)\bigcap {{\rm B}}_{{\rm a}}\left({\rm x}\right).$ An ${\rm a-}$complex ${\rm A}\left({\rm X,a}\right)$ is simplicial complex with vertex set ${\rm X}$ such that ${{\rm x}}_{{{\rm i}}_0}{\rm ,\dots }$, ${{\rm x}}_{{{\rm i}}_{{\rm n}}}$ forms an n ${\rm -}$ simplex in ${\rm A}\left({\rm X,a}\right)$, iff $\bigcap^{{\rm n}}_{{\rm k=0}}{{\rm V}\left({\rm X,}{{\rm x}}_{{{\rm i}}_{{\rm k}}}\right)\ne \emptyset }$. Process of construction of Alpha-comples represented on Fig. \ref{Fig444}. Computation complexity O(${{\rm 2}}^{\left\lceil {\rm d/2}\right\rceil }{\rm )}$ with ${\rm n=}\left|{\rm X}\right|$ and ${\rm X}\subset {{\mathbb R}}^{{\rm n}}$.


Because calculation of Alpha-complex became extremely hard with increase of dimension, we can consider approximation methods for calculation of Cech complex by applied calculation to subset of less dimension. Example of such complexes is Vietoris-Rips (VR) complex.

The VR complex is an abstract simplicial complex defined on a finite metric space, where each simplex corresponds to a subset of points whose diameter is smaller that some given threshold. Varying the threshold, we can also see the Rips complex as a filtration of the $(n-1)$-dimensional simplex, where the filtration value of each simplex is the diameter of the corresponding subset of points. This filtered complex is most often used as an approximation of the Cech complex. After re-scaling (Rips using the length of the edges and Cech the half-length), they share the same 1-skeleton and are multiplicatively 2-interleaved or better. While it is slightly bigger, it is also much easier to compute, \cite{Ghris17, SilvaCarlson, Zomo10}. 

The paper  \cite{Gu18,CaGa20} utilized topological data analysis to gain insights into the computations performed by convolutional deep neural networks. By analyzing the internal states of these networks, they were able to modify their computations to improve speed and generalization across different datasets.

As a result of their analysis, they also produced a new geometry for sets of features on image datasets (MNIST, CIFAR-10, SVHN, ImageNet and etc). This allowed them to develop a methodology for constructing analogues of CNNs for many other geometries, including graph structures constructed by topological data analysis.

Their work builds on previous research, such as the study by Carlsson et al., \cite{CaGa20}, which showed that local patches in images tend to concentrate around a primary circle and can be modeled by algebraically defined functions. This insight has led to the construction of a set of features for images that admit a circular geometry.

In addition to circular geometries, it is also possible to construct Klein bottle-based sets of features and corresponding geometries \ref{fig:images001}, ~\ref{Fig444} . These could potentially have applications in modeling complex systems and optimizing their performance using topological data analysis.

 \begin{figure}
\centering
\includegraphics[width=5in]{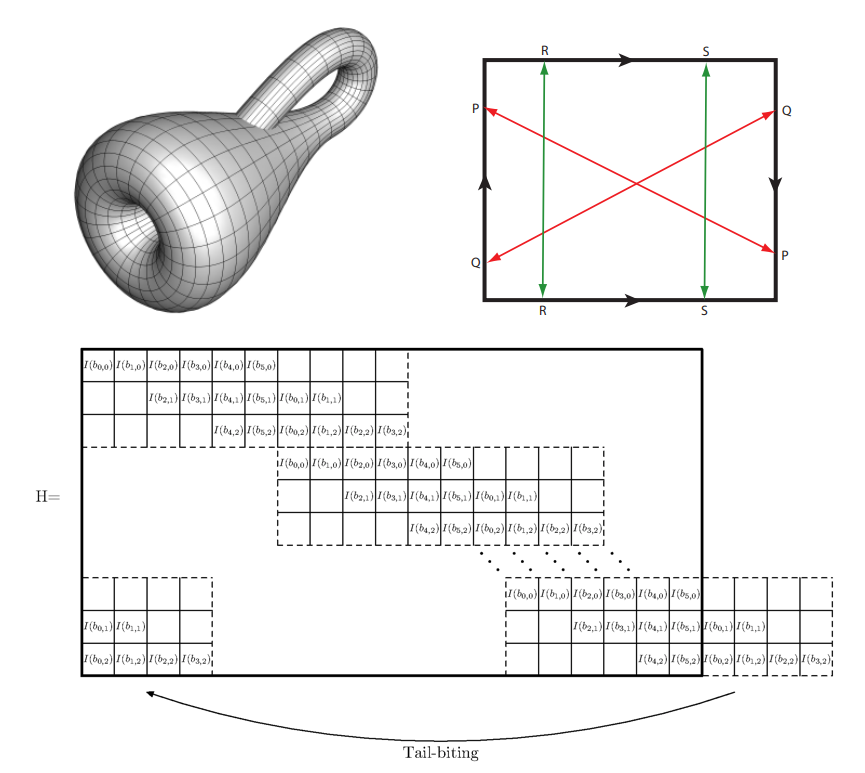} 
  \caption{Klein bottle and Zigangirov's Spatially-Coupled Convolutional QC-LDPC code \cite{Zig}  as half-twisted Hyperbolic Torical Topology}
  \label{Fig333}
\end{figure}
 \begin{figure}
\centering
\includegraphics[width=3.1in]{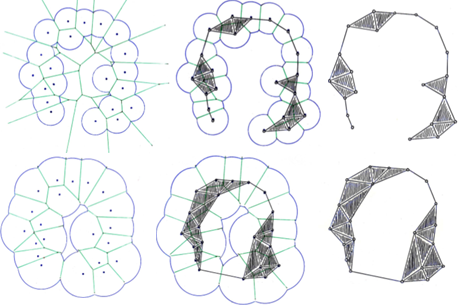} 
  \caption{Alpha-complex construction}
  \label{Fig444}
\end{figure}
\begin{figure*}[t]
	\begin{center}
            \includegraphics[width=4in]{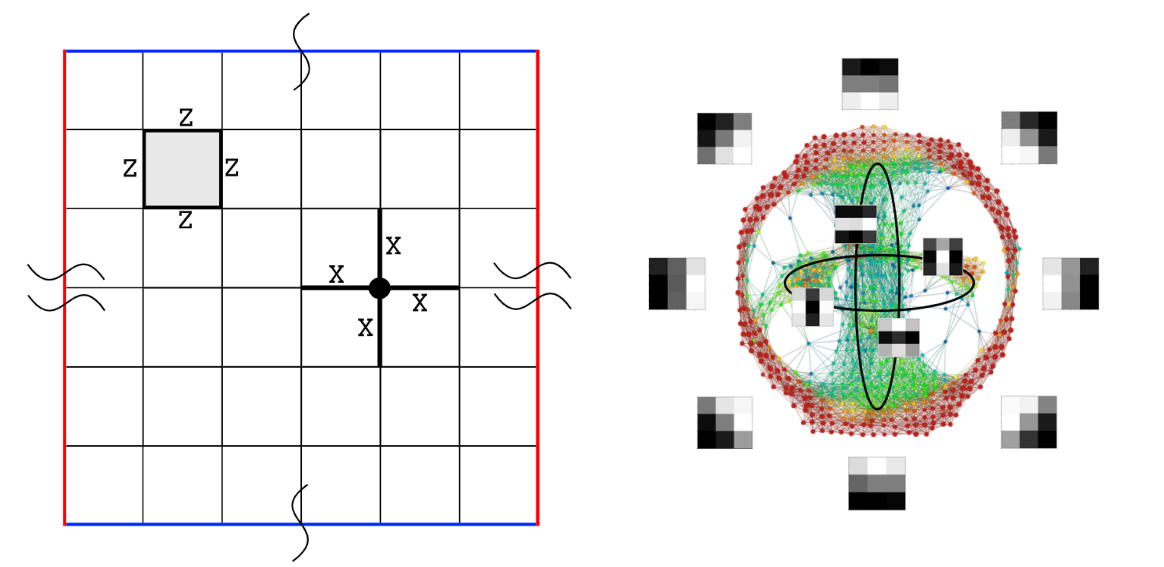}
	\end{center}
	\caption { Left side: Kitaev's Toric 2D  code \cite{Kitaev03}. Right side: Natural image Topology First layer, CIFAR-10,\cite{CaGa20}  }
	\label{Fig555}
\end{figure*}

The Klein bottle is a non-orientable surface that cannot be embedded in three-dimensional Euclidean space without self-intersections. Although the Klein bottle is not directly related to toric hyperbolic topology, it is possible to construct a torus with a Klein bottle hole using toric hyperbolic geometry. Toric hyperbolic topology involves studying the properties of geometric structures on surfaces of constant negative curvature, such as hyperbolic surfaces. These geometric structures can be used to model complex systems and optimize their performance. One way to use toric hyperbolic geometry to construct a torus with a Klein bottle hole is by representing the torus as a quotient space of the hyperbolic plane under a suitable group action. Specifically, the torus can be obtained by identifying opposite sides of a rectangular fundamental domain in the hyperbolic plane. To introduce a Klein bottle hole into the torus, we need to modify the identification process so that the opposite sides are identified with a half-twist. This results in a torus with a Klein bottle hole that cannot be realized in Euclidean space without self-intersections.

Spatially-Coupled MET QC-LDPC code, \cite{Zig} constructed by coupling together a series of $L$ disjoint, or uncoupled, block MET QC-LDPC codes into a single coupled chain by means of an edge spreading operation, Fig. ~\ref{Fig444}. A $(W,C,N,L)-$ tail-bitten spatial-couple QC-LDPC code $C(H)$ of length $W$ multiple  \textit{L} multiple \textit{N}  is defined by a parity-check base matrix 

\begin{equation} 
H=\left[\begin{array}{cccc} {A_{0,0} } & {A_{1,0} } & {...} & {A_{WL-1,0} } \\ {A_{0,1} } & {A_{1,1} } & {} & {A_{WL-1,1} } \\ {\vdots } & {\vdots } & {\ddots } & {\vdots } \\ {A_{0,CL-1} } & {A_{1,CL-1} } & {...} & {A_{WL-1,CL-1} } \end{array}\right],
\end{equation} 
where $0\le i\le CL-1$,$1\le j\le WL-1$ and CPM $A_{i,j} $ represents either $N\times N$ zero matrix $Z$ or the $N\times N$ circulant permutation matrix $I\left(p_{i,j} \right)$ obtained by cyclically right-shifting the $N\times N$ identity matrix $I\left(0\right)$ by $p_{i,j} $ positions.  Denote by CPM-column the i-th column of  $A_{0,i},A_{1,i},\dots ,A_{CL-1,i}$.

For a specific Tail-bitten Spatial-couple MET QC-LDPC code we define the corresponding CPM-shifts matrix  as the matrix of circulant shift that defines the QC-LDPC code:
\begin{equation} 
B=\left[\begin{array}{cccc} {b_{0,0} } & {b_{1,0} } & {...} & {b_{W-1,0} } \\ {b_{0,1} } & {b_{1,1} } & {} & {b_{W-1,1} } \\ {\vdots } & {\vdots } & {\ddots } & {\vdots } \\ {b_{0,C-1} } & {b_{1,C-1} } & {...} & {b_{W-1,C-1} } \end{array}\right].
\end{equation} 

Define vector D of shifts: $D^{T}=\left[ {d_{0} },  {d_{1} },  {\cdots },  {d_{C-1} }\right] ,$ with conditions $d_0=0,\ d_i<\ d_j\ {\rm iff\ }i<j,\ \forall i\ d_i<W$. If  $\frac{W}{C}\ $ is integer, vector $D$ can be defined as $d_i=\frac{i\ W}{C}{\rm \ }\ $.

For $0\le i\le CL-1$ and $0\le j\le WL-1$, CPMs $A_{i,j}$ in parity-check matrix $H$  can be calculated:
if $i'\ {\rm mod\ }\ WL\ \le i\ <\left(i'+W\right)\ {\rm mod\ }\ WL$, then $A_{i,j}=I(b_{i\ {\rm mod\ W}, j\ {\rm mod\ C\ }})$, else  $A_{i,j}=Z.$

\begin{figure}[t]
	\begin{center}
            \includegraphics[width=6in]{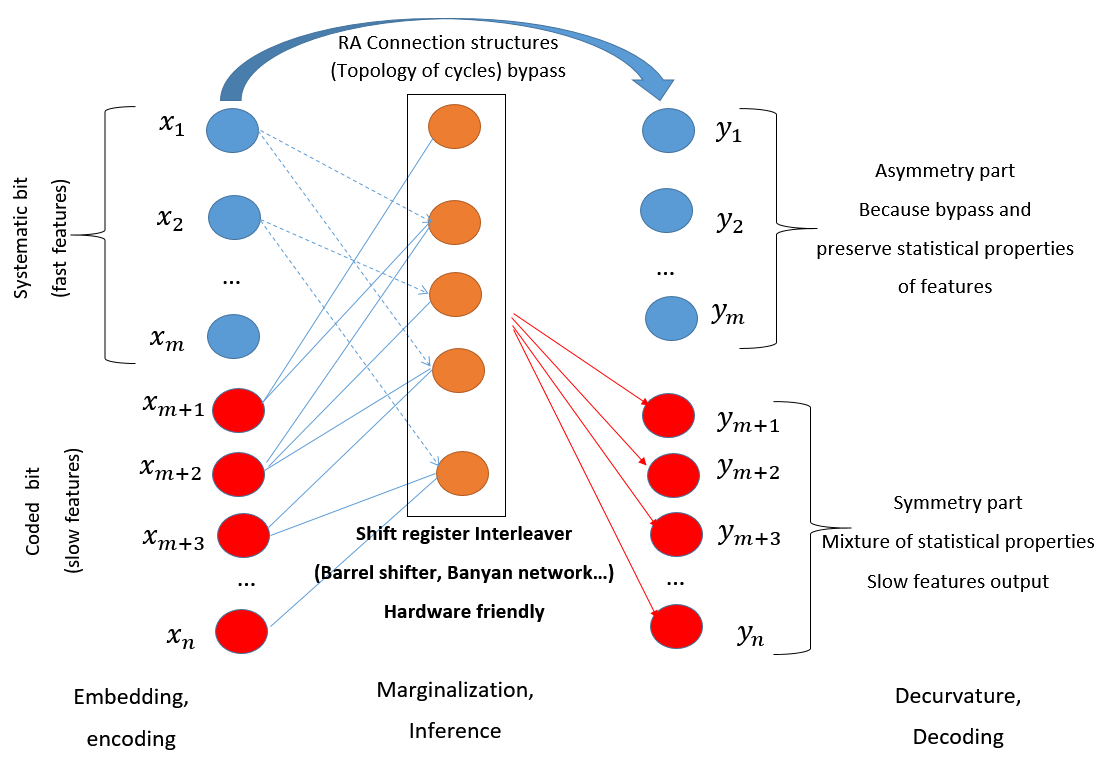}
	\end{center}
	\caption {Repeat Accumulate Code on the graph non-linear low dimensional toric/spherical (hyperbolic) topology embedding as combination of symmetry and asymmetry(loss surface lattice/Hessian), \textbf{fast features avoid non-linear transformation} }
	\label{fig:images0099999}
\end{figure}

In this article, a novel embedding methodology is proposed for torical (spherical) topology codes on the graph, specifically QC and QC-LDPC codes. The main objective is to encode the properties of the input tensor, ensuring alignment between the data topology and the DNN topology. Within neural network architectures, this encoding process is crucial, Fig. \ref{fig:images0099999} .

Let's consider example of LDPC code parity-check matrix $H$ application to encoding (embedding) message (Tensor) $m$ in error correction. Mention that nonlinear mapping and encoding represented embedding in ML.
\begin{equation} 
x=H\times m,
\end{equation} 
where $m$ - transmitted message, $H$- parity-check matrix, $G$ - generator matrix, $x$-encoded message.

Encoding satisfied zero syndrome condition:
\begin{equation} 
G=H^{-1},    H\times x=0.
\end{equation} 

If the generator matrix of the code, by definition satisfied $G=H^-1$ than: 
\begin{equation} 
H \times G^T=0.
\end{equation} 

An important note, in classical coding the number of rows and columns are not equal (not square matrices), but in quantum codes this is solved using the Calderbank-Shor-Steane (CSS) stabilizer code construction \cite{CalShor96,Kitaev03,Gleb21}. Relationship with ground states of Ising model proved in previous subsections.

The QC-LDPC codes parity-check matrix of Repeat Accumulated (RA) codes can be represented as follows:

\begin{equation} 
H=	[ \, H_1 | H_2  ] \, ,
\end{equation}

where $H_1$ - parity-check part which multiplied by systematic bits (features at DNN) which directly embedding by fast connection, $H_2$ - special structures RA submatrix.

\begin{figure}[t]
	\begin{center}
            \includegraphics[width=6in]{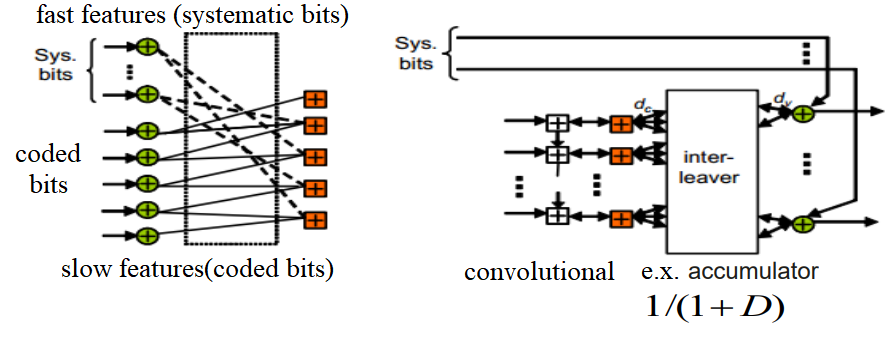}
	\end{center}
	\caption {Repeat Accumulate Code on the graph structures, fast part systematic bits(features for DNN) and repeat accumulated coded bits, \cite{Li05,RiUr01,DivMcel98,Sarah10}  }
	\label{fig:images00888}
\end{figure}

\begin{figure}[t]
	\begin{center}
            \includegraphics[width=6in]{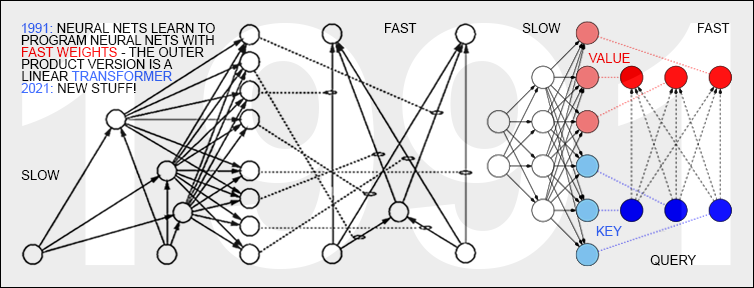}
	\end{center}
	\caption {The dynamic approach combines: convolutional and block codes by incorporating attention and convolution within a deep neural network (DNN) proposed by Schmidhube at 1991, \cite{Sch91}. RA codes example of such approach }
	\label{fig:images00999}
\end{figure}

For example if RA parity-check matrix defined as follow: 
\begin{equation} 
H_{2} =\left[\begin{array}{cccccc} {I} & {I} & {0} & {...} & {0} & {0} \\ {0} & {I} & {I} & {...} & {0} & {0} \\ {0} & {0} & {I} & {...} & {0} & {0} \\ {\vdots } & {0} & {0} & {...} & {0} & {0} \\ {\vdots } & {\vdots } & {\vdots } & {...} & {\vdots } & {\vdots } \\ {\vdots } & {\vdots } & {\vdots } & {...} & {I} & {0} \\ {0} & {0} & {0} & {...} & {I} & {I} \\ {0} & {0} & {0} & {...} & {0} & {I} \end{array}\right]  ,
\end{equation} 

where $I$-entity circulant matrix of size $L \times L$.
The inverse matrix $H_{2}^{-1}$ became triangular and encoding became $G \times x$ described by use of convolutional accumulator Fig. \ref{fig:images00888}, \cite{RiUr01,DivMcel98,Sarah10}, or under DNN embedding done by dynamic convolution (involution and convolution features)    \ref{fig:images00999}, \cite{Sch91,WuFa19,LiHu21}. 

Moreover such RA codes on the graph structures can be generalized irregular RA (GeIRA), as we show on the example of Transformer MEGA architecture, which represented by GeIRA codes on the graph,  \cite{DivMcel98, Li05}, \ref{Mega}.

\textbf{Proposition 3} Based on the research paper \cite{Ta04}, a connection has been identified between QC block codes and spatially coupled QC convolutional codes (terminated with limited memory). This connection, observed under the ising model, is founded on the klein bottle twisted topology found in block QC codes with a torical (hyperbolic) topology, as we show at \ref{Fig333}. By utilizing this relationship, it is possible to create a dynamic convolution that relies on the input tensor and aligns with the aforementioned topology (low complexity non-linear multiplicative correlation), fast and slow feature, systematical bits (fast feature) and coded bits in channel codes using extremely efficient convolutions accumulator (Repeat Accumulate codes) processing widely used at wireless easy encoders QC-LDPC codes, \cite{Li05,DivMcel98,RiUr01,Sarah10}. This aligns with the findings of TDA analysis by Carlsson \cite{CaGa20}, as proposed by Schmidhuber(1991) and described in papers \cite{Sch91,WuFa19,LiHu21}. The integration of QC-LDPC block and convolutional codes , along with their relationship to Polar and Turbo codes, following serial processing on Klein bottle and Torus Topology, plays a crucial role in enhancing the quality of structured code on the graph models (high automorphism) associated with diverse problems ranging from material science and complexity theory to efficient implementations of deep neural networks. The core metric for gauge quality of any dynamic system is Trapping sets enumerator: codewords and pseudocodewords. Asserts that \underline{the Trapping sets enumerator is a fundamental metric used to evaluate the gauge}, which define loss surface of spherical and toroidal dynamic systems, including Ising models, DNNs, control systems, differential equations, social networks, other related topology graph models and tensor networks.

\textbf{Proposition 4} It is possible to construct SC codes with topologies similar to that of a Klein bottle by choosing an appropriate graph structure. This can be done by taking a torus and identifying opposite edges with a half-twist, resulting in a Klein bottle, \ref{Fig444}.  The connection between block and convolutional QC-LDPC codes is evident in a  paper  \cite{Ta04}. Similarly, the connection of Polar codes after sequential processing using successive cancellation scheduler to LDPC code is discussed in papers \cite{Fer13,Fos15} . Additionally, a showcases the connection between Turbo code and QC-LDPC codes under serial processing, \cite{Ji06}. Similar approach code be used to choise optimatimal embedding using TDA.
Non-linear sparse factorization can be considered as a low-dimensional embedding that captures the underlying structure of high-dimensional data. By applying methods from topology data analysis and information geometry, we can analyze the curvature of the data and create models that capture this curvature. This allows us to reduce the complexity of classification, interpolation, and regression tasks, making them more efficient and accurate. Additionally, by using stochastic dynamic systems, we can account for the inherent uncertainty in real-world data and make predictions that are robust to noise and other sources of variability.  The spectral gap is closely associated with certain topological properties, such as the Bethe tree or lattice, Bethe permanent (pseudocodewords TS(a,b)) \cite{Vo10,Sma13}, and permanent (codewords TS(a,0)) \cite{Mac01,Sma12}. Additionally, the graph spectral bound \cite{Ta01,Mi05,Lu06} is also related to these properties. Neural networks that have a minimal clustering coefficient tend to outperform others. Entangled nets (code on the graph with optimized Trapping set enumerator) models possess a significantly low clustering coefficient due to the presence of only large loops, with high symmetry topology. Consequently, they are considered promising candidates for constructing artificial neural networks with high capacity and excellent performance \cite{Lu06}. Improvements in spectral properties, particularly the bethe permanent and permanent, lead to enhancements in the enumerators of Trapping sets in graph models. This, in turn, results in increased capacity and excellent performance DNN.

Overall, the use of topological data analysis, information geometry and Codes on the Graph (channel coding) in deep learning has shown promise in improving performance, gaining understanding of optimal architecture (shift register-automorphism, sparsity, quantization using aprior graph model mutual information) and the computations performed by neural networks. The development of new geometries and methodologies for constructing codes on the graph (channel codes) and quantum system analogues of DNN: CNNs, RNNs, Transformers  could lead to even further advancements in this field.

\newpage

\section{Application of LDPC Codes Matrix Factorization as a Special Case of Code on the Graph Embedding}

Matrix factorization can be considered as a special case of (Ising spin-glass) embedding, low dimension projections, \cite{CamMez23,CamMez23_etend}. In matrix factorization, a matrix is decomposed into two or more matrices that represent latent factors or features. These latent factors can be seen as embeddings, which capture the underlying relationships and structure in the data. By representing data as embedding, matrix factorization techniques allow for dimensionality reduction, pattern extraction, and recommendation systems, among other applications. Therefore, matrix factorization can be viewed as a specific form of Codes on the graph based Ising model embedding where a matrix is decomposed into lower-dimensional representations.

Low-rank matrix factorization is a technique used to approximate a given matrix by decomposing it into two or more lower-dimensional matrices that are multiplied together. The goal of this approach is to identify the underlying patterns and structures in large datasets by reducing the complexity of the original matrix.

Truncated Singular Value Decomposition (TSVD) and Principal Component Analysis (PCA) are among the most commonly used low-rank matrix factorization techniques. TSVD, for example, approximates a given matrix by selecting the k largest singular values and their corresponding singular vectors, where k is a user-defined parameter that determines the rank of the approximation, \cite{Halko09}. PCA, on the other hand, identifies the principal components of the dataset by analyzing its covariance matrix and represents the original matrix as a linear combination of these components. There are several ways in which the TSVD can be improved or extended to handle more complex data structures and noise models. Here are a few examples:

\begin{itemize}
\item Regularization: One way to improve the TSVD is to add a regularization term to the optimization problem that penalizes large singular values. This can help to reduce the impact of noise and outliers in the data, and can provide a more stable and robust approximation.

\item Adaptive rank selection: Another way to improve the TSVD is to use adaptive rank selection techniques that automatically determine the optimal rank of the approximation based on the properties of the data. For example, one approach is to use the ratio of the largest to smallest singular value as a criterion for selecting the rank.

\item Non-negative matrix factorization: The TSVD is based on the singular value decomposition (SVD), which assumes that the matrix being approximated has both positive and negative singular values. However, in some applications, it may be desirable to have a non-negative approximation, which can be achieved using non-negative matrix factorization (NMF) techniques.

\item Bayesian methods: Bayesian methods provide a way to incorporate prior knowledge about the properties of the data into the low-rank approximation problem. For example, one approach is to use a hierarchical Bayesian model that places a Gaussian prior on the singular values and a Laplacian prior on the singular vectors.

\item Structured low-rank approximations: In some applications, it may be desirable to impose additional structure on the low-rank approximation, such as sparsity or block-diagonal structure. This can be achieved using structured low-rank approximation techniques, such as the structured total least squares method or the structured low-rank approximation by principal component pursuit.

\end{itemize}

We'll combine 3 approaches in our factorization: Bayesian methods, Structured low-rank approximations,  Adaptive rank selection. Adaptive rank selection reached by adaptive rate construction at parity-check matrix using Simulated Annealing EMD and Trapping Set Spectrum improving (Hamming distance, codewords TS(a,0) and pseudocodewords TS(a,b))  maximization for parity-check submatrices (nested rates codes), \cite{USA18,USA_UK19,EmdSpectrum}.  Structured low-rank approximations shall be used by application of Quasi-cyclic LDPC matrix for QC-LDPC codes, and hidden authomorphism which we incorporate by ACE maximization at PEG construction method, ACE maximization increase symmetry of subgraph (weaker that EMD maximization but still improve structures compare to random PEG method), \cite{USA18,USA23}. Bayesian methods applied by using Topology results of TDA analysis (\cite{Gu18}) and KAM theorem (torical topology, \cite{Laz93}), proofs about probability of Hyperbolic Topology under stochastic dynamic process  \cite{Gromov87}.

The conventional approach to matrix approximation involves factorizing a large square matrix $X$ into a small number of low-rank matrices, such as  $X\approx \hat{X}=WH$ or $X\approx WSH$   where $W\in {\rm R}^{N\times r} $, $S\in {\rm R}^{r\times r} $, $H\in {\rm R}^{r\times N} $ and   $r\ll N$

Different trifactor decomposition methods exist, such as Nystrum decomposition~(\cite{Chr2000}), where  ~$W$ and $H$ are calculated using normal kernel function, and  ~$S$ is learned, and CUR decomposition~(\cite{MichaMa09} ), where  ~$W$ contains~$r$~columns of~$X$,~$H$~contains~$r$ rows of~$X$, and~~$S$ is learned.

The accuracy of the matrix approximation can be evaluated using different measures or divergences, such as the squared Frobenius norm. The squared Frobenius norm is a common measure used to assess the difference between two matrices and is defined as the sum of the squared differences between the corresponding elements of the two matrices. In the case of matrix approximation, the squared Frobenius norm can be used to quantify the error between the original matrix and its approximation. By minimizing the squared Frobenius norm between the original matrix and its approximation, the low-rank factorization can be optimized to achieve the best possible approximation of the original matrix.
The squared~Frobenius norm:~

\begin{align}
\label{eq:Frobenius}
   D\left(X\parallel \hat{X}\right)=\left\| X-\hat{X}\right\| _{F}^{2} =\sum _{ij}\left(X_{ij} -\hat{X}_{ij} \right) ^{2} .
\end{align}

According to paper \cite{Ech1936}, minimizing the squared Frobenius norm 
 ~$\left\| X-WH\right\| _{F}^{2} $ with respect to  ~$W$and~$H$~ can be solved using Truncated Singular Value Decomposition (TSVD). This involves constructing a diagonal matrix ~$\Lambda $ with the largest  $r$ singular values and using the corresponding left and right singular vectors as columns to form the matrices ~$U$ and~$V$, respectively. Then, the minimum is obtained by setting ~$W=U$~and~$H=\Lambda V^{T} $. Scaling can be moved from ~$W$~to~$H$~or vice versa.\textbf{}

\begin{figure*}[t]
	\begin{center}
		\includegraphics[width=\textwidth]{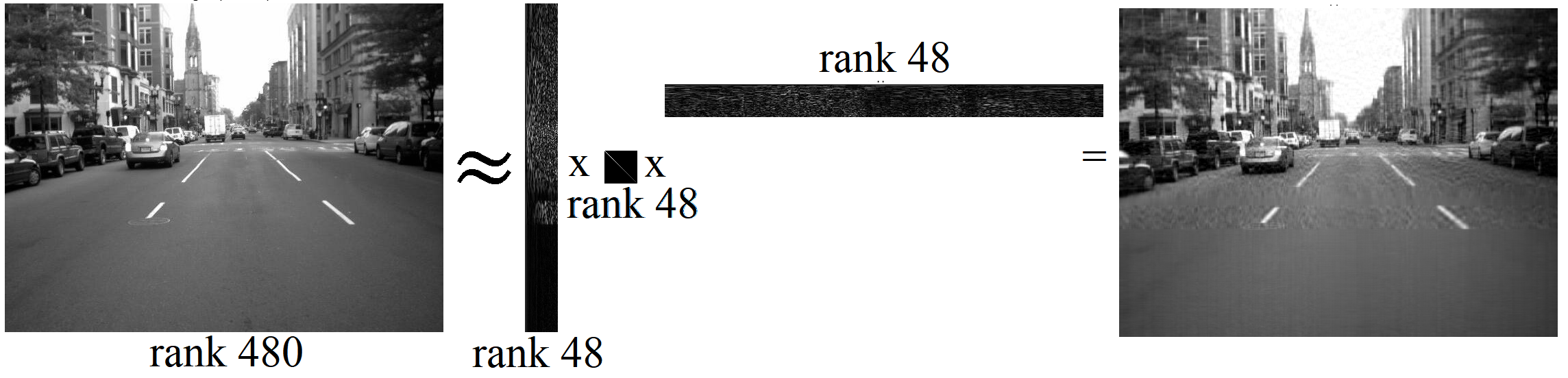}
	\end{center}
	\caption{Examples of low-rank matrix factorization using  SVD, svdsketch function from Matlab}
	\label{fig:factorize}
\end{figure*}

Matrix factorization involving more than two factors cannot achieve a lower approximation error than TSVD. In general, we can write $\hat{X}=\prod _{m=1}^{M}W^{(m)}$, where $W^{\left(m\right)} \in {\rm R}^{r_{m}, r_{m+1}}$ and $r_{1}=r_{M+1}=N$. We can always reduce a multi-factor case to the two-factor form by grouping $L=\prod _{m=1}^{m'-1}W^{(m)}$ and $R=\prod _{m=m'}^{M}W^{(m)}$, where $m'=\arg \min_{m}\left(\left\{r_m\right\}_{m=2}^{M}\right)$ and $X\approx LR$. It is necessary to ensure that $r_{m'} \ll N$ when approximating large square matrices; otherwise, the factorizing matrices may still be too large. The bottleneck of the grouping process is illustrated in Fig. \ref{fig:factorize}.

Several low-rank matrix factorization methods impose constraints on the factorizing matrices, including nonnegative matrix factorization (\cite{Lee99}) and binary matrix factorization (
\cite{MartSla2013, Zhang2010}). However, these constraints limit the solution space and often result in a higher approximation error. Other techniques use penalty terms on the factorizing matrices, such as Sparse Dictionary Learning at Compressed Sensing (\cite{Donoho06}) or Matrix Completion (\cite{Kim2010}, Funk matrix factorization (\cite{Simon2006}). Despite their specific applications, these methods generally produce a higher approximation error than TSVD (in terms of Frobenius norm) due to compromises made to satisfy the extra constraints or penalties.

\subsection{Sparse factorization}

We propose a new matrix approximation method that is not limited by the low-rank constraint. Our approach involves approximating a given matrix with a number ($M$) of square and sparse factorizing matrices. The approximation remains computationally efficient if the total number of non-zero entries is much smaller than $N^{2}$.

There are various ways to specify the sparse structure, i.e., the non-zero positions of the factorizing matrices. A good specification should ensure that the product of the factorizing matrices, denoted as $\hat{X}$, is a full matrix, thus avoiding any always-zero entries in the approximation. In addition, we consider a secondary requirement that each factorizing matrix has the same sparse structure, which provides better symmetry in the approximation.

We propose using the Hyperbolic Toric Topology Ising-model ground state solution LDPC codes parity-check matrix to decurve the data space and make it as close as possible to a Euclidean regular grid. The use of such matrices can help to describe the center of mass for input processing data, which facilitates efficient clustering and classification of data in Euclidean space. By transforming the original data into a form that is as close as possible to a regular Euclidean grid, we can simplify the data analysis process and improve the accuracy of data forecasting using simple hypersurfaces.

The goal of using Hyperbolic Toric Topology-based LDPC codes is to enable effective data analysis by enforcing a certain degree of sparsity on the factorizing matrices, which reduces computational complexity while maintaining sufficient accuracy in the approximation process. This approach can be particularly useful when dealing with high-dimensional or large-scale datasets where traditional low-rank matrix factorization methods may not be effective.

The LDPC codes parity-check matrix of rate zero specifies the non-zero positions in each factorizing matrix. This approach allows for processing data with different levels of noise, where the sparsity structure of the factorizing matrices can be adapted to match the level of noise or temperature in the input data. In channel coding, such codes are referred to as rate-adaptive codes and can correct errors at different Signal-to-Noise Ratio (SNR) levels.

In the context of neural network optimization, applying such a parity-check matrix can allow for choosing the rate or weight of the matrix based on the features of the input data. This approach can help to improve the accuracy and efficiency of the neural network by adapting to the level of noise in the input data, which could vary across different feature sets. Ultimately, this can lead to more accurate predictions and better performance of the neural network in various applications.

The construction of the proposed parity-check matrix is based on statistical physics theory and the properties of the input data probability distribution. Specifically, we use techniques from channel coding theory to improve the cycle topology of the matrix and enhance its sparsity structure. These techniques involve using a Hyperbolic Toric Topology Ising-model ground state solution LDPC codes parity-check matrix of rate zero, which has been shown to be effective in constructing sparse parity-check matrices for channel coding.

The approach involves modeling the input data as a probability distribution and using this distribution to design the parity-check matrix. By selecting the appropriate probability distribution and utilizing the underlying statistical properties of the data, we can create a matrix that is both sparse and effective in approximating the original data. This approach offers a powerful tool for data analysis and enables efficient processing of large-scale datasets with varying degrees of noise or uncertainty. Overall, the proposed construction method provides an innovative approach to designing factorizing matrices that can improve the performance of low-rank matrix factorization methods.  Thus every factorizing matrix has~$N\log _{2} N$~stored non-zero entries.

\begin{figure*}[t]
	\begin{center}
		\includegraphics[width=\textwidth]{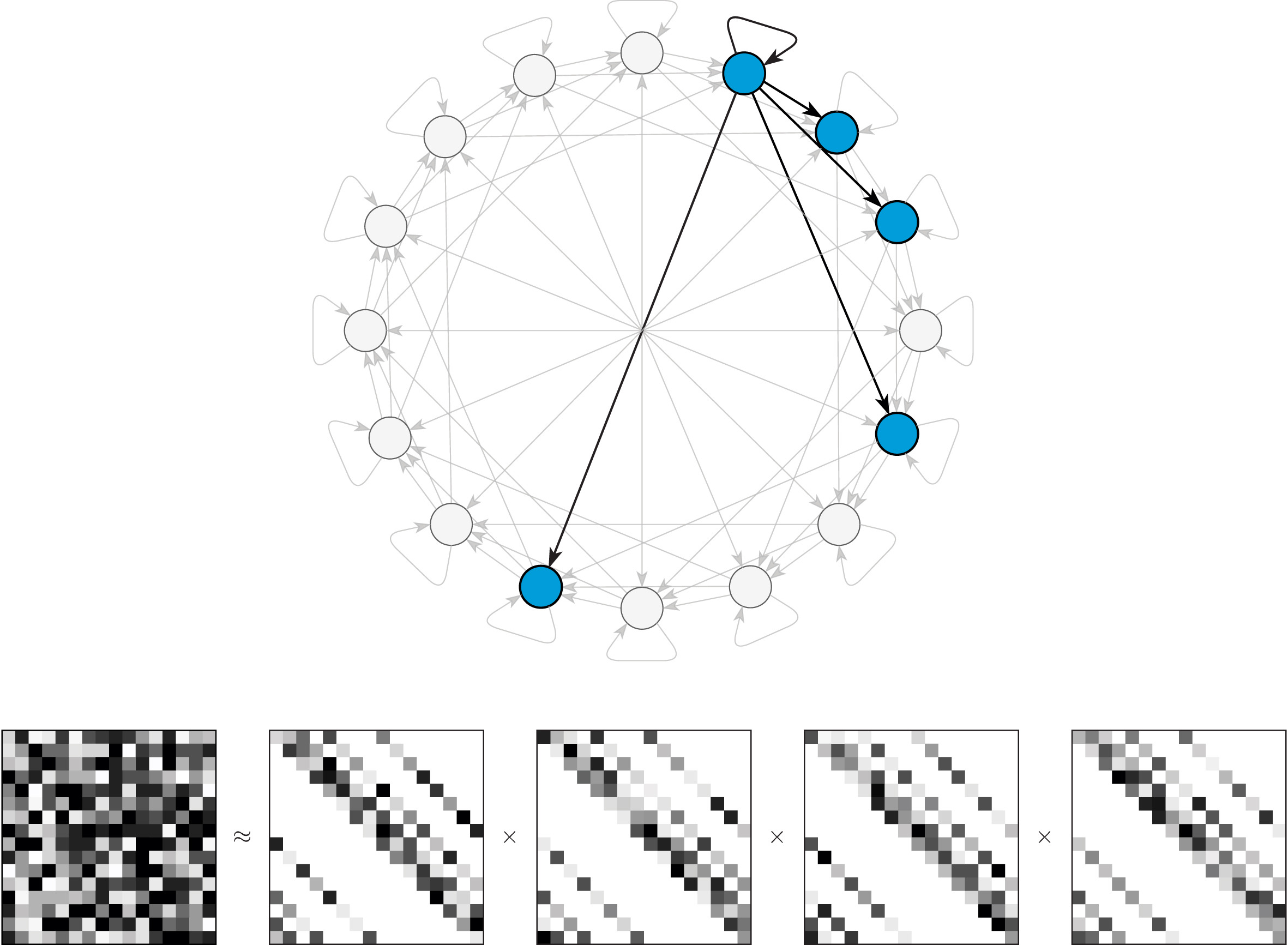}
	\end{center}
	\caption {Illustration of (top) the Chord protocol and (bottom) sparse factorization of a square matrix for $N=16$. Grayscale squares in the factorizing matrices represent stored entries (non-zeros) and explicit zeros), and completely white squares represent non-stored entries (implicit zeros), from \cite{Ru22}.}
	\label{fig:chord}
\end{figure*}

Both methods aim to approximate large square matrices using sparse factorizing matrices. However, the details of the approaches differ, with the proposed method using Hyperbolic Toric Topology Ising-model ground state solution LDPC codes parity-check matrices while the Chord factorization method proposed at paper ~\cite{Kha22} employs a chordal decomposition approach related to cage (distance graph) graph based LDPC codes ~\cite{Ma07}, which is based on the P2P Chord protocol described in \cite{Stoica2001}, detailed described previously in  \ref{Chord}. Additionally, the proposed method introduces a rate-adaptive aspect to adapt to the varying levels of noise or uncertainty in the input data, while the Chord method does not appear to incorporate such adaptation.

A comparison of the two methods could involve evaluating their respective approximation errors, computational complexities, and suitability for different types of datasets. It would be interesting to see how these methods perform on various benchmarks and applications and to identify their respective strengths and weaknesses. Overall, comparing the proposed method with existing state-of-the-art techniques such as the Chord Sparse Factorization method can provide valuable insights into the development of effective sparse factorization approaches for large square matrices.

The product of the factorizing matrices corresponds to the connections in the circular graph after multiple hops. We can set the number of factorizing matrices to~$M=\log _{2} N$, which corresponds to the number of hops, and the resulting~matrix product~becomes a full matrix with high probability~(see~\cite{Stoica2001}, ~Theorem 2). In total, there are~$N\left(\log N\right)^{2} $ non-zero entries, still much smaller than~$N^{2}$ for a large~ $N$.

The Sparse Factorization (SF) can thus be formulated as the following optimization problem:

\begin{align}
\label{eq:CHORD3}
\mathop{\min }\limits_{W^{(1)} ,\ldots ,W^{(M)} } \left\| X-\prod _{m=1}^{M}W^{(M)}  \right\| _{F}^{2}
\end{align}

where~$W^{(M)} $'s are sparse square matrices with non-zero positions specified by the Chord protocol (SF Chord, \cite{Stoica2001,Ru22}), LDPC codes parity-check  using PEG+ACE \cite{Hu05,Ti04} and QC-LDPC codes parity-check matrix, MET QC-LDPC codes with circulant more than 1 and product multigraph MET QC-LDPC codes parity-check  using SA+EMD, Simulated Annealing with exact cycle extrinsic message degree optimization (EMD) \cite{USA18}.

The minimization \ref{eq:CHORD3}  can be implemented by any existing solvers for unconstrained smooth optimization (we shall use BFGS method, \cite{Bro1970,Fle1970,Gold1970,Shanno1970}), where the cost of computing the gradient is~$O\left(NMK\right)$. Once the approximation error is minimized, we obtain the factorizing matrices~$W^{(1)} ,\ldots W^{(M)} .$. We call the approach non-parametric Sparse factorization under different structured matrices from code on the graph models as we directly optimize over the factorizing matrices \cite{Ru22}.

\subsection{Experiments settings}
\label{sec:exps}
We use different data for Sparse Factorization optimization problem from paper \cite{Ru22}, detailed described in Appendix\ref{sec:data}. SF Chord method and TSVD methods results taken from paper \cite{Ru22}.  We run  LDPC code construction methods: Regular LDPC constructed using Progressive edge grown (PEG) with Approximate cycle extrinsic message (ACE) degree optimization  \cite{Hu05,Ti04}, Simulated Annealing with exact cycle extrinsic message degree optimization (EMD) \cite{USA18} for QC-LDPC codes (including multi-edge type) and applied it in  experiments on a Workstation with 
AMD Ryzen 9 3950X @ 4.0GHz processors, with 128GB DDR4 2400 Mhz (dual channel Kingston SK Hynix)  of system memory, except non Parametrical optimization of Covariance "Mfeat" problem using BFGS Matlab which require 137 Gb of RAM  for which we use 2 CPU Intel Xeon E5-2696 V4 with 256 Gb DDR4 2133 Mhz (8 channels Cisco SK Hynix). Source code of platform with constructed codes available at github repository \cite{USA_UK_TopoML23}.

\subsection{Non-parametric Sparse Factorisation results}

There are various types of square matrices, each with its own unique properties and characteristics. The Truncated Singular Value Decomposition (TSVD) is a well-known method for approximating matrices that are close to low rank in terms of the Frobenius norm.

To evaluate the effectiveness of our proposed non-parametric Hyperbolic Toric Topology LDPC method, we conducted an empirical study (1) to compare its performance with that of the SF Chord and TSVD methods. Our results show that the LDPC method can outperform both SF Chord and TSVD in terms of the F-norm even when using the same number of non-zero entries. We also used equation \ref{eq:CHORD3} to identify the types of square matrices that are more suitable for our method.

To compare the performance of different methods for approximating large square matrices, we used the Matlab fminunc optimizer to solve the problem described in equation \ref{eq:CHORD3}. The non-zero entries in the factorizing matrices were initialized to random numbers between $\left[K^{-1}, K^{-1}+10^{-2}\right]$. The total number of non-zero entries for SF, QC-LDPC SA+EMD, Regular LDPC PEG+ACE, and TSVD were set to $N(\log_2 N)^2$ and $2Nr+r$, respectively. To ensure a fair comparison, we set $r = \left\lceil (\log_2 N)^2/2 \right\rceil$ in TSVD so that it has nearly the same number and no fewer non-zeros than SF, LDPC.

\begin{figure*}[t]
	\begin{center}
            \includegraphics[width=\textwidth]{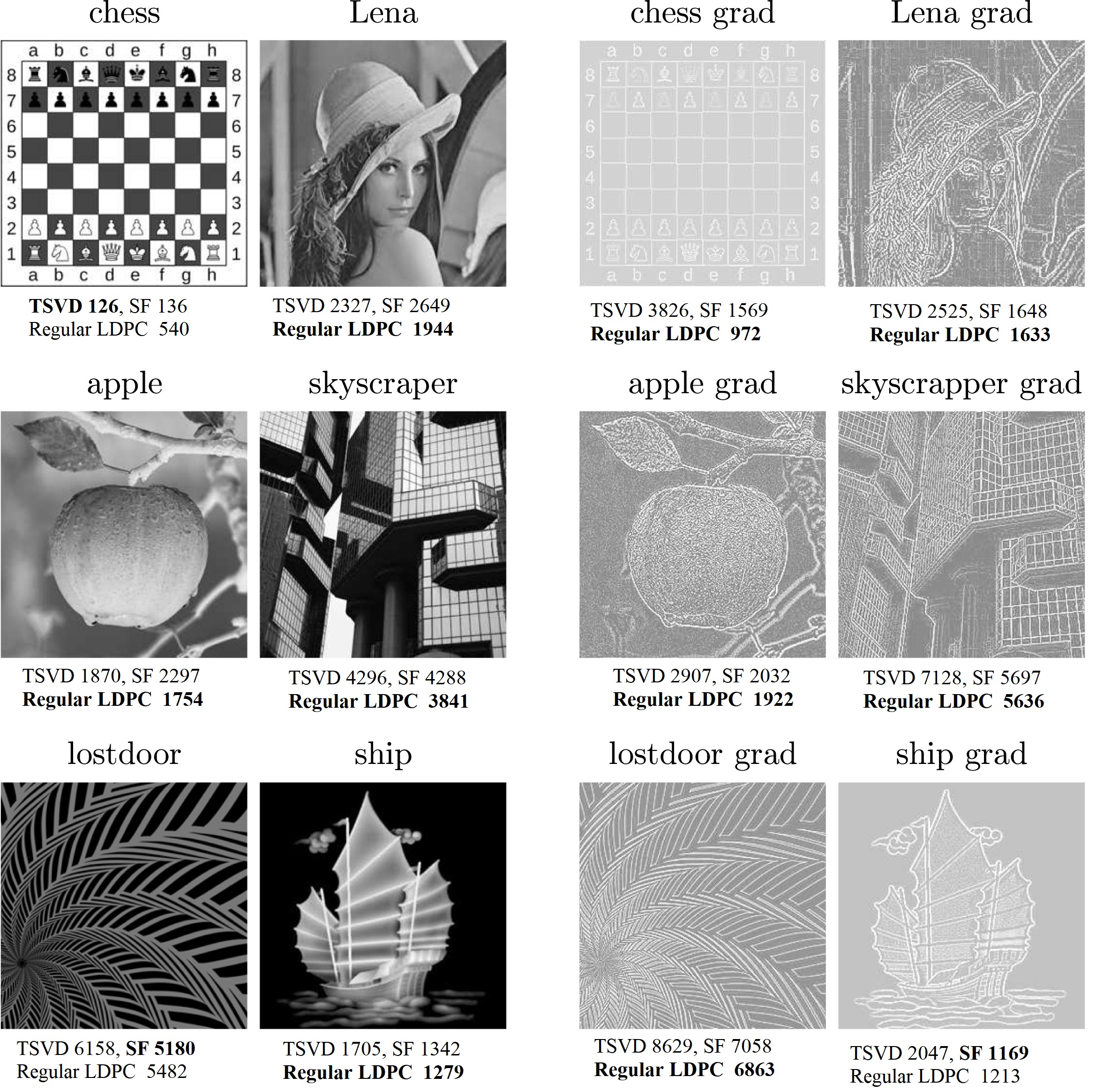}
	\end{center}
	\caption {Example square matrices: (left) original images and (right) the gradient magnitude images (displayed after histogram equalization for better visibility). Approximation errors by using TSVD, Regular LDPC PEG+ACE and SF with the same number of non-zeros are shown below the images. Boldface font indicates the winner for each case.}
	\label{fig:images}
\end{figure*}

We chose grayscale images with dimensions of 256~$\times$~256 as the approximated matrices for visual clarity. Figure \ref{fig:images} shows six typical square images, along with the resulting TSVD, LDPC and SF Chord approximation errors in F-norm below each image. This approach allows us to directly compare the quality of the approximations produced by each method and identify their respective strengths and weaknesses. By evaluating the performance of different methods on various types of datasets and applications, we can gain insight into the most effective approaches for sparse factorization of large square matrices.
 The chess image is close to low-rank since it consists of a black-and-white chessboard which has only two ranks. Therefore, TSVD performs better for this image as expected. However, for the other images that contain rich high-frequency details such as lines and corners, TSVD is not as good as LDPC, indicating a matrix type where LDPC can outperform TSVD and SF. To further verify this, we computed the gradient magnitudes of the images (shown in Figure \ref{fig:images} right). This method sets intensities in constant areas to zero, leaving only the remaining non-zero values which mainly represent high-frequency details. We observed that LDPC gives a lower approximation error than TSVD for all such matrices, which confirms that LDPC is more advantageous for approximating matrices with rich high-frequency details.

In summary, our results demonstrate that the performance of TSVD and SF Chord does not always surpass that of LDPC when using the same number of non-zeros for approximating square matrices. The winning cases indicate that LDPC is often superior to SF and TSVD when the approximated matrix is sparse, intrinsically high-rank, or contains rich high-frequency details. By identifying the types of matrices that are most suitable for each method, we can optimize the performance of sparse factorization approaches for different types of datasets and applications.

\begin{table*}[t]
\centering
\caption{ Approximation errors in F-norm by using TSVD \cite{Halko09}, Regular LDPC constructed using Progressive edge grown (PEG) with Approximate cycle extrinsic message (ACE) degree optimization  \cite{Hu05,Ti04}, Simulated Annealing with exact cycle extrinsic message degree optimization (EMD) \cite{USA18,EmdSpectrum} and Sparse Factorization (SF) using Chord Protocol \cite{Ru22,Stoica2001} for different types of square matrices. Boldface font indicates the winner for each case.}
\label{tab:result}
\begin{tabular}{|c|c|c|c|c|c|} \hline 
Data type & Data name & TSVD & SF Chord &  LDPC PEG+ACE & QC-LDPC SA+EMD (MET)\\  \hline 
Dense graph & AuraSonar & 8.54E+00 & 8.68E+00 & \textbf{7.01E+00} \\   \hline 
Dense graph & Protein & 1.17E+01 & 1.09E+01 & \textbf{0.74E+01} \\ \hline 
Dense graph & Yeast & 3.72E+01 & 3.61E+01 & \textbf{3.15E+01} \\ \hline 
Dense graph & Voting & \textbf{8.07E-04} & 1.71E+01 & 1.47E-02* &2.1E-02   \\ \hline 
Network & Sawmill & 3.24E+00 & 1.03E+00 & \textbf{0.26E+00} \\ \hline 
Network & Scotland & 5.90E+00 & 3.76E+00 & 3.99E+00* &  \textbf{2.42E+00} $\dagger$  \\ \hline 
Network & A99 m & 1.47E+01 & 1.01E+01 & 1.04E+01 & \textbf{0.99E+01} \\  \hline 
Network & Mexican Power & 3.85E+00 & 1.71E+00\textbf{} & \textbf{0.51E+00} \\ \hline 
Network & Strike & 2.73E+00 & 1.04E+00 & \textbf{0.20E+00} \\ \hline 
Network & Webkb Cornell & 6.98E+00 & 4.80E+00 & 5.48E+00 & 5.09E+00 (w3), \textbf{4.36E+00} $\dagger$ \\ \hline 
Network & Worldtrade & 8.65E+04 & 4.47E+04 & \textbf{3.54E+04}*  \\ \hline 
Surface mesh & Mesh1e1 & 1.87E+01 & 9.82E+00 & \textbf{8.45E+00} \\ \hline 
Surface mesh & Mesh2e1 & 2.48E+02 & 3.47E+02 & \textbf{1.86E+02} \\ \hline 
Surface mesh & OrbitRaising & 9.37E+01 & 8.53E+01 & \textbf{7.83E+01}* \\ \hline 
Surface mesh & Shuttle Entry & 2.73E+03 & 1.86E+03 & \textbf{ 1.85E+03} \\ \hline 

Surface mesh & AntiAngiogenesis & 5.85E+01 & 3.29E+01 & 4.36E+01  & \textbf{2.03E+01} $\dagger$    \\ \hline 
Covariance & Phoneme & 2.80E+01 & 5.27E+01 & \textbf{2.34E+01} \\ \hline 
Covariance & MiniBooNE & \textbf{1.04E+00} & 6.36E+03 & 4.81E+05 &   1.90E+03 (weight 2) \\ \hline 
Covariance & Covertype & 8.22E-02 & 1.90E-02 & \textbf{1.50E-02} \\ \hline 
Covariance & Mfeat & 1.11E+03 & 4.01E+05 &  3.36E+03 & \textbf{0.72E+03}* \\ \hline 
Covariance & OptDigits & 3.28E+01 & 7.01E+01 & \textbf{1.23E+01} \\ \hline 
Covariance & PenDigits & 4.00E+02 & 1.87E+02 & \textbf{6.00E-07} \\ \hline 
Covariance & Acoustic & 1.36E-02 & 1.11E-02 & 5.00E-03 & \textbf{4.55E-03}\\ \hline 
Covariance & IJCNN & 5.24E-02 & 3.03E-02 & \textbf{5.10E-03} \\ \hline 
Covariance & Spam Ham & 1.07E-01 & 4.97E-02 & \textbf{4.70E-02} \\ \hline 
Covariance & TIMIT & 9.64E+01& 1.56E+02 &   \textbf{8.85E+01} \\ \hline 
Covariance & Votes & 4.00E-01 & 1.70E-01 & \textbf{2.12E-05} \\ \hline 
\end{tabular}

\end{table*}

Besides square images, we have also compared TSVD, SF Chord and LDPC on several other types of square matrices.~Table~\ref{tab:result} ~shows the comparison results. The data types include~affinity matrices~of dense graphs (dense graph), affinity matrix of sparse networks (network), affinity matrix of surface mesh over 3D objects (surface mesh), and~covariance matrix~of vectorial data (covariance).  Table \ref{tab:result} shows the approximation errors in F-norm for different types of square matrices using TSVD, SF Chord, regular LDPC PEG+ACE and MET (weight of circulant >1) QC-LDPC codes. The winner for each case is indicated in boldface font. The table includes data types such as dense graph, network, surface mesh, and covariance.  For factorization train used 20 000 iterations (5 times less than used for SF Chord at original article \cite{Ru22}) and for category with * use 100 000 iteration similar as used in SF Chord optimization. For category with  $\dagger$   we construct SF Chord like multigraph Product code using QC-LDPC SA+EMD optimization.

The parity-check matrix of the multigraph product code used for data named "AntiAngiogenesis" is denoted by $H$, circulant $L=205$ and it is created using the QC-LDPC SA+EMD method, \cite{USA18}:
\begin{equation} \label{MultigraphProduct)} 
H=\left( I^0+I^2+I^3+I^{65}+I^{70}+I^{85}+I^{97}+I^{154}	   \right) 
\end{equation} 

 The parity-check matrix of the MET QC-LDPC codes used for data named "Webkb Cornell" is denoted by $H$, circulant $L=65$ and it is created using the QC-LDPC SA+EMD method, \cite{USA18}:
\begin{equation} 
H_2=\left(\begin{array}{ccccc} {I^{1} +I^{26} +I^{50} } & {I^{2}+I^{19} +I^{49} } &{I^{5}+I^{13}+I^{42} } \\ {I^{5} +I^{58} } & {I^{5} +I^{60} } & {I^{4}+I^{60}}   \\ {I^{4}+I^{18}+I^{48}} & {I^{23}+I^{28}+I^{61}} & {I^{1}+I^{4}+I^{53} }   \end{array}\right)
\end{equation} 

These results suggest that regular LDPC PEG+ACE and MET QC-LDPC codes, MET QC multigraph product codes are generally effective for reducing approximation errors in a variety of matrix types, but the choice of algorithm may depend on the specific characteristics of the data being analyzed. The ongoing work on constructing MET QC-LPDC codes is progressing, and future versions of the article will include additional information in the table  \ref{tab:result} and \cite{USA_UK_TopoML23}.

\section{Conclusion}
\label{sec:conclusion}
This paper showcases how quantum Ising models of fermions and bosons on spherical, (hyperbolic) toric topologies can be equivalently represented as structural (block) codes on graphs, specifically cyclic and MET QC-LDPC codes. The study also highlights the connection between modern transformer architectures, renowned for their exceptional performance in the Long Range Arena, and established LDPC codes graph models. Additionally, the paper explores the correlation between the structure of symmetric and asymmetric intersections of cycles, Trapping sets, in the graph-based code and the optimization algorithm landscapes. By considering factorization as a specialized form of embedding, the proposed methods for constructing codes on a graph are applied to a wide range of data, outperforming the Chord Protocol-based factorization method previously employed in the superior transformer architecture, ChordMixed, across the majority of datasets.  This research has the potential to make significant contributions to various areas, including Information Theory, DNN architecture design (through sparse and structured prior graph topology), efficient hardware design for Quantum and Classical Deep Neural Network Processing Unit/Tensor Processing Unit  (using aprior graph, quantize and shift register architecture), as well as Materials Science. In particular, it utilizes the KAM-theory method to gauge Trapping sets for Symmetric/AntiSymmetric wavefunctions, enabling the embedding of finite dimension coded based torus into the infinite dimensional Hilbert space and beyond. In conclusion, as the use of development tools incorporating a priori models of Code on Graph Data Manifold, characterized by sparsity, block structure (quasi-cyclic and etc), and low accuracy of soft metric representation (few bits quantization), becomes more prevalent, we can anticipate a decrease in reliance on hard computation (power consumption's) GPUs/TPU devices for neural network inference and a significant reduction in their role in DNN training. Instead of blind and inefficient optimization of the architecture of neural networks (ML in general), we will optimize the graph representations of the topology of our data Manifold in order to approximate the data channel with a given reconstruction error (distortion due residual curvature).

\clearpage
\begin{center}
\textbf{\large Appendix A. Data description}
\end{center}
\setcounter{equation}{0}
\setcounter{figure}{0}
\setcounter{table}{0}
\setcounter{page}{1}
\makeatletter
\renewcommand{\theequation}{A\arabic{equation}}
\renewcommand{\thefigure}{A\arabic{figure}}
\renewcommand{\thetable}{A\arabic{table}}

\section{Non-parametric Experiments} \label{sec:data}  
We present a comparison between different matrix factorization methods - TSVD, SF Chord, Regular LDPC PEG with ACE, QC-LDPC Simulated Annealing with EMD optimization - for approximating various types of square matrices using the same number of non-zeros in our paper (as displayed in Table \ref{tab:result}). The sources and statistics of these square matrices are also provided.

Here is a list of datasets with their respective sizes (N) and descriptions:
\begin{itemize}
\item \textbf{AuralSonar} (N = 100): This dataset contains Aural Sonar data from the study by Philips et al. (2006) investigating human ability to distinguish different types of sonar signals by ear. The data has been presented in Chen et al. (2009).

\item \textbf{Protein} (N = 213): This dataset, also presented in Chen et al. (2009), contains the radial basis function (RBF) kernel between 213 proteins.

\item \textbf{Voting} (N = 435): This dataset, also presented in Chen et al. (2009), contains dissimilarities between 435 voting records with 16 scaled voting attributes.

\item \textbf{Yeast} (N = 200): This dataset, from the same repository as AuralSonar in Chen et al. (2009), converts the pairwise Smith-Waterman similarities $s_{ij}$ (Lanckriet et al., 2004; Xu et al., 2014) to dissimilarities by $d_{ij}=\sqrt{s_{ii}+s_{jj}-s_{ji}-s_{ij}}.$ 

Lanckriet, G., Deng, M., Cristianini, N., Jordan, M., Noble, W., 2004. Kernel-
based data fusion and its application to protein function prediction in yeast.
150 Biocomputing 2004, Proceedings of the Pacific Symposium, Hawaii, USA , 300--311.
Xu, W., Hancock, E.R., Wilson, R.C., 2014. Ricci flow embedding for rectifying
non-euclidean dissimilarity data. Pattern Recognition 47, 3709--3725.

\item \textbf{Sawmill} (N = 36): This dataset is a sparse matrix with 124 non-zero entries representing the Sawmill communication network from the Pajek data sets. Data available at http://vlado.fmf.uni-lj.si/pub/networks/data/

\item \textbf{Scotland} (N = 108): This dataset is about corporate interlocks in Scotland from 1904-5. It is a sparse matrix with 644 non-zero entries.

\item \textbf{A99m} (N = 234): This dataset is about the characters and their relations in the long-running German soap opera called `Lindenstrasse'. It is a sparse matrix with 510 non-zero entries.

\item \textbf{Mexican power} (N = 35): This dataset contains the core of the Mexican political elite: the presidents 40 and their closest collaborators. It is a sparse matrix with 117 non-zero entries.

\item \textbf{Strike} (N = 24): This dataset is a social network about informal communication within a sawmill on strike. It is a sparse matrix with 38 non-zero entries.

\item \textbf{Webkb Cornell} (N = 195): This dataset is about citations among 195 publications from Cornell. It is a sparse matrix with 304 non-zero entries. Data available at https://linqs.soe.ucsc.edu/data

\item \textbf{WorldTrade} (N = 80): This dataset is about world trade in miscellaneous manufactures of metal, 1994. It is a sparse matrix with 998 non-zero entries.

\item \textbf{Mesh1e1} (N = 48): This dataset is originally from NASA, collected by Alex Pothen. It is a sparse matrix with 306 non-zero entries.  Data available at   https://sparse.tamu.edu/

\item \textbf{Mesh2e1} (N = 306): This dataset is also originally from NASA, collected by Alex Pothen. It is a sparse matrix with 2018 non-zero entries.

\item \textbf{OrbitRaising} (N = 442): This dataset was from an optimal control problem. It is a sparse matrix with 2906 non-zero entries.

\item \textbf{Shuttle Entry} (N = 560): This dataset was also from an optimal control problem. It is a sparse matrix with 6891 non-zero entries.

\item \textbf{AntiAngiogenesis} (N = 205): This dataset was also from an optimal control problem. It is a sparse matrix with 1783 non-zero entries.

\item \textbf{Phoneme} (N = 256): This dataset contains the covariance matrix of the Phoneme data set accompanied with the Elements of Machine Learning book (Hastie et al., 2001). The original data has 4508 instances of 256 dimensions.  Data available at  https://web.stanford.edu/~hastie/ElemStatLearn/data.html

    Hastie, T., Tibshirani, R., Friedman, J., 2001. The Elements of Statistical Learning. Springer New York Inc.

\item \textbf{MiniBooNE} (N = 50): This dataset contains the covariance matrix of the MiniBooNE particle identification data set in the UCI Repository. The original data has 130064 instances of 50 dimensions. Data available at  https://archive.ics.uci.edu/ml/

\item \textbf{Covertype} (N = 54): This dataset contains the covariance matrix of the Covertype data set in the UCI Repository. The original data has 581012 instances of 54 dimensions.

\item \textbf{Mfeat} (N = 649): This dataset contains the covariance matrix of the Multiple Features data set in the UCI Repository. The original data has 2000 instances of 649 dimensions.

\item \textbf{OptDigits} (N = 64): This dataset contains the covariance matrix of the Optical Recognition of Handwritten Digits data set in the UCI Repository. The original data has 5620 instances of 64 dimensions.

\item \textbf{PenDigits} (N = 16): This dataset contains the covariance matrix of the Pen-Based Recognition of Handwritten Digits data set in the UCI Repository. The original data has 10992 instances of 16 dimensions

\item \textbf{Acoustic} (N = 50): This dataset contains acoustic features from a vehicle sound signal, which can be used to classify the type of vehicle. It is a dataset commonly used in machine learning research, and has been made available by the LIBSVM Classification data collection. Data available at https://www.csie.ntu.edu.tw/~cjlin/libsvmtools/datasets/multiclass.
html

\item \textbf{IJCNN} (N = 22): This dataset contains features from the ijcnn data set, which is also commonly used in machine learning research. It consists of binary classification problems with 22 features, and has been made available by the LIBSVM Classification data collection. Data available at https://www.csie.ntu.edu.tw/~cjlin/libsvmtools/datasets/binary.html

\item \textbf{Spam Ham} (N = 448): This dataset is used for email classification practice, with the goal of determining whether an email is spam or ham. It contains 10000 instances with 448 features.

\item \textbf{TIMIT} (N = 390): This dataset is used in speech recognition research, with the goal of identifying spoken words. It contains 151290 instances, each with 390 features that represent Mel Frequency Cepstral Coefficients (MFCCs) calculated over 10 consecutive 30ms windows. Data available at https://catalog.ldc.upenn.edu/LDC93S1

\item \textbf{Votes} (N = 16): This dataset contains voting records from the US Congress, and is often used as a benchmark for supervised learning algorithms. It contains 435 instances with 16 dimensions representing different attributes related to congressional voting.
\end{itemize}


\newpage



\begin{thebibliography}{Xyz12}


\bibitem{G65} Gallager R.G., Low-density parity-check codes'', IRE Trans. Inform.Theory, vol. IT-8, pp. 21-28, Jan. 1962.
\bibitem{ZyabPins75}  V. V. Zyablov, M. S. Pinsker, “Estimation of the error-correction complexity for Gallager low-density codes”, Problems Inform. Transmission, 11:1 (1975), 18–28 2. 

\bibitem{Tanner81} Tanner R.M., ``A recursive approach to low complexity codes'', IEEE Trans. Inform. Theory, IT-27, pp. 533-547, September 1981.


\bibitem{Marg82}  Margulis, G. Explicit construction of graphs without short cycles and low density codes. 1982. Combinatorica. 2. 71-78. 10.1007/BF02579283. 


\bibitem{Isa81} Isakov S. B. Stability of the Fermi surface: Theory and Application", Comm. Math. Phys. 79 (1981), no. 3, pp. 303-312.



\bibitem{Ru22} Ruslan Khalitov, Tong Yu, Lei Cheng, Zhirong Yang, Sparse factorization of square matrices with application to neural attention modeling, Neural Networks, Volume 152, 2022, Pages 160-168


\bibitem{RR08}   Richardson T.,   Urbanke R.  Modern Coding Theory  Cambridge University Press 2008

\bibitem{RyanShu09} Ryan W., Shu Lin Channel Codes. Classical and Modern Cambridge University Press June 2009



\bibitem{Vasic09} Vasic B., et al. “Trapping set ontology”, 2009 47th Annual Allerton Conf. on Comm. Control and Comp., Monticello, IL, 2009, pp. 1-7.

\bibitem{Ti04} Tian T., Jones C. R. , Villasenor J. D. and Wesel R. D., ”Selective avoidance of cycles in irregular LDPC code construction,” in IEEE Trans. on Commun., vol. 52, no. 8, pp. 1242-1247, Aug. 2004.

\bibitem{DMS03} I. Dumer, D. Micciancio and M. Sudan, "Hardness of approximating the 
minimum distance of a linear code," in IEEE Transactions on Information Theory, vol. 49, no. 1, pp. 22-37, Jan. 2003

\bibitem{VeSuWo22} Velasquez A., Subramani K. , Wojciechowski P. , On the complexity of and solutions to the minimum stopping and trapping set problems,Theoretical Computer Science, Volume 915, 2022, Pages 26-44



\bibitem{Ji05} Jinghu C., R. M. Tanner, C. Jones and Yan Li, ”Improved min-sum decoding algorithms for irregular LDPC codes,” International Symp. on Inform. Theory, ISIT 2005., Adelaide, SA, 2005, pp. 449-453.

\bibitem{Hu05}  Xiao-Yu Hu, Eleftheriou E., Arnold D. M.  "Regular and irregular progressive edge-growth tanner graphs," in IEEE TiT, vol. 51, no. 1, pp. 386-398, Jan. 2005, Implementation available at:  https://github.com/Lcrypto/classic-PEG-



\bibitem{Di16} Diouf M., Declercq D., Fossorier  M., S. Ouya, B. Vasic, "Improved PEG construction of large girth QC-LDPC codes", 9th International Symposium on Turbo Codes and Iterative Information Processing (ISTC), pp. 146-150,
2016.


\bibitem{USA18} Usatyuk V. , Vorobyev I. "Simulated Annealing Method for Construction of High-Girth QC-LDPC Codes," 2018 41st International Conference on Telecommunications and Signal Processing (TSP), Athens,
Greece, 2018, pp. 1-5 Implementation available at: https://github.com/Lcrypto/Simulated-annealing-lifting-QC-LDPC


\bibitem{USA23} Usatyuk V. S. , "Low Error Floor QC-LDPC Codes Construction Using
Modified Cole’s Trapping Sets Enumerating," 2023 25th International Conference on Digital Signal Processing and its Applications (DSPA), Moscow, Russian Federation, 2023 Implementation available at: https://github.com/Lcrypto/trapping-sets-enumeration

\bibitem{USA_UK19} Usatyuk V. S., Egorov S., Svistunov G. Construction of Length and Rate Adaptive MET QC-LDPC Codes by Cyclic Group Decomposition. IEEE East-West Design \& Test Symposium (EWDTS), Batumi, Georgia, 2019, pp. 1-5,

\bibitem{Laz93}  Lazutkin V.F. KAM-theory and semiclassical approximations to eigenfunctions
 Berlin — Heidelberg — New York: Springer, 1993. — 388 p. 



\bibitem{Gromov87} M. Gromov, Hyperbolic groups in: Essays in Group Theory, Springer, pp. 75-265, 1987.



 \bibitem{Ri02}  Richardson T. J.,  Urbanke R. L. "Multi-edge type LDPC codes," in Workshop honoring Prof. Bob McEliece on his 60th birthday, California Institute of Technology, Pasadena, California, 2002.

 \bibitem{Di05}  Divsalar D.,  Jones C.,  Dolinar S. ,  Thorpe J. "Protograph based LDPC codes with minimum distance linearly growing with block size," IEEE Global Telecomm. Conf., 2005., St. Louis, MO, 2005, pp. 5 

 \bibitem{AjiMcE00}   Aji S.M., McEliece R.J. "The generalized distributive law". Information Theory, IEEE Transactions on. 46 (2), Mar 2000, 325-343 


 \bibitem{Waerd67}  B. L. van der Waerden, editor, Sources of Quantum Mechanics, North- Holland, Amsterdam (1967), page 167 

 \bibitem{Kita97} A. Y. Kitaev, “Quantum computations: algorithms and error correction”, Russian Mathematical Surveys 52, 1191 (1997) 

 \bibitem{Kita297}  A. Yu. Kitaev, “Quantum Error Correction with Imperfect Gates”, Quantum Communication, Computing, and Measurement 181 (1997) 

 \bibitem{Kita03} A. Yu. Kitaev, “Fault-tolerant quantum computation by anyons”, Annals of Physics 303, 2 (2003) arXiv:quant-ph/9707021 DOI

 \bibitem{Freed98}  M. H. Freedman and D. A. Meyer, “Projective plane and planar quantum codes”, (1998) arXiv:quant-ph/9810055

 \bibitem{Bravyi98}  S. B. Bravyi and A. Yu. Kitaev, “Quantum codes on a lattice with boundary”, (1998)

  \bibitem{Gleb21} Panteleev P.,  Kalachev G. Degenerate Quantum LDPC Codes With Good Finite Length Performance Quantum 5,  volume 5, page 585 (2021). arXiv:1904.02703v3





 \bibitem{Kout2022}   D. Koutný, L. Ginés, M. Moczała-Dusanowska, S. Höfling, C. Schneider, A. Predojević, and M. Ježek, "Deep Learning of Quantum Entanglement," in Quantum 2.0 Conference and Exhibition, Technical Digest Series (Optica Publishing Group, 2022), paper QTh2A.4.

\bibitem{LiHuTay18} Hao Li, Zheng Xu, Gavin Taylor, Christoph Studer and Tom Goldstein. Visualizing the Loss Landscape of Neural Nets. NIPS, 2018. https://arxiv.org/abs/1712.09913


\bibitem{Choro14} Anna Choromanska, Mikael Henaff, Michael Mathieu, Gérard Ben Arous, Yann LeCun, The Loss Surfaces of Multilayer Networks. International Conference on Artificial Intelligence and Statistics, 2014





 

 \bibitem{KoB16} Korb M.,  Blanksby A. ”Non-binary LDPC codes over finite division near rings,” Intern. Conf. on Telecommunications
(ICT), 2016, pp. 1-7.

 \bibitem{For01}  Forney G. D.  "Codes on graphs: normal realizations," in IEEE Transactions on Information
Theory, vol. 47, no. 2, pp. 520-548, Feb 2001


 \bibitem{Fe14} Ferris A.J., Poulin D. Tensor Networks and Quantum Error Correction;Phys. Rev. Lett. 113, 030501 – Published 16 July 2014 

 \bibitem{Me18} Meshulam R. “Graph codes and local systems,” Mar. 2018. Available: https://arxiv.org/abs/1803.05643

  \bibitem{For18} Forney G. D. "Codes on Graphs: Models for Elementary Algebraic Topology and Statistical Physics," in IEEE Transactions on Information Theory, vol. 64, no. 12, pp. 7465-7487, 2018 

\bibitem{Fer13}  Ferris A., Poulin D. Branching MERA codes: a natural extension of polar codes. 2013. https://arxiv.org/abs/1312.4575



  \bibitem{Fos15} Fossorier M.  "Polar Codes: Graph Representation and Duality," in IEEE Communications Letters, vol. 19, no. 9, pp. 1484-1487, Sept. 2015



\bibitem{Ji06} Jiang F., Psota E.,   Perez L. C. "The Generator and Parity-Check Matrices of Turbo Codes," 2006 40th Annual Conference on Information Sciences and Systems, Princeton, NJ, USA, 2006, pp. 1451-1454


\bibitem{USA_UK20} V. Usatyuk, "Wireless Channels Topology Invariant as Mathematical Foundation of Neural Network Channel Estimation Transfer Learning Properties," 2020 TSP, Milan, Italy, 2020, pp. 105-111
\bibitem{USA_UK23}
V. S. Usatyuk and S. I. Egorov, "2D/3D ResNet Deep Neural Network for 4G and 5G NR Wireless Channel Estimation," 2023 25th International Conference on Digital Signal Processing and its Applications (DSPA), Moscow, Russian Federation, 2023, pp. 1-4, doi: 10.1109/DSPA57594.2023.10113403.



\bibitem{Ro19}  Robeva E., Seigal A., Duality of Graphical Models and Tensor Networks, Inform. and Inference: A Journal of the IMA, V. 8(2), 2019, p. 273-288]



\bibitem{Am09}  A. Amraoui, A. Montanari, T. Richardson, and R. Urbanke, “Finite-length scaling for iteratively decoded LDPC ensembles,” IEEE Trans. Inform. Theory, vol. 55, no. 2, pp. 473–498, Feb. 2009.

\bibitem{Sti16} M. Stinner, L. Barletta and P. M. Olmos, "Finite-length scaling based on Belief Propagation for spatially coupled LDPC codes,"  IEEE ISIT, Barcelona, Spain, 2016, pp. 2109-2113






\bibitem{Cole0Wi}C. A. Cole, et al., "Regular {4, 8} LDPC Codes and Their Low error Floors," MILCOM, 2006, pp. 1-7


\bibitem{Cole06}C. A. Cole, et al. "Analysis and Design of Moderate Length Regular LDPC Codes with Low Error Floors,"  40th CISS, 2006, pp. 823-828

 \bibitem{Cole08} C. A. Cole, "Error floor analysis for an ensemble of easily implementable irregular (2048, 1024) LDPC codes," MILCOM08, pp. 1-5








\bibitem{CheSte07} M. Chertkov,  M. Stepanov, "Searching for low weight pseudo-codewords,"  Inform. Theory and Appl.  Workshop, 2007, pp. 94-100

\bibitem{CheSte07_2} M. Chertkov and M. Stepanov, "Pseudo-codeword Landscape," 2007 IEEE International Symp. on Information Theory, 2007, pp. 1546-1550



\bibitem{CCSV09} S.K. Chilappagari, et al., Instanton-based techniques for analysis and reduction of error floors of LDPC codes, IEEE J. Sel. Areas Commun. Special Issue in Capacity Approaching Codes 27(6) (2009) 855 –865.


\bibitem{CheSte11} M. Chertkov, M. Stepanov, "Polytope of correct (linear programming) decoding and low-weight pseudo-codewords," ISIT, 2011, pp. 1648-1652











\bibitem{VasCNP09} Vasic B., et al. ``Trapping set ontology'', 47th Annual Allerton Conf. Comm., Control Comp, 2009, pp. 1-7.


\bibitem{KaBa12} M. Karimi, A. H. Banihashemi, "Efficient Algorithm for Finding Dominant Trapping Sets of LDPC Codes," in IEEE Transactions on Information Theory, vol. 58, no. 11, pp. 6942-6958, 2012















\bibitem{RaDeVa20}N. Raveendran, et al "A Sub-Graph Expansion-Contraction Method for Error Floor Computation," in IEEE Transactions on Communications, vol. 68, no. 7, pp. 3984-3995,  2020



\bibitem{AbDeDiRy10} S. Abu-Surra, et al., “Trapping set enumerators for specific LDPC codes,” in Proc. Inf. Theory Appl. Workshop, 2010, pp. 1–5.



\bibitem{ToBa14} S. Tolouei, A. H. Banihashemi, "Fast and Accurate Error Floor Estimation of Quantized Iterative Decoders for Variable-Regular LDPC Codes," in IEEE Communications Letters, vol. 18(8), pp. 1283-1286, 2014

\bibitem{BuSi14}B. K. Butler, Siegel P. H. "Error Floor Approximation for LDPC Codes in the AWGN Channel," in ITIT, vol. 60, no. 12, pp. 7416-7441, 2014 


\bibitem{VeSuDra18} Velasquez A., et al Finding Minimum Stopping and Trapping Sets: An Integer Linear Programming Approach. ISCO 2018. pp. 402–415




\bibitem{PaShiChu12}   H. Park, et al "On the combinatorial substructures of LDPC codes causing error floors in the AWGN channel,"  ICTC, 2012, pp. 420-42


\bibitem{KiMyJe15} K. Kim, S. Myung, H. Jeong, "Lowering error floors by removing dominant trapping sets of low-density parity-check codes for broadcasting systems," 2015 IEEE Intern. Symp. on BMSB, 2015, pp. 1-3


\bibitem{KaBa20} B. Karimi, A. H. Banihashemi, "Construction of QC LDPC Codes With Low Error Floor by Efficient Systematic Search and Elimination of Trapping Sets,"  IEEE Trans. on Commun., v. 68(2), pp. 697-712, 2020




\bibitem{Faba21} A. Farsiabi, A. H. Banihashemi, "Error Floor Analysis of LDPC Row Layered Decoders," in ITIT, vol. 67(9), pp. 5804-5826, 2021 

\bibitem{RoYt09}E. Rosnes and Ø. Ytrehus, "An Efficient Algorithm to Find All Small-Size Stopping Sets of Low-Density Parity-Check Matrices," in IEEE Transactions on Information Theory, vol. 55(9), pp. 4167-4178, 2009




	

\bibitem{Pe88} Pearl, J., Probabilistic Reasoning in Intelligent Systems: Networks of Plausible Inference, Morgan Kaufmann Publishers, Inc., 1988


\bibitem{Lu97}   M. Luby, M. Mitzenmacher, A. Shokrollahi, D. Spielman, and V. Stemann. “Practical loss-resilient codes,” in Proc. the 29th annual ACM Symposium on Theory of Computing, 1997, pp. 150-159

\bibitem{Ri06}  T. Richardson, "Error-floors of LDPC codes", Proc. 41st Annu. Allerton Conf., pp. 1426-1435, Oct. 2003. Cole, et al. A general method for finding low error rates of LDPC codes CoRR [Online]. Available: arxiv.org/abs/cs/0605051. 2006

\bibitem{LRA21}  Tay Yi, Dehghani Mostafa, Abnar Samira, Shen Yikang, Bahri Dara, Pham Philip, Rao Jinfeng, Yang Liu, Ruder Sebastian, and Metzler Donald. 2021. Long range arena: A benchmark for efficient transformers. Proc. of ICLR (2021).


\bibitem{Ma22}  Ma X., Zhou, et al  Mega: Moving Average Equipped Gated Attention. 2022 10.48550/arXiv.2209.10655.

\bibitem{Li05}   Liva G. et al.'s "Simple Reconfigurable Low-Density Parity-Check Codes" in IEEE COMMUNICATIONS LETTERS, VOL. 9, NO. 2, FEBRUARY 2005

\bibitem{Kha22}  Khalitov R., et al.. ChordMixer: A Scalable Neural Attention Model for Sequences with Different Lengths.   Neural Networks, Volume 152, 2022  2022 10.48550/arXiv.2206.05852. 


\bibitem{Stoica2001}  Stoica Ion,  Morris Robert Karger, David, Kaashoek, M., Balakrishnan, Hari. . Chord: A Scalable Peer-to-Peer Lookup Service for Internet Applications. ACM SIGCOMM Computer Communication Review, vol. 31. 31. 2001. 10.1145/964723.383071. 


\bibitem{Ma07}  Malema, G., Liebelt, M. High Girth Column-Weight-Two LDPC Codes Based on Distance Graphs. J Wireless Com Network 2007, https://doi.org/10.1155/2007/48158 


\bibitem{Lei23}  Lei Cheng, Ruslan Khalitov, Tong Yu, Jing Zhang, Zhirong Yang, Classification of long sequential data using circular dilated convolutional neural networks, Neurocomputing,Volume 518,2023, Pages 50-59,

\bibitem{Xi21} X. Xiao, N. Raveendran, B. Vasić, S. Lin and R. Tandon, "FAID Diversity via Neural Networks," 2021 11th International Symposium on Topics in Coding (ISTC), Montreal, QC, Canada, 2021, pp. 1-5



\bibitem{Rao45} Rao C. R. "Information and Accuracy Attainable in the Estimation of Statistical Parameters". Bulletin of the Calcutta Mathematical Society. 1945 37: 81–91. Reprinted in Breakthroughs in Statistics. Springer. 1992. pp. 235–247.



\bibitem{Rao83}   Shun'ichi A. "A foundation of information geometry". Electronics and Communications in Japan. 66 (6), 1983: 1–10. doi:10.1002/ecja.4400660602.
\bibitem{Rao00} Amari, Shun'ichi; Nagaoka, Hiroshi  Methods of Information Geometry. Translations of Mathematical Monographs. Vol. 191. American Mathematical Society. 2000




\bibitem{Egu18} Eguchi, S. On the launch of the journal Information Geometry. Info. Geo. 1, 1 (2018). 

\bibitem{Ra18} Rao C.R. Congratulatory message. Info. Geo. 1, 3 (2018). 

 \bibitem{Kha19} Knauf, A. Nihat Ay, Jürgen Jost, Hông Vân Lê, Lorenz Schwachhöfer: “Information Geometry”. Jahresber. Dtsch. Math. Ver. 121, 297–302 (2019). 

 
 \bibitem{Fra18} Nielsen F. . "An Elementary Introduction to Information Geometry". Entropy. 22 (10). 2018
 
\bibitem{FraBar21} Ed. Frank Nielsen, Frédéric Barbaresco Geometric Science of Information  5th International Conference, GSI 2021, Paris, France, July 21–23, 2021, Proceedings Lecture Notes in Computer Science

\bibitem{NielBo10}  Boissonnat, JD., Nielsen, F. , Nock, R. Bregman Voronoi Diagrams. Discrete Comput Geom 44, 281–307 (2010). 



\bibitem{Pe18} Pennec Xavier. “Barycentric Subspace Analysis on Manifolds.” The Annals of Statistics, vol. 46, no. 6A, 2018, pp. 2711–46. JSTOR




\bibitem{Gu19} R. Brüel Gabrielsson,  Carlsson G. "Exposition and Interpretation of the Topology of Neural Networks," 18th IEEE Intern. Conference On Machine Learning And Applications (ICMLA), Boca Raton, FL, USA, 2019, pp. 1069-1076






\bibitem{EdAn75}Edwards  S. F. und Anderson P.W. Theory of spin glasses. Cavendish Laboratory, Cambridge. UK. Received 14 October 1974. in final form 13 February 1975


\bibitem{Ge96} Gérard Ben Arous. Raphaël Cerf. "Metastability of the Three Dimensional Ising Model on a Torus at Very Low Temperatures." Electron. J. Probab. 1 1 - 55, 1996. 

\bibitem{Me84} M. Mézard, G. Parisi, N. Sourlas, G. Toulouse, M.A. Virasoro, Replica symmetry breaking and the nature of the spin-glass phase, J. de Physique 45 (1984) 843

\bibitem{Sour89} Sourlas, N. Spin-glass models as error-correcting codes. Nature, 1989, 339, pp. 693–695 



\bibitem{Sour94}Sourlas N. (1994). Statistical Mechanics and Error-Correcting Codes. In: Grassberger, P., Nadal, JP. (eds) From Statistical Physics to Statistical Inference and Back. NATO ASI Series, vol 428. Springer, Dordrecht. 


\bibitem{Sourlas07}  Sourlas, N.. (2007). Spin Glasses, Error-Correcting Codes and Finite-Temperature Decoding. EPL (Europhysics Letters). 25. 159. 
\bibitem{Monta09} Mézard M.,  Montanari A. Information, Physics, and Computation Oxford Graduate Texts, 2009, 569 pages



\bibitem{Yu22}  Yurtsever, A., Birdal, T., Golyanik, V. (2022). Q-FW: A Hybrid Classical-Quantum Frank-Wolfe for Quadratic Binary Optimization. In: Avidan, S., Brostow, G., Cissé, M., Farinella, G.M., Hassner, T. (eds) Computer Vision – ECCV 2022. Lecture Notes in Computer Science, vol 13683. Springer, Cham. 


\bibitem{Ble23}   Blekos K. et al. A Review on Quantum Approximate Optimization Algorithm and its Variants arXiv:2306.09198v2 2023

\bibitem{Pont23}  Pontus Vikstål, Laura García-Álvarez, Shruti Puri, and Giulia Ferrini. Quantum Approximate
Optimization Algorithm with Cat Qubits, May 2023. https://arxiv.org/abs/2305.05556 


\bibitem{Lu23}   Lucas A. Ising formulations of many NP problems arXiv:1302.5843v3 2023



\bibitem{Ta20} Tabi X. et al. Quantum Optimization for the Graph Coloring Problem with Space-Efficient Embedding, arXiv:2009.07314v1 


\bibitem{Glo19}  Glover et al., Quantum Bridge Analytics II: Network Optimization and Combinatorial Chaining for Asset Exchange arXiv:1911.03036v2  2019



\bibitem{Mo22} Mohseni, N., McMahon, P.L., Byrnes, T. Ising machines as hardware solvers of combinatorial optimization problems. Nat Rev Phys 4, 363–379 (2022).


\bibitem{Da19} Date, P., Patton, R., Schuman, C. et al. Efficiently embedding QUBO problems on adiabatic quantum computers. Quantum Inf Process 18, 117 (2019). https://doi.org/10.1007/s11128-019-2236-3


\bibitem{Fa22}Farhi E. et al. The Quantum Approximate Optimization Algorithm and the Sherrington-Kirkpatrick Model at Infinite Size   Quantum 6, 759 (2022). 2022-07-07, volume 6, page 759 arXiv:1910.08187v4

\bibitem{Wibb96} Wiberg N., ``Codes and Decoding on General Graphs'', Ph.D. dissertation, Linkoping University, 1996.


 \bibitem{Ta001}R. M. Tanner, D. Sridhara, and T. E. Fuja, “A class of group-structured LDPC codes,” in Proc. Int. Symp. Communication Theory and Applications, Ambleside, U.K., July 2001


\bibitem{Me87} Mezard M ., Parisi G., Virasoro M., "Spin Glass Theory and Beyond", World Scientific, 1987





\bibitem{Bai17} Baik, Jinho , Lee, Ji. Free energy of bipartite spherical Sherrington--Kirkpatrick model. Annales de l'Institut Henri Poincaré, Probabilités et Statistiques. 56.  201710.1214/20-AIHP1062. 


\bibitem{Spa04} Spavieri G., Gillies G. T., Rodriguez M. Physical implications of Coulomb's Law Metrologia, Volume 41, Issue 5, pp. S159-S170 (2004).



\bibitem{KZ90} Lagarias, J. C.; Lenstra, H. W. Jr.; and Schnorr, C. P. "Korkin-Zolotarev Bases and Successive Minima of a Lattice and Its Reciprocal Lattice." Combinatorica 10, 333-348, 1990.

\bibitem{DeRy71} B. N. Delone, S. S. Ryshkov, “Extremal problems of the theory of positive quadratic forms”, Proc. Steklov Inst. Math., 112 (1971), 211–231

\bibitem{Kor04} N. M. Korobov, Theoretic-Numerical Methods in Approximate Analysis, 2nd. revised and extended edition, MTsNMO, Moscow, 2004, p. 288  





\bibitem{Sc22} Luis Scoccola, et. al Toroidal Coordinates: Decorrelating Circular Coordinates With Lattice Reduction To appear in proceedings of 39th International Symposium on Computational Geometry 2022 arXiv:2212.07201


\bibitem{Ta04} R. M. Tanner, D. Sridhara, A. Sridharan, T. E. Fuja and D. J. Costello, "LDPC block and convolutional codes based on circulant matrices," in IEEE Transactions on Information Theory, vol. 50, no. 12, pp. 2966-2984, 2004

\bibitem{RiUr01}  Richardson T., R. Urbanke. Efficient encoding of low-density parity-check codes. In IEEE Trans. Inform. Theory,  volume 47, pages 638–656,2001

\bibitem{DivMcel98}  D. Divsalar, H. Jin, and R. J. McEliece. "Coding theorems for ‘turbo-like’ codes." Proc. 36th Allerton Conf. on Communication, Control and Computing, Allerton, Illinois, Sept. 1998, pp. 201–210.

\bibitem{Sarah10}   D. F. Hayes, S. J. Johnson and S. R. Weller, "Irregular repeat-accumulate-like codes with improved error floor performance," 2010 IEEE Information Theory Workshop, Dublin, Ireland, 2010, pp. 1-5, 


\bibitem{Sch91} Schmidhuber J. Learning to control fast-weight memories: An alternative to recurrent nets. Technical Report FKI-147-91, Institut für Informatik, Technische Universität München, 26 March 1991. https://people.idsia.ch/~juergen/FKI-147-91ocr.pdf
illustration https://people.idsia.ch/~juergen/fast-weight-programmer-1991-transformer.html


\bibitem{CamMez23} Camilli F., Mezard M. Matrix factorization with neural networks, American Physical Society, Phys. Rev. E, v. 107(6), 2023.

\bibitem{CamMez23_etend}  Camilli F., Mezard M. The Decimation Scheme for Symmetric Matrix Factorization. 2023,	arXiv:2307.16564

\bibitem{WuFa19}  Wu Felix, Fan Angela, Baevski Alexei, Dauphin Yann, Auli Michael. Pay Less Attention with Lightweight and Dynamic Convolutions. (2019).arXiv:1901.10430


\bibitem{LiHu21}  Li Duo,  Hu Jie, Wang Changhu, Li Xiangtai, She, Qi, Zhu, Lei,  Zhang, Tong, Chen, Qifeng. Involution: Inverting the Inherence of Convolution for Visual Recognition.  (2021).	arXiv:2103.06255


\bibitem{Fos01} Y. Kou, S. Lin and M. P. C. Fossorier, "Low-density parity-check codes based on finite geometries: a rediscovery and new results," in~\textit{IEEE Transactions on Information Theory}, vol. 47, no. 7, pp. 2711-2736, Nov. 2001

\bibitem{Xio23} Xiong R., Nie J., Brantly S. Hays P., Sailus R.,  Watanabe K., Taniguchi T. Tongay S.,  Jin C. Correlated insulator of excitons in WSe2/WS2 moiré superlattices. Science (New York, N.Y.). 380. 2023  

\bibitem{Sege76} L. Segerlind, Applied Finite Element Analysis, John Willey and sons, New Yorl 1976 pp 47-53







\bibitem{ChanUrRic01}  Changyan Di, R. Urbanke and T. Richardson, "Weight distributions: how deviant can you be?," Proceedings. 2001 IEEE International Symposium on Information Theory, Washington, DC, USA, 2001, pp. 50-




\bibitem{Lore21} Lorenzo Dall'Amico et al Nishimori meets Bethe: a spectral method for node classification in sparse weighted graphs J. Stat. Mech. (2021) 093405

\bibitem{Nishimori80}  H. Nishimori, J. Phys. C: Solid State Phys. 13(21), 4071 (1980)
\bibitem{Nishimori81}   H. Nishimori, Prog. Theor. Phys. 66(4), 1169 (1981)
\bibitem{Nishimori86}  H. Nishimori, Prog. Theor. Phys. 76, 305~1986


\bibitem{Yed03} J. S. Yedidia, W. T. Freeman, and Y. Weiss, “Understanding belief propagation and its generalizations,” Exploring artificial intelligence in the new millennium, pp. 239–269, 2003
\bibitem{Von10}  P. O. Vontobel, "The Bethe permanent of a non-negative matrix," 2010 48th Annual Allerton Conference on Communication, Control, and Computing (Allerton), Monticello, IL, USA, 2010, pp. 341-346, doi: 10.1109/ALLERTON.2010.5706926.
\bibitem{Smarandache13}  R. Smarandache, 
 "Pseudocodewords from Bethe permanents," 2013 IEEE International Symposium on Information Theory, Istanbul, Turkey, 2013, pp. 2059-2063
\bibitem{Von20}   S. Vatedka and P. Vontobel, "Modified Bethe Permanent of a Nonnegative Matrix," 2020 International Conference on Signal Processing and Communications (SPCOM), Bangalore, India, 2020, pp. 1-5, doi: 10.1109/SPCOM50965.2020.9179492.

\bibitem{Von23} Y. Huang and P. O. Vontobel, "Bounding the Permanent of a Non-negative Matrix via its Degree- M Bethe and Sinkhorn Permanents," 2023 IEEE International Symposium on Information Theory (ISIT), Taipei, Taiwan, 2023, pp. 2774-2779, doi: 10.1109/ISIT54713.2023.10206878.





\bibitem{Li18}  Li H., et al. Visualizing the Loss Landscape of Neural Nets NIPS 2018 arXiv:1910.08187v4


 

\bibitem{Per02} Perelman, Grigori. "The entropy formula for the Ricci flow and its geometric applications" . 2002, arXiv:math.DG/0211159.

\bibitem{Per03}  Perelman Grigori  "Ricci flow with surgery on three-manifolds". 2003, arXiv:math.DG/0303109.
\bibitem{Per031}   Perelman  Grigori "Finite extinction time for the solutions to the Ricci flow on certain three-manifolds". 2003, arXiv:math.DG/0307245
 
\bibitem{Be2010} L. Bessieres, G. Besson, M. Boileau, S. Maillot, J. Porti, 'Geometrisation of 3-manifolds', EMS Tracts in Mathematics, volume 13. European Mathematical Society, Zurich, 2010

\bibitem{Fossorier04} Fossorier  M.P.C., "Quasi-cyclic low-density parity-check codes from circulant permutation matrices", ITIT, vol. 50(8), pp. 1788--1793, 2004.

\bibitem{Gu18} Rickard Bruel Gabrielsson and Gunnar Carlsson "Exposition and Interpretation of the Topology of Neural Networks" arXiv:1810.03234, 2018

\bibitem{CaGa20}  Carlsson, G., Gabrielsson, R.B. Topological Approaches to Deep Learning. In: Baas, N., Carlsson, G., Quick, G., Szymik, M., Thaule, M. (eds) Topological Data Analysis. Abel Symposia, vol 15. 2020 Springer


\bibitem{Ege2009} J.H. Eggert, et al.  "Melting temperature of diamond at ultrahigh pressure". Nature Physics. 6: 40–43. Nov 8, 2009. 


\bibitem{Frenkel04}    Frenkel E. Lectures on the Langlands Program and Conformal Field Theory Les Houches School "Number Theory and Physics" in March of 2003  arXiv:hep-th/0512172


\bibitem{Knapp04} Bernstein, J.; Gelbart, S., An Introduction to the Langlands Program Birkhäuser Boston 2004,  281 pages



\bibitem{AscFeZu08} Aschieri P., Ferrara S., Zumino B. Three lectures on electric-magnetic duality, Jul. 2008, Published in: SFIN A 1 (2009), pp. 1-42


\bibitem{Rob14}Robinson M. Topological Signal Processing. Springer, 2014, 208 p.


\bibitem{RobGhris12} Robinson M.,  Ghrist R., “Topological localization via signals of opportunity,” IEEE Trans. Signal Processing, 60(5), 2362-2373, 2012.

\bibitem{Ghris17} R. Ghrist, “Homological Algebra and Data", in The Mathematics of Data, IAS/Park City Mathematics, V. 25, 2017, 273-325.





\bibitem{SilvaCarlson} de Silva, G. Carlsson. Topological estimation using witness complexes. In Eurographics Symposium on Point-Based Graphics, 2004. 



\bibitem{Zomo10} Zomorodian A. Fast construction of the Vietoris-Rips complex.  Computers \& Graphics, 2010, 34(3), pp. 263–271  

\bibitem{Zig}Jimenez Felstrom  A., Zigangirov K. S. "Time-varying periodic convolutional codes with low-density parity-check matrix," in IEEE Transactions on Information Theory, vol. 45, no. 6, pp. 2181-2191,1999

\bibitem{CalShor96} Robert Calderbank ,  Peter Shor . "Good quantum error-correcting codes exist". Physical Review A. 54 (2): 1098–1105. 1996 arXiv:quant-ph/9512032. doi:10.1103/PhysRevA.54.1098.

\bibitem{Kitaev03}  Kitaev A. Yu.  “Fault-tolerant quantum computation by anyons”, Annals of Physics 303, 2 (2003) arXiv:quant-ph/9707021 DOI


\bibitem{Ta01} R. M. Tanner, "Minimum-distance bounds by graph analysis," in IEEE Transactions on Information Theory, vol. 47, no. 2, pp. 808-821, Feb 2001

\bibitem{Mi05}  Min-Ho Shin, Joon-Sung Kim and Hong-Yeop Song, "Generalization of Tanner's minimum distance bounds for LDPC codes," in IEEE Communications Letters, vol. 9, no. 3, pp. 240-242, March 2005

\bibitem{Vo10} P. O. Vontobel, "The Bethe permanent of a non-negative matrix," 2010 48th Annual Allerton Conference on Communication, Control, and Computing (Allerton), Monticello, IL, USA, 2010, pp. 341-346

\bibitem{Sma13} R. Smarandache, "Pseudocodewords from Bethe permanents," 2013 IEEE International Symposium on Information Theory, Istanbul, Turkey, 2013, pp. 2059-2063, doi: 10.1109/ISIT.2013.6620588.

\bibitem{Mac01} D. J. MacKay and M. C. Davey, “Evaluation of Gallager codes for short block length and high rate applications,” Proc. of the IMA Workshop on Codes, System and Graphical Models, 1999. Springer-Verlag 2001, pp. 113–130

\bibitem{Sma12} R. Smarandache and P. O. Vontobel, “Quasi-cyclic LDPC codes: Influence of proto- and Tanner-graph structure on minimum Hamming distance upper bounds,” IEEE Trans. Inf. Theory, vol. 58, no. 2, pp. 585–607, Feb. 2012


\bibitem{CheTan05} Chen J., Tanner R. M., Jones C., Li Y. "Improved min-sum decoding algorithms for irregular LDPC codes," ISIT, 2005, pp. 449-453


\bibitem{Lu06}  Luca Donetti et al. Optimal network topologies: expanders, cages, Ramanujan graphs, entangled networks and all that  J. Stat. Mech. (2006)  arXiv:cond-mat/0605565v2.


\bibitem{Halko09} Halko Nathan, Martinsson Per-Gunnar, Tropp Joel Finding Structure with Randomness: Stochastic Algorithms for Constructing Approximate matrix Decompositions. Preprint. 2009 arXiv:0909.4061


\bibitem{McK96}  MacKay D.J.C., Neal R.M., ``Near Shannon limit performance of low-density parity-check codes'', Elec. Lett., vol. 32, pp. 1645-1646, 1996.
\bibitem{SipSpiek96} Sipser M.,  Spielman D.A., Expander codes, IEEE Trans. Inform. Theory, vol. 42, pp. 1710-1722, Nov. 1996.

\bibitem{Am} Amari, Si. Information geometry. Jpn. J. Math. 16, 1–48 (2021).

\bibitem{Ri00} T. Richardson, "The geometry of turbo-decoding dynamics," in IEEE Transactions on Information Theory, vol. 46, no. 1, pp. 9-23, Jan. 2000


\bibitem{ike05}  S. Ikeda, "Information geometry of turbo and LDPC codes," Proceedings. ICASSP '05, pp. 1029-1032 Vol. 5

\bibitem{Ma05} J. H. Manton, "On the role of differential geometry in signal processing," Proceedings. (ICASSP '05)., 2005, pp. v/1021-v/1024 Vol. 5

\bibitem{Ko06} L. Kocarev, et al "Nonlinear dynamics of iterative decoding systems: analysis and applications," ITIT, vol. 52, no. 4, pp. 1366-1384, 2006


\bibitem{NiCr17} Nielsen F., Critchley F., Dodson C. T. J. Computational Information Geometry For Image and Signal Processing, Springer, 2017, pp 299


\bibitem{LiTeRo17}Lin, H.W., Tegmark, M., Rolnick, D. Why Does Deep and Cheap Learning Work So Well?. J Stat Phys 168, 1223–1247 (2017).


\bibitem{Di2001WeightDO} Di C., Urbanke, R., Richardson T. (2001). Weight distributions: how deviant can you be?. ISIT 2001.935913. 



\bibitem{AmMoRiUr09}Amraoui A.,  et al , "Finit-Length Scaling for Iteratievly Decoded LDPC 
Ensembles," in IEEE ITIT, vol. 55, no. 2, pp. 473-498, Feb. 2009.


\bibitem{ScZa10}Schlegel  C.,Zhang S. On the Dynamics of the Error Floor Behavior in (Regular) LDPC Codes, ITIT, vol. 56(7), pp. 3248-3264, 2010 
\bibitem{Richardson2003ErrorFO} Richardson T., Error-floors of LDPC codes, Proc. 41st Annu. All. Conf., pp. 1426-1435, Oct. 2003.

\bibitem{ColeWiHaGi06}  Cole, et al. A general method for finding low error rates of LDPC codes CoRR [Online]. Available: arxiv.org/abs/cs/0605051. 2006

\bibitem{Jer84} M. Jeruchim, “Techniques for estimating the bit error rate in the simulation of digital communication systems,” IEEE Journal on selected areas in communications, vol. 2, no. 1, pp. 153–170, 1984

\bibitem{Yedi00} Yedidia  J .S., Freeman W., Weiss Y. Generalized Belief Propagation Advances in Neural Information Processing Systems. MIT Press,13, 2000

\bibitem{Yedi05} J. S. Yedidia, W. T. Freeman and Y. Weiss, "Constructing free-energy approximations and generalized belief propagation algorithms," in IEEE Transactions on Information Theory, vol. 51, no. 7, pp. 2282-2312, 2005



\bibitem{VoKo03}R. Koetter, P. O. Vontobel, “Graph covers and iterative decoding of finite-length codes,” 3d ISTC and Related
Topics, pp. 75–82, 2003.


\bibitem{Voko11} Vontobel P. O. The Bethe Permanent of a Non-Negative Matrix 2011
 arXiv:1107.4196v3 

 \bibitem{Sm11} Smarandache R. Pseudocodewords from Bethe Permanents  arXiv:1112.4625v2 2011


 \bibitem{MiSe19}A. Minja, Senk V., "Quasi-Analytical Simulation Method for Estimating the Error Probability of Star Domain Decoders," ITC, vol. 67, no. 5, pp. 3101-3113, May 2019



\bibitem{USAVO18} V. Usatyuk, I. Vorobyev, "Simulated Annealing Method for Construction of High-Girth QC-LDPC Codes," 41st Intern. Conf. TSP, 2018, pp. 1-5









\bibitem{Zha-Sch13}S. Zhang, C. Schlegel, "Controlling the Error Floor in LDPC Decoding," in IEEE Transactions on Comm., vol. 61(9), 2013, pp. 3566-3575

\bibitem{CTanJoLi05} Chen J., Tanner R. M., Jones C., Li Y. "Improved min-sum decoding algorithms for irregular LDPC codes," ISIT, 2005, pp. 449-453







 \bibitem{MCGregMi07} McGregor A., Milenkovic O. On the Hardness of Approximating Stopping and Trapping Sets in LDPC Codes// IEEE ITW, 2007, pp. 248-253




\bibitem{TiJoViWe}Tao Tian, et al, "Selective avoidance of cycles in irregular LDPC code construction," Trans. on Comm., vol. 52, no. 8, pp. 1242-1247, 2004
 \bibitem{EmdSpectrum}Exact cycle extrinsic message degree Spectrum calculation tool, EMD-
Spectrum https://github.com/Lcrypto/EMD-Spectrum-LDPC






\bibitem{BeVa} Berrou  C.,  Vaton S. "Computing the minimum distance of linear codes by the error impulse method", Proc. IEEE ISIT, July 2002.


 \bibitem{HuFoEl}  Xiao-Yu Hu,et al. On the computation of the minimum distance of low-density parity-check codes," 2004 IEEE Intern. Conf. on Comm. , 2004, v. 2 pp. 767-771
 






\bibitem{Chr2000} Williams C.,  Seeger M. Using the Nyström Method to Speed Up Kernel Machines Part of Advances in Neural Information Processing Systems 13 (NIPS 2000)
 
 
\bibitem{MichaMa09}  Mahoney M.W.,  Drineas P.  CUR matrix decompositions for improved data analysis Ed. by Jon Kleinberg, Cornell University, Ithaca, NY, 106 (3) 697-702,  January 20, 2009 



\bibitem{Ech1936} Eckart, C., Young G.  The Approximation of One Matrix by Another of Lower Rank. Psychometrika, 1, 211-218 1936



\bibitem{Lee99}  Lee D., Seung H. Learning the parts of objects by non-negative matrix factorization. 1999. Nature 401, 788–791 

\bibitem{MartSla2013} Martin Slawski, Matthias Hein, Pavlo Lutsik  Matrix factorization with binary components Part of Advances in Neural Information Processing Systems 26 (NIPS 2013)

\bibitem{Zhang2010}  Zhang, ZY., Li, T., Ding, C. et al. Binary matrix factorization for analyzing gene expression data. Data Min Knowl Disc 20, 2010 28–52 


\bibitem{Donoho06}  D. L. Donoho, "Compressed sensing," in IEEE Transactions on Information Theory, vol. 52, no. 4, pp. 1289-1306, April 2006, doi: 10.1109/TIT.2006.871582.

\bibitem{Kim2010} B. -H. Kim, A. Yedla, H. D. Pfister, "IMP: A message-passing algorithm for matrix completion," 2010 6th International Symposium on Turbo Codes  Iterative Information Processing, Brest, France, 2010, pp. 462-466, 

\bibitem{Simon2006}  Simon Funk’s Netflix Recommendation System using Funk Factorization
https://sifter.org/~simon/journal/20061211.html





\bibitem{Bro1970}   Broyden, C. G. (1970), "The convergence of a class of double-rank minimization algorithms", Journal of the Institute of Mathematics and Its Applications, 6: 76–90

\bibitem{Fle1970}   Fletcher, R. (1970), "A New Approach to Variable Metric Algorithms", Computer Journal, 13 (3): 317–322

\bibitem{Gold1970}  Goldfarb, D. (1970), "A Family of Variable Metric Updates Derived by Variational Means", Mathematics of Computation, 24 (109): 23–26, doi:10.1090/S0025-5718-1970-0258249-6
\bibitem{Shanno1970}   Shanno, David F. (July 1970), "Conditioning of quasi-Newton methods for function minimization", Mathematics of Computation, 24 (111): 647–656, doi:10.1090/S0025-5718-1970-0274029-X, MR 0274029

\bibitem{USA_UK_TopoML23} Usatyuk V. Matlab Platform for Sparse Factorization using Hyperbolic Torical and Spherical Topology LDPC codes and MET QC-LDPC codes, PEG+ACE and QC-LDPC MET SA+EMD https://github.com/Lcrypto/Classical-and-Quantum-Topology-ML-toric-spherical



\end{thebibliography}
\end{document}